\documentclass{article}
\pdfoutput=1
\usepackage{arxiv}

\usepackage{booktabs}   
\usepackage{subcaption} 

\usepackage{thmtools} 
\usepackage{thm-restate}

\usepackage{natbib}
\usepackage{color}
\usepackage{stmaryrd}
\usepackage{graphicx}
\usepackage{amsmath}
\usepackage{amssymb}
\usepackage{mathtools}
\usepackage{mathpartir}
\usepackage{bbm}
\usepackage{mathrsfs}
\usepackage{amsthm}
\usepackage{enumitem}
\usepackage{multirow}
\usepackage[T1]{fontenc}
\usepackage{url}
\usepackage{hyperref}
\usepackage{color}
\usepackage{pifont}
\usepackage{tikz}
\usetikzlibrary{calc,decorations.pathmorphing,positioning,shapes}
\usepackage{xspace}
\usepackage[capitalize]{cleveref}
\usepackage[noend]{algpseudocode}
\usepackage[noend]{algorithm}
\usepackage{algorithmicx}

\usepackage{thm-restate}
\usepackage{cancel}

\colorlet{colorDEPS}{violet}
\colorlet{colorPO}{darkgray!80!black}
\colorlet{colorRF}{green!60!black}
\colorlet{colorJF}{blue}
\colorlet{colorEW}{brown}
\colorlet{colorMO}{orange}
\colorlet{colorFR}{purple}
\colorlet{colorECO}{red!80!black}
\colorlet{colorSYN}{green!40!black}
\colorlet{colorHB}{blue}
\colorlet{colorWB}{blue}
\colorlet{colorPPO}{magenta}
\colorlet{colorPB}{olive}
\colorlet{colorRMW}{olive!70!black}
\colorlet{colorRSEQ}{green!40!black}
\colorlet{colorSC}{violet}
\colorlet{colorPSC}{violet}
\colorlet{colorREL}{olive}
\colorlet{colorCONFLICT}{olive}
\colorlet{colorRACE}{olive}

\tikzset{
   every path/.style={>=stealth},
   po/.style={->,color=colorPO,thin,shorten >=-0.5mm,shorten <=-0.5mm},
   sw/.style={->,color=colorSYN,shorten >=-0.5mm,shorten <=-0.5mm},
   rf/.style={->,color=colorRF,dashed,,shorten >=-0.5mm,shorten <=-0.5mm},
   jf/.style={->,color=colorJF,dashed,,shorten >=-0.5mm,shorten <=-0.5mm},
   ew/.style={<->,color=colorEW,dashed,,shorten >=-0.5mm,shorten <=-0.5mm},
   fr/.style={->,color=colorFR,dashed,,shorten >=-0.5mm,shorten <=-0.5mm},
   hb/.style={->,color=colorHB,thick,shorten >=-0.5mm,shorten <=-0.5mm},
   mo/.style={->,color=colorMO,dotted,very thick,shorten >=-0.5mm,shorten <=-0.5mm},
   no/.style={->,dotted,thick,shorten >=-0.5mm,shorten <=-0.5mm},
   deps/.style={->,color=colorDEPS,dotted,thick,shorten >=-0.5mm,shorten <=-0.5mm},
   rmw/.style={->,color=colorRMW,thick,shorten >=-0.5mm,shorten <=-0.5mm},
   rseq/.style={->,color=colorRSEQ,thick,dotted,shorten >=-0.5mm,shorten <=-0.5mm},
   xrf/.style={->,color=colorRF,shorten >=-0.5mm,shorten <=-0.5mm},
}

\newcounter{sarrow}

\newtheorem{lemma}{Lemma}
\newtheorem{definition}{Definition}

\crefname{section}{{\S\!}}{\S\S} 
\crefname{subsection}{{\S\!}}{\S\S} 
\crefname{subsubsection}{{\S\!}}{\S\S} 
\crefformat{section}{#2\S{}#1#3}
\crefformat{subsection}{#2\S{}#1#3}
\crefformat{subsubsection}{#2\S{}#1#3}
\Crefformat{section}{#2\S{}#1#3}
\Crefformat{subsection}{#2\S{}#1#3}
\Crefformat{subsubsection}{#2\S{}#1#3}

\colorlet{colorDEPS}{violet}
\colorlet{colorPO}{darkgray!80!black}
\colorlet{colorRF}{green!60!black}
\colorlet{colorJF}{blue}
\colorlet{colorEW}{brown}
\colorlet{colorCA}{brown}
\colorlet{colorMO}{orange}
\colorlet{colorCO}{orange}
\colorlet{colorFR}{purple}
\colorlet{colorECO}{red!80!black}
\colorlet{colorSW}{green!40!black}
\colorlet{colorHB}{blue}
\colorlet{colorWB}{blue}
\colorlet{colorPPO}{magenta}
\colorlet{colorPB}{olive}
\colorlet{colorRMW}{olive!70!black}
\colorlet{colorRSEQ}{green!40!black}
\colorlet{colorSC}{violet}
\colorlet{colorPSC}{violet}
\colorlet{colorREL}{olive}
\colorlet{colorCONFLICT}{olive}
\colorlet{colorRACE}{olive}
\colorlet{colorRMW}{brown}

\tikzset{
   every path/.style={>=stealth},
   po/.style={->,color=colorPO,thin,shorten >=-0.5mm,shorten <=-0.5mm},
   sw/.style={->,color=colorSW,shorten >=-0.5mm,shorten <=-0.5mm},
   rf/.style={->,color=colorRF,dashed,shorten >=-0.5mm,shorten <=-0.5mm},
   jf/.style={->,color=colorJF,dashed,shorten >=-0.5mm,shorten <=-0.5mm},
   ew/.style={<->,color=colorEW,dashed,shorten >=-0.5mm,shorten <=-0.5mm},
   fr/.style={->,color=colorFR,dashed,shorten >=-0.5mm,shorten <=-0.5mm},
   hb/.style={->,color=colorHB,thick,shorten >=-0.5mm,shorten <=-0.5mm},
   mo/.style={->,color=colorMO,dotted,very thick,shorten >=-0.5mm,shorten <=-0.5mm},
   no/.style={->,dotted,thick,shorten >=-0.5mm,shorten <=-0.5mm},
   deps/.style={->,color=colorDEPS,dotted,thick,shorten >=-0.5mm,shorten <=-0.5mm},
   rmw/.style={->,color=colorRMW,thick,shorten >=-0.5mm,shorten <=-0.5mm},
   rseq/.style={->,color=colorRSEQ,thick,dotted,shorten >=-0.5mm,shorten <=-0.5mm},
   xrf/.style={->,color=colorRF,shorten >=-0.5mm,shorten <=-0.5mm},
}

\newcommand{\tup}[1]{\langle#1\rangle}

\newcommand{\MOna}{\textsc{na}}
\newcommand{\MOrlx}{\textsc{rlx}}
\newcommand{\MOacq}{\textsc{acq}}

\newcommand{\MOrel}{\textsc{rel}}
\newcommand{\MOsc}{\textsc{sc}}

\newcommand{\behav}{\mathsf{Behavior}}

\newcommand{\irreflexive}{\mathsf{irreflexive}}

\newcommand\getmf{\mathsf{getG}}

\newcommand{\MOat}{\textsc{a}}

\newcommand\oncyc{\mathsf{OnCyc}}

\newcommand\ex{\mathsf{X}}

\newcommand\rloc[1]{{{#1}|_\lloc}}
\newcommand\rnloc[1]{{#1}|_{\neq\lloc}}

\newcommand\F{\mathsf{F}}

\newcommand{\dobcc}{\mathsf{dobcc}}
\newcommand{\ndobcc}{\mathsf{ndobcc}}

\newcommand\E{\mathsf{E}}

\newcommand\N{\mathsf{N}}

\newcommand\mpairs{\mathsf{mpairs}}

\newcommand\ordered{\mathsf{Ordered}}

\newcommand\mayalias{\mathsf{mayAlias}}
\newcommand\mustalias{\mathsf{mustAlias}}

\newcommand\cfg{\mathsf{G}}

\newcommand\mbbp{\mathbb{P}}

\newcommand\srcp{\mbbp_{\!\sf src}}
\newcommand\tgtp{\mbbp_{\!\sf tgt}}

\newcommand\srcx{\ex_{\sf s}}
\newcommand\tgtx{\ex_{\sf t}}

\newcommand\xp{\mbbp_{\!\sf x86}}
\newcommand\armp{\mbbp_{\!\sf ARMv8}}
\newcommand\oarmp{\mbbp_{\!\sf ARMv7}}
\newcommand\oarmpmca{\mbbp_{\!\sf ARMv7\text{-}mca}}

\newcommand{\denot}[1]{[\![{#1}]\!]}

\newcommand\inarr[1]{\begin{array}{@{}l@{}}#1\end{array}}

\newcommand\inparII[2]{\begin{array}{@{}l@{~~}||@{~~}l@{}}\inarr{#1}&\inarr{#2}\end{array}}

\newcommand\inparIV[4]{\begin{array}{@{}l@{~~}||@{~~}l@{~~}||@{~~}l@{~~}||@{~~}l@{}}
\inarr{#1}&\inarr{#2}&\inarr{#3}&\inarr{#4}\end{array}}
\newcommand\inparV[5]{\begin{array}{@{}l@{~~}||@{~~}l@{~~}||@{~~}l@{~~}||@{~~}l@{~~}||@{~~}l@{}}
\inarr{#1}&\inarr{#2}&\inarr{#3}&\inarr{#4}&\inarr{#5}\end{array}}

\newcommand\imm{\mathsf{imm}}

\newcommand\getnfs{\textsc{getNFs}}

\newcommand\fdelete{\textsc{FDelete}}

\newcommand\xfelim{\textsc{x86FElim}}
\newcommand\armafelim{\textsc{ARMv8FElim}}
\newcommand\armsfelim{\textsc{ARMv7FElim}}

\newcommand{\moe}{\mathsf{\color{colorMO}moe}}

\newcommand\epo{\mathsf{epo}}

\newcommand{\ii}{\mathsf{ii}}
\newcommand{\ic}{\mathsf{ic}}
\newcommand{\ci}{\mathsf{ci}}
\newcommand{\cc}{\mathsf{cc}}
\newcommand{\rdw}{\mathsf{rdw}}
\newcommand{\detour}{\mathsf{detour}}

\newcommand\Wlab {\mathsf{St}}
\newcommand\Rlab {\mathsf{Ld}}
\newcommand\Ulab {\mathsf{U}}

\newcommand\WUlab {\mathcal{W}}
\newcommand\RUlab {\mathcal{R}}

\newcommand\Flab {\mathsf{F}}
\newcommand\Elab {\mathcal{E}}

\newcommand\skips{\mathsf{skip}}
\newcommand\A{\mathsf{A}}
\newcommand\V{\mathsf{V}}

\newcommand\co{\mathsf{\color{colorMO}co}}

\newcommand\rels{\mathsf{Rel}}
\newcommand\acqs{\mathsf{Acq}}

\newcommand\irr{\mathsf{irr}}
\newcommand\poloc{\rloc \lpo}

\newcommand\awr{\mathsf{WR}}
\newcommand\aww{\mathsf{WW}}

\newcommand\isawra{\mathsf{isWR}}

\newcommand{\kw}[1]{\textsf{\bfseries #1}}

\newcommand{\Cif}[1]{\kw{if}\,({#1})\ }

\newcommand\dmb {\F}

\newcommand\wo {\mathsf{wo}}

\newcommand\mov{\texttt{MOV}}
\newcommand\movld{\texttt{RMOV}}
\newcommand\movst{\texttt{WMOV}}

\newcommand{\mfence}{\texttt{MFENCE}}

\newcommand\irmw{\texttt{RMW}}

\newcommand\ldrex {\texttt{LDREX}}
\newcommand\strex {\texttt{STREX}}

\newcommand\ldr{\texttt{LDR}}
\newcommand\str{\texttt{STR}}
\newcommand\ldar{\texttt{LDAR}}
\newcommand\stlr{\texttt{STLR}}

\newcommand\ldxr {\texttt{LDXR}}
\newcommand\stxr {\texttt{STXR}}
\newcommand\ldaxr {\texttt{LDAXR}}
\newcommand\stlxr {\texttt{STLXR}}

\newcommand\idmbfull{\texttt{DMBFULL}}
\newcommand\idmbld{\texttt{DMBLD}}
\newcommand\idmbst{\texttt{DMBST}}

\newcommand\idmb {\texttt{DMB}}
\newcommand\isb {\texttt{ISB}}

\newcommand\cbisb {\texttt{CBISB}}

\newcommand{\xarch}{\mathsf{x86}}
\newcommand{\arma}{\mathsf{ARMv8}}
\newcommand{\arms}{\mathsf{ARMv7}}
\newcommand{\armsmca}{\mathsf{ARMv7\text{-}mca}}

\newcommand{\isync}{\mathsf{isync}}
\newcommand{\fence}{\mathsf{fence}}

\newcommand{\mo}{\mathsf{\color{colorMO}mo}}
\newcommand{\ca}{\mathsf{\color{colorCA}ca}}

\newcommand{\fr}{\mathsf{\color{colorFR}fr}}

\newcommand{\fre}{\mathsf{\color{colorFR}fre}}
\newcommand{\fri}{\mathsf{\color{colorFR}fri}}

\newcommand{\SC}{\mathsf{SC}}

\newcommand{\lid}{\mathsf{id}}
\newcommand{\ltid}{\mathsf{tid}}

\newcommand{\llab}{\mathsf{lab}}
\newcommand{\lloc}{\mathsf{loc}}

\newcommand{\lpo}{\mathsf{po}}
\newcommand{\lrf}{\mathsf{\color{colorRF}rf}}

\newcommand{\lhb}{\mathsf{\color{colorHB}hb}}

\newcommand{\xhb}{\mathsf{\color{colorHB}xhb}}

\newcommand{\rfe}{\mathsf{\color{colorRF}rfe}}

\newcommand{\rVal}{\mathsf{rval}}
\newcommand{\wVal}{\mathsf{wval}}

\newcommand{\po}{\mathsf{po}}
\newcommand{\ppo}{\mathsf{{\color{colorPPO}ppo}}}
\newcommand{\xppo}{\mathsf{{\color{colorPPO}xppo}}}

\newcommand{\op}{\mathsf{op}}

\newcommand{\event}{e}

\newcommand{\rmw}{\mathsf{\color{colorRMW}rmw}}
\newcommand{\frange}{\mathsf{range}}

\newcommand{\Locs}{\mathsf{Locs}}
\newcommand{\regs}{\mathsf{Reg}}

\newcommand{\Val}{\mathsf{val}}

\newcommand{\br}{\mathbf{br}}

\newcommand{\eco}{\mathsf{\color{colorECO}eco}}

\newcommand\coi{\mathsf{\color{colorCO}coi}}
\newcommand\rfi{\mathsf{\color{colorRF}rfi}}
\newcommand\coe{\mathsf{\color{colorCO}coe}}
\newcommand\ctrl{\mathsf{ctrl}}
\newcommand\data{\mathsf{\color{colorPPO}data}}
\newcommand\addr{\mathsf{\color{colorPPO}addr}}

\newcommand\sbprg{\mathsf{SB}}

\newcommand\aob{\mathsf{aob}}
\newcommand\bob{\mathsf{bob}}
\newcommand\dob{\mathsf{dob}}
\newcommand\obs{\mathsf{obs}}
\newcommand\ob{\mathsf{ob}}
\newcommand\obx{\mathsf{obx}}

\newcommand\obi{\mathsf{obi}}

\newcommand\Alab{\mathsf{A}}
\newcommand\Qlab{\mathsf{Q}}
\newcommand\Llab{\mathsf{L}}
\newcommand\Nlab{\mathsf{N}}

\newcommand\dmbfull{\F}
\newcommand\dmbld{\F_{\textsc{ld}}}
\newcommand\dmbst{\F_{\textsc{st}}}

\newcommand{\parCtx}[1]{
\textcolor{blue!70!green}{\inarr{~~\textbf{Context:}\\\left[\inarr{\\[-2mm]
\inparII{-}{#1}
\\[-2mm]~}\right]}}}

\newcommand{\dom}{\mathsf{dom}}
\newcommand{\codom}{\mathsf{codom}}

\newcommand{\set}[1]{\{#1\}}

\newcommand{\hbppc}{\mathsf{\color{colorHB}ahb}}
\newcommand{\prop}{\mathsf{{\color{colorPB}prop}}}
\newcommand{\ahb}{\mathsf{\color{colorHB}ahb}}

\newcommand{\cmm}{\mathsf{C11}}

\newcommand{\codecomment}[1]{\color{teal}\text{ // }{#1}}

\newcommand{\acy}{\mathsf{acy}}
\newcommand{\total}{\mathsf{total}}

\newcommand{\yes}{{\color{green!70!black}\ding{51}}}
\newcommand{\no}{{\color{red!70!black}\ding{55}}}

\newcommand{\isst}{\mathsf{isSt}}
\newcommand{\isld}{\mathsf{isLd}}
\newcommand{\isrel}{\mathsf{isRel}}
\newcommand{\isacq}{\mathsf{isAcq}}
\newcommand{\isll}{\mathsf{isLL}}
\newcommand{\issc}{\mathsf{isSC}}

\newcommand{\isw}{\mathsf{isW}}
\newcommand{\isr}{\mathsf{isR}}

\newcommand{\srels}{\mathsf{Lcoi}}
\newcommand{\relacq}{\mathsf{RA}}

\newcommand{\true}{\mathsf{true}}
\newcommand{\false}{\mathsf{false}}

\newcommand{\uord}{O}

\newcommand{\reach}{\mathsf{Reach}}
\newcommand{\reachwo}{\mathsf{ReachWO}}

\newcommand{\lbreak}{\mathsf{break}}

\newcommand{\nelim}{{\mathsf{nfs}}}

\newcommand{\onelim}{{\mathsf{NFS}}}

\newcommand{\opath}{\mathsf{Path}}

\algnewcommand\algorithmicforeach{\textbf{for each}}
\algdef{S}[FOR]{ForEach}[1]{\algorithmicforeach\ #1\ \algorithmicdo}

\colorlet{dark-green}{green!70!black}
\colorlet{dark-red}{red!80!black}
\colorlet{dark-blue}{blue!80!black}

\newenvironment{observation}[1]{\par\noindent{\bf Observation.}\space#1}{}

\tikzset{
   every path/.style={>=latex},
   lightarrow/.style={->,thick,>=triangle 60},
   longarrow/.style={->,thick,shorten >=-1mm,shorten <=-1mm},
   box1/.style={minimum height=3mm, minimum width=4mm, rounded corners,rectangle,draw=gray, 
                thick,inner xsep=1mm,inner ysep=1mm},
   src/.style={rounded corners,rectangle,draw=gray,fill=gray!10,inner sep=1.5mm}
}

\newcommand{\Case}[1]{\bigskip\noindent\textbf{Case {#1}:}}
\newcommand{\Subcase}[1]{\bigskip\noindent\textbf{Subcase {#1}:}}
\newcommand{\Subsubcase}[1]{\bigskip\noindent\textbf{Subsubcase {#1}:}}

\algrenewcommand\algorithmicindent{1em}%

\date{} 

\begin{document}

\title{On Architecture to Architecture Mapping for Concurrency}

\author{Soham Chakraborty \\
  Department of Computer Science and Engineering \\
  IIT Delhi \\
  Delhi 110016, India \\
  \texttt{soham@cse.iitd.ac.in}
}


%


\maketitle


\begin{abstract}

Mapping programs from one architecture to another plays a key role in technologies 
such as binary translation, decompilation, emulation, virtualization, and application migration.
Although multicore architectures are ubiquitous,
the state-of-the-art translation tools do not handle concurrency primitives correctly.
Doing so is rather challenging because of the subtle differences in the concurrency models between architectures. 

In response, we address various aspects of the challenge. 
First, we develop correct and efficient translations between the concurrency models
of two mainstream architecture families: x86 and ARM (versions 7 and 8).
We develop direct mappings between x86 and ARMv8 and ARMv7,
and fence elimination algorithms to eliminate redundant fences after direct mapping.
Although our mapping utilizes ARMv8 as an intermediate model for mapping between x86 and ARMv7,
we argue that it should not be used as an intermediate model in a decompiler
because it disallows common compiler transformations.

Second, we propose and implement a technique for inserting memory fences
for safely migrating programs between different architectures.
Our technique checks robustness against x86 and ARM,
and inserts fences upon robustness violations.
Our experiments demonstrate that in most of the programs both our techniques
introduce significantly fewer fences compared to naive schemes
for porting applications across these architectures.

\end{abstract}


\section{Introduction}
\label{sec:introduction}

Architecture to architecture mapping is the widely applicable concept 
of converting an application that runs over some architecture $X$
to run over some different architecture $Y$.
For example, \emph{binary translators} \cite{notaz:2014,Chernoff:1998},
which recompile machine code from one architecture to another in a semantic preserving manner. 
Such translation is facilitated by decompilers \cite{dagger,mcsema,Yadavalli:2019,retdec,Shen:2012},
which lift machine code from a source architecture to an intermediate representation (IR) and compile to a target architecture.
\emph{Emulators} implement a guest architecture on a host architecture.
For instance, QEMU \cite{qemu} emulates a number of architectures (including x86 and ARM) over other architectures,
the Android emulator \cite{androidx} runs x86 images on ARM, 
while Windows 10 on ARM emulates x86 applications \cite{msxa}.

Architecture to architecture mapping is essential for application migration and compatibility. 
An application written for an older architecture may need upgraded to execute on latest architectures,
while an application primarily targeting a later architecture 
may need to preserve backward compatibility with respect to older one. 
For example, \citet{arm57} discusses the required measures to port an application
from ARMv5 to ARMv7 including synchronization primitives. 
Besides its practical uses, formally mapping between architectures is helpful
in the design process of future processors and architectures, as it allows one
to compare and relate subtle features like concurrency,
which vary significantly from one architecture to another.


A key feature that has been overlooked in these mappings is \emph{concurrency},
which is crucial for achieving good performance with modern multicore processors.
To emulate or port a concurrent application correctly requires us to map 
the concurrency primitives of the source to those of the target,
taking into account the subtle differences in their concurrency models. 
Such semantic differences appear not only between architectures (e.g., x86 and ARM),
but also between different versions of the same architecture (e.g., ARMv7 and ARMv8 \cite{Pulte:2018}).

In this paper, we address the challenge of developing correct and efficient translations
between relaxed memory concurrency models of x86, ARMv8, and ARMv7.
We approach the problem from multiple angles.

First, we develop correct mapping schemes between these concurrency models,
using the ARMv8 model as an efficient intermediate concurrency model
for mapping between x86 and ARMv7.

This naturally leads to the question whether ARMv8 model can also 
serve as a concurrency model for IR in a decompiler.
Decompilers typically (1) raise the source machine code to an IR,
(2) optimize the IR, and (3) generate the target code. 
Thus, If the IR follows the ARMv8 concurrency model,
steps (1) and (3) can be performed efficiently
to facilitate translations between x86 and ARM concurrent programs. 
For step (2), we evaluate common optimizations on ARMv8 concurrency 
and observe that a number of common transformations are unsound. 
The result demonstrates that to achieve correct and efficient mapping 
by all steps (1,2,3) we require to come up with a different concurrency model. 
We leave the exploration for such a model for future research.

Next, we focus on optimizing the direct mappings further.
The issue is that for correctness direct mappings introduce fences in translating stronger accesses 
to weaker ones.
The introduced fences can be often redundant in certain memory access sequences and can be eliminated safely. 
We identify conditions of safe fence elimination, prove safe fence elimination, and based on these conditions we propose fence elimination algorithms.   
 
In addition to fence elimination, we apply memory sequence 
analysis to check and enforce robustness for a class of concurrent programs. 
Robustness analysis checks whether a program running model demonstrate 
only the behaviors which are allowed by a stronger model. 
The behaviors of a robust program are indistinguishable on stronger model from an 
weaker model and therefore the program can seamlessly migrate from one architecture to another 
as far as concurrent behaviors are concerned. 
If a program is not robust we insert fences to enforce robustness against a stronger model.
It is especially beneficial in application porting and migration \cite{Barbalace:2020,Barbalace:2017} 
where it is crucial to preserve the observable behaviors of a running application.

\begin{center}
\begin{figure}
\centering
\begin{tikzpicture}[yscale=0.8]
  \node[draw, rectangle, align=center,minimum width=2cm,minimum height=2cm] (A) at (0,0)  {{\Large x86}};
  \node[draw, rectangle, align=center,minimum width=2cm,minimum height=2cm] (B) at (4,0)  {ARMv8};
  \node[draw, rectangle, align=center,minimum width=4cm,minimum height=2cm] (C) at (9,0)  {};
  \node (D) at (10,0)  {ARMv7};
  \node[draw, dashed, rectangle, align=center,minimum width=2cm,minimum height=1.8cm] (E) at (8,0)  {\begin{tabular}{c}ARMv7- \\ mca \end{tabular}};
  \draw[po] ($(A.east) - (0,0.7)$) to node[below]{$\cmm$ (\cref{sec:cxarm}) } ($(B.west) - (0,0.7)$);
  \draw[po] ($(A.east) - (0,0.2)$) to node[above]{(\cref{sec:x86toarmv8}) } ($(B.west) - (0,0.2)$);
  \draw[po] ($(B.west) + (0,0.7)$) to node[above]{ (\cref{sec:armv8tox86})} ($(A.east) + (0,0.7)$);
  \draw[po] ($(B.east) - (0,0.7)$) to node[below]{$\cmm$ (\cref{sec:carmv8v7}) } ($(E.west) - (0,0.7)$);
  \draw[po] ($(B.east) - (0,0.2)$) to node[above]{(\cref{sec:armv8toarmv7}) } ($(E.west) - (0,0.2)$);
  \draw[po] ($(E.west) + (0,0.7)$) to node[above]{ (\cref{sec:armv7toarmv8})} ($(B.east) + (0,0.7)$);
\end{tikzpicture}
\caption{Correct and efficient mapping schemes between x86, ARMv8, and ARMv7/ARMv7-mca.}
\label{fig:mappingschemes}
\end{figure}
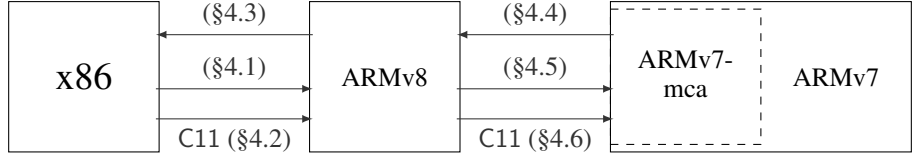
\end{center}
\vspace{-5mm}
\textit{Contributions \& Results.}
Now we discuss the specific contributions and obtained results.
\begin{itemize}[leftmargin=*]
\item In \cref{sec:mappings} we propose the mapping schemes ($\mapsto$) between x86 and ARMv8, and between ARMv8 and ARMv7 as shown in \cref{fig:mappingschemes}.
We do not propose any direct mapping between x86 and ARMv7, instead we consider ARMv8 as an intermediate model. 
We achieve x86 to ARMv7 mapping by combining x86 to ARMv8 and ARMv8 to ARMv7 mapping. 
Similarly, ARMv7 to x86 mapping is derived by combining ARMv7 to ARMv8 and ARMv8 to x86 mappings.
We show that the direct mapping schemes would be same as these two step mappings through ARMv8. 

We also show that these mapping schemes are efficient; each of the leading and/or trailing fences used in mapping with the memory accesses are required to preserve correctness.

\item We show that multicopy-atomicity (MCA) (a write operation is observable to all other threads at the same time) does not affect the mapping schemes between ARMv8 and ARMv7 though it is a major difference between ARMv8 and ARMv7 \cite{Pulte:2018} as ARMv7 allows non-MCA behavior unlike ARMv8. 
To demonstrate the same we propose ARMv7-mca in \cref{sec:mappings} which restricts non-MCA behaviors in ARMv7 
and show that the mapping scheme from ARMv8 to ARMv7-mca is same as ARMv8 to ARMv7 (\cref{tab:armv8v7}) and the mapping scheme of ARMv7-mca to ARMv8 is same as ARMv7 to ARMv8 mapping (\cref{tab:armv7v8}) respectively.   

\item In \cref{sec:cxarm}, \cref{sec:carmv8v7}, and in \cref{sec:armv8toarmv7mca} we propose alternative schemes for x86 to ARMv8 and ARMv8 to ARMv7 mapping where the respective x86 and ARMv8 programs are generated from C11 concurrent programs. 
In these schemes we exploit the catch-fire semantics of C11 concurrency \cite{cstandard,cppstandard}. 
We do not generate additional fences for the load or store accesses 
generated from non-atomic loads or stores unlike the x86 to ARMv8 and ARMv8 to ARMv7 mappings.

\item In \cref{sec:armaopt} we study the reordering, elimination, and access strengthening transformations 
in ARMv8 model. 
We prove the correctness of the safe transformations and provide counter-examples 
for the unsafe transformations.

\item The mapping schemes introduce additional fences while mapping the memory accesses. 
These fences are required to preserve translation correctness in certain scenarios and otherwise redundant. 
In \cref{sec:fenceopt} we identify the conditions when the fences are redundant and prove that eliminating the 
fences are safe. 
Based on these conditions we define fence elimination algorithms to eliminate redundant fences without affecting 
the transformation correctness.
%

\item  
We define the conditions for robustness for an 
(i) ARMv8 program against sequential consistency (SC) and x86 model, 
(ii) ARMv7/ARMv7-mca program against SC, x86, and ARMv8 model, and 
(iii) ARMv7 program against ARMv7-mca model in \cref{sec:robustness} 
and prove their correctness in \cref{app:robustness}. 
We also introduce fences to enforce robustness for a stronger model againts 
a weaker model. 
To the best of our knowledge we are the first to check and enforce 
robustness for ARM programs as well as for non-SC models.

\item In \cref{sec:experiment} we discuss our experimental results. 
We have developed a compiler based on LLVM to capture the effect of 
mappings between x86, ARMv8, and ARMv7.
Next, we have developed fence elimination passes based on \cref{sec:fenceopt}.
The passes eliminate significant number of 
redundant fences in most of the programs 
and in some cases generate more efficient program than LLVM mappings.
 
We have also developed analyzers to check and enforce robustness in x86, ARMv8, and ARMv7.
For a number of x86 programs the result of our SC-robustness checker matches the results from Trencher\cite{Bouajjani:2013} 
which also checks SC-robustness against TSO model.
Moreover, we enforce robustness with significantly less number of fences 
compared to naive schemes which insert fences without robustness information. 
\end{itemize}
In the next section we informally explain the overview of the proposed approaches. 
Next, in \cref{sec:models} we discuss the axiomatic models of the respective architectures 
which we use in later sections. 
The proofs and additional details are in the supplementary material.

\section{Overview}
\label{sec:observations}

%
%
%
%
%
%

In this section we discuss the overview of our proposed schemes, related observations, and the analysis techniques. 

\begin{center}
\begin{figure}
\begin{subfigure}{0.95\linewidth}
\[
\inparIV{X[1] = 1; \ }{
a=X[1]; 
\\ b = Y[a]; 
}{
c=Y[1]; 
\\ d=Z[c]; 
}{ Y[1]=1; \ }
\ \mapsto \
\inparIV{X[1] = 1; \ }{
a=X[1]; 
\\ \cbisb
\\ b = Y[a]; 
\\ \cbisb
}{
c=Y[1]; 
\\ \cbisb
\\ d=Z[c]; 
\\ \cbisb
}{ Y[1]=1; \ }
\]
\caption{Initially $X[1]=Y[1]=0$ and behavior in question: $a=c=1, \ b=d=0$.}
\label{fig:map:iriwv87}
\end{subfigure}
\\[1.2ex]
\begin{subfigure}{0.49\linewidth}
\centering
\begin{tikzpicture}[yscale=0.8]
  \node (t11) at (-1,0)  {$\Wlab(X[1],1)$};
  \node (t41) at (1,0) {$\Wlab(Y[1],1)$};
  \node (t21) at (-2,-1.5)  {$\Rlab(X[1],1)$};
  \node (t22) at (-2,-3) {$\Rlab(Y[1],0)$};
  \node (t31) at (2,-1.5)  {$\Rlab(Y[1],1)$};
  \node (t32) at (2,-3) {$\Rlab(X[1],0)$};
  \draw[po] (t21) to  node[right]{$\addr$} (t22);
  \draw[po] (t31) to node[left]{$\addr$} (t32);  
  \draw[rf,bend right=20] (t11) to (t21);
  \draw[rf,bend left=20] (t41) to (t31);
  \draw[fr,bend left=0] (t22.east) to (t41);
   \draw[fr,bend left=0] (t32.west) to (t11);
\end{tikzpicture}
\caption{Disallowed in ARMv8}
\label{fig:iriwv8}
\end{subfigure}\hfill
\begin{subfigure}{0.49\linewidth}
\centering
\begin{tikzpicture}[yscale=0.8]
  \node (t11) at (-1,0)  {$\Wlab(X[1],1)$};
  \node (t41) at (1,0) {$\Wlab(Y[1],1)$};
  \node (t21) at (-2,-1.5)  {$\Rlab(X[1],1)$};
  \node (t23) at (-2,-3) {$\Rlab(Y[1],0)$};
  \node (t24) at (-2,-3.7)    {};  
  \node (t31) at (2,-1.5)  {$\Rlab(Y[1],1)$};
  \node (t33) at (2,-3) {$\Rlab(X[1],0)$};
  \node (t34) at (2,-3.7)    {};  
  \draw[po] (t21) to  node[left]{$R$} (t23);
  \draw[po] (t23) to (t24);
  \draw[po] (t31) to node[right]{$R$} (t33);
  \draw[po] (t33) to (t34);
  \draw[rf,bend right=20] (t11) to (t21);
  \draw[rf,bend left=20] (t41) to (t31);
  \draw[fr,bend left=0] (t23.east) to (t41);
  \draw[fr,bend left=0] (t33.west) to (t11);
\end{tikzpicture}
\caption{Allowed in ARMv7. $R=\ctrl_{\mathsf{isb}} \cup \addr$}
\label{fig:iriwv7}
\end{subfigure}
\caption{$\ldr \mapsto \ldr;\cbisb$ in ARMv8 to ARMv7 mapping is unsound.}
\end{figure}
\end{center}

\begin{center}
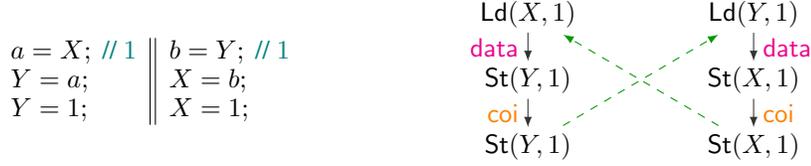
\begin{figure}
\begin{subfigure}{0.49\textwidth}
\[
\inparII{
a = X; \codecomment{1}
\\ Y = a; 
\\ Y = 1; 
}{
b = Y; \codecomment{1}
\\ X = b; 
\\ X = 1; 
}
\]
\end{subfigure}
\begin{subfigure}{0.49\textwidth}
\begin{tikzpicture}[yscale=0.8]
  \node (t11) at (-1.5,0)  {$\Rlab(X,1)$};
  \node (t12) at (-1.5,-1.1) {$\Wlab(Y,1)$};
  \node (t13) at (-1.5,-2.2)  {$\Wlab(Y,1)$};
  \node (t21) at (1.5,0) {$\Rlab(Y,1)$};
  \node (t22) at (1.5,-1.1)  {$\Wlab(X,1)$};
  \node (t23) at (1.5,-2.2) {$\Wlab(X,1)$};
%
  \draw[po] (t11) to  node[left]{$\data$} (t12);
  \draw[po] (t12) to   node[left]{$\coi$} (t13);  
  \draw[po] (t21) to  node[right]{$\data$} (t22);
  \draw[po] (t22) to   node[right]{$\coi$} (t23);    
  \draw[rf,bend right=0] (t13) to (t21);
  \draw[rf,bend left=0] (t23) to (t11);
\end{tikzpicture}
\end{subfigure}
\caption{$a=b=1$ is disallowed in ARMv8 but allowed in ARMv7-mca for $\ldr \mapsto \ldr$ mapping.}
\label{fig:armamcaldrldr}
\end{figure}
\end{center}
\vspace{-8mm}
\subsection{Alternatives in x86 to ARMv8 mapping}

In x86 to ARMv8 mapping we considered two alternatives for mapping loads and stores:
(1) x86 store and load to ARMv8 release-store ($\movst \mapsto \stlr$) and acquire-load ($\movld \mapsto \ldar$) respectively. (2) x86 store and load to ARMv8 regular store and load accesses with respective leading and trailing fences as proposed in \cref{tab:xarm}, that is, $\movst \mapsto \idmbst;\str$ and $\movld \mapsto \ldr;\idmbld$ respectively. 
We choose (2) over (1) for following reasons. 
\begin{itemize}[leftmargin=*]
\item \textit{Reordering is restricted.} 
x86 allows the reordering of independent store and load operations accessing different locations \cite{Lahav:fm16}. 
ARMv8 also allows reordering of different-location store-load pairs, but restricts  
the reordering of a pair of release-store and acquire load operation as it 
violates \emph{barrier-ordered-by} ($\bob$) order \cite{Pulte:2018}. 
Thus scheme (1) is more restrictive than (2) considering reordering flexibility after mapping.
%

\item \textit{Further optimization.} 
(2) generates certain fences which are redundant in certain scenarios and can be removed safely. 
Consider the mappings below; the generated $\idmbst$ is redundant in (2) and can be eliminated safely unlike mapping (1). 
\\[0.5ex]
(1) $\movld;\movst \mapsto \ldar;\stlr$ \hfill (2) $\movld;\movst \mapsto \ldr;\idmbld;\idmbst;\str \leadsto \ldr;\idmbld;\str$ 

\item \textit{$\xarch \mapsto \arma \mapsto \arms$ would introduce additional fences}
To map x86 to ARMv7, if we use ARMv8 as intermediate step then scheme (1) 
introduces additional fences unlike (2) as follows.
\\[0.5ex]
(1) $\movst \mapsto \stlr \mapsto \idmb;\str;\idmb$ \hfill (2) $\movst \mapsto \idmbst;\str \mapsto \idmb;\str$  
\end{itemize}

\subsection{ARMv8 to ARMv7 mapping: ARMv7 $\ldr$ is significantly weaker than ARMv8 $\ldr$}
\label{sec:case:armv87}
ARMv8 to ARMv7 mapping in \cref{tab:armv8v7} 
%
introduces a trailing $\idmb$ fence 
for ARMv8 $\ldr$ and $\ldar$ accesses as introducing a trailing control fence ($\cbisb$) is not enough for correctness.
%
Consider the mapping of the program from ARMv8 to ARMv7 in \cref{fig:map:iriwv87}. 
The execution is disallowed in ARMv8 as it creates an \emph{observed-by} ($\ob$) cycle as shown in \cref{fig:iriwv8}. 
However, whie mapping to ARMv7, if we map $\ldr \mapsto \ldr;\cbisb$ and rest of the instructions are mapped 
following the mapping scheme in \cref{tab:armv8v7} then the execution would be allowed as shown in \cref{fig:iriwv7}.  
Therefore $\ldr \mapsto \ldr;\cbisb$ is too weak and 
we require a $\idmb$ fence after each load as well as $\irmw$ for the same reason. 

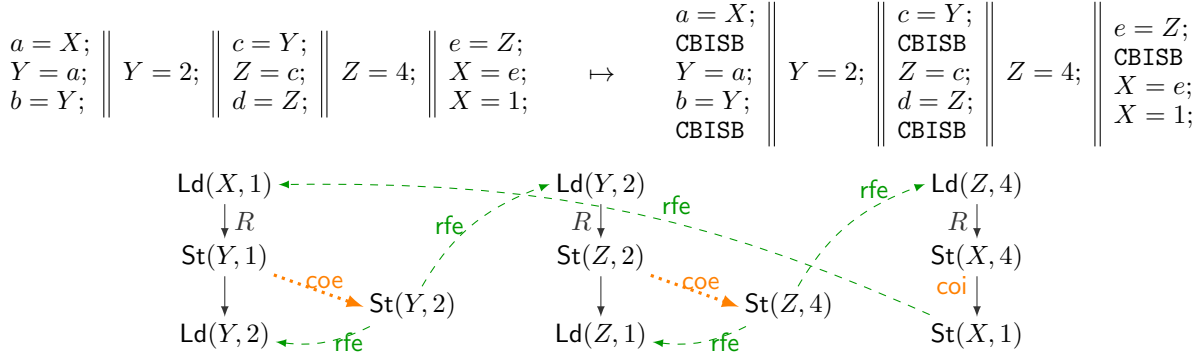
\begin{figure}
\begin{subfigure}{0.95\linewidth}
\[
\inparV{
a = X; 
\\ Y=a;
\\ b = Y; 
}{
Y=2;
}{
c = Y; 
\\ Z=c;
\\ d = Z; 
}{
Z=4;
}{
e = Z; 
\\ X=e;
\\ X=1;
}
\qquad \mapsto \qquad
\inparV{
a = X; 
\\ \cbisb
\\ Y=a;
\\ b = Y; 
\\ \cbisb
}{
Y=2;
}{
c = Y; 
\\ \cbisb
\\ Z=c;
\\ d = Z; 
\\ \cbisb
}{
Z=4;
}{
e = Z; 
\\ \cbisb
\\ X=e;
\\ X=1;
}
\]
\end{subfigure}
\\[1.2ex]
\begin{subfigure}{0.95\linewidth}
\centering
\begin{tikzpicture}[yscale=0.8]
  \node (t11) at (-5,0)  {$\Rlab(X,1)$};
  \node (t12) at (-5,-1.2) {$\Wlab(Y,1)$};
  \node (t13) at (-5,-2.5) {$\Rlab(Y,2)$};
  \node (t21) at (-2.5,-2)  {$\Wlab(Y,2)$};
  \node (t31) at (0,0)  {$\Rlab(Y,2)$};
  \node (t32) at (0,-1.2) {$\Wlab(Z,2)$};
  \node (t33) at (0,-2.5) {$\Rlab(Z,1)$};

  \node (t41) at (2.5,-2)  {$\Wlab(Z,4)$};
  \node (t51) at (5,0)  {$\Rlab(Z,4)$};
  \node (t52) at (5,-1.2) {$\Wlab(X,4)$};
  \node (t53) at (5,-2.5) {$\Wlab(X,1)$};
  \draw[po] (t11) to  node[right]{$R$} (t12);
  \draw[po] (t12) to  (t13);
  \draw[po] (t31) to  node[left]{$R$} (t32);
  \draw[po] (t32) to  (t33);
  \draw[po] (t51) to  node[left]{$R$} (t52);
  \draw[po] (t52) to node[left,pos=0.15]{$\coi$} (t53);
  \draw[rf,bend right=15] (t53) to node[above,pos=0.4]{$\rfe$} (t11);
  \draw[rf,bend left=20] (t21) to node[left]{$\rfe$} (t31.west);
  \draw[rf,bend left=20] (t41) to node[above,pos=0.7]{$\rfe$} (t51.west);
  \draw[mo,bend left=0] (t12) to node[right,pos=0.2]{$\coe$} (t21.west);
  \draw[mo,bend left=0] (t32) to node[right,pos=0.2]{$\coe$} (t41.west);
  \draw[rf,bend left=20] (t21) to node[right]{$\rfe$} (t13);
  \draw[rf,bend left=20] (t41) to node[right]{$\rfe$} (t33);
\end{tikzpicture}
\end{subfigure}
\caption{Behavior $a=1,b=c=2,d=e=4$ is disallowed in ARMv8 but allowed in ARMv7-mca for $\ldr \mapsto \ldr;\cbisb$ mapping. In the execution $R = \data$ in ARMv8 and $R = \data \cup \ctrl_{\mathsf{isb}}$ in ARMv7-mca.}
\label{fig:ldrmcaisync}
\end{figure}

\subsection{Multicopy atomicity does not change the mapping from ARMv8}
\label{sec:case:mca}
In \cref{fig:consistency} we strengthen the ARMv7 model to ARMv7-mca to exclude non-multicopy atomic 
behaviors. 
However, even with such a strengthening an $\ldr$ mapping requires a trailing $\idmb$ fence.
\paragraph{Load access mapping without trailing fence is unsound in ARMv8 to ARMv7-mca mapping.}
Consider the example in \cref{fig:armamcaldrldr} where the ARMv8 to ARMv7-mca mapping does not introduce trailing fence 
for a load access and therefore we analyze the same execution in ARMv8 and ARMv7-mca.
The shown behavior is not allowed in ARMv8 as there is a \emph{dependency-ordered-befoe} ($\dob$) relation from the reads to the respective writes due to $\data;\coi$ relation. 

In this case there is no \emph{preserved-program-order} ($\ppo$) relation in ARMv7a-mca from the reads to the respective writes as $\data;\coi \not\subseteq \ppo$. 
Therefore the execution is ARMv7-mca consistent and the mapping introduces a new outcome.
Hence $\ldr \mapsto \ldr$ in ARMv8 to ARMv7-mca mapping is unsound.

\paragraph{Trailing control fence  is not enough.}
Consider the example mapping and 
the execution in \cref{fig:ldrmcaisync}. 
The execution in ARMv8 has \emph{ordered-by} ($\ob$) cycle and hence not consistent. 
The $\ldr \mapsto \ldr;\cbisb$ mappings would result in respective $\ppo$ relations in the execution, but these $\ppo$ relations 
do not restrict such a cycle. 
As a result, the execution is ARMv7 or ARMv7-mca consistent. 
Hence $\ldr \mapsto \ldr;\cbisb$ is unsound in ARMv8 to ARMv7-mca mapping.

\subsection{Mapping schemes for programs generated from C11}
\label{sec:cmap}
In \cref{sec:cxarm}, \cref{sec:carmv8v7}, and \cref{sec:armv8toarmv7mca} we study $\cmm \!\mapsto\! \xarch \!\mapsto\! \arma$, $\cmm \!\mapsto\! \arma \!\mapsto\! \arms$, and $\cmm \!\mapsto\! \arma \!\mapsto\! \armsmca$ mapping schemes respectively. 
In these mappings from stronger to weaker models, we consider that the source architecture 
program is generated from a C11 program following the mapping in \cite{mappings}. 
We use this information to categorize the accesses in architectures as non-atomic ( $\MOna$) and atomic ($\MOat$), and 
exploit two aspects of C11 concurrency; 
first, a program with data race on non-atomic access results in undefined behavior. Second,  C11 uses atomic accesses to achieve synchronization and avoid data race on non-atomics.
Considering these properties we introduce leading or trailing fences in mapping particular atomic accesses 
and we map non-atomics to respective accesses without any leading or trailing fence.  

\smallskip
\textit{Pros and Cons}
$\cmm \mapsto \xarch \mapsto \arma$ scheme has a tradeoff; 
in case of non-atomics it is more efficient than $\xarch \mapsto \arma$ as it does not introduce 
additional fences whereas an atomic store mapping requires a leading full fence 
or a pair of $\idmbld$ and $\idmbst$ fences. 
Consider the mapping of the sequence: $\Rlab_{\MOna};\Wlab_{\MOna};\Wlab_{\MOrel} \ \ \mapsto \ \ \movld_{\MOna};
\movst_{\MOna};\movst_{\MOat} \ \ \mapsto \ \ \ldr;\str;\idmbfull;\movst_{\MOat}$.

In this case the C11 non-atomic memory accesses cannot be moved after the release write access. 
Hence we introduce a leading $\idmbfull$ with $\movst_{\MOat}$ in $\cmm \mapsto \xarch \mapsto \arma$ to preserve the same order. 
Consider the C11 to x86 to ARMv8 mapping of the program below. 
\[
\inparII{
a = X_\MOna; 
\\ Y_\MOna = 1;
\\ Z_\MOrel = 1;
}{
r = Z_\MOacq; 
\\ \Cif{r==1} \{
\\ \quad X_\MOna = 2;
\\ \quad Y_\MOna = 2;
\\ \quad c = Y_\MOna; 
\\ \}
}
\quad \mapsto \quad
\inparII{
a = X_\MOna; 
\\ Y_\MOna = 1;
\\ Z_\MOat = 1;
}{
r = Z_\MOat; 
\\ \Cif{r==1} \{
\\ \quad X_\MOna = 2;
\\ \quad Y_\MOna = 2;
\\ \quad c = Y_\MOna; 
\\ \}
}
\quad \mapsto \quad
\inparII{
a = X; 
\\ Y = 1;
\\ \idmbfull
\\ Z = 1;
}{
r = Z; 
\\ \idmbld
\\ \Cif{r==1} \{
\\ \quad X = 2;
\\ \quad Y=2;
\\ \quad c = Y; 
\\ \}
}
\] 
The C11 program is data race free as it is well-synchronized 
by release-acquire accesses on $Z$ and 
the outcome $a=2,r=c=1$ is disallowed in the program. 
The generated ARMv8 program disallows the outcome, however, 
without the $\idmbfull$ in the first thread the outcome would be possible. 
It is because a $\idmbld$ or $\idmbfull$ fence is required to 
preserve $\bob$ relation between 
$\Rlab(X,2)$ and $\Wlab(Z,1)$ events.
Note that a $\idmbld$ is not sufficient to establish $\bob$ relation between 
$\Wlab(Y,1)$ and $\Wlab(Z,1)$ and hence we require a $\idmbst$ or $\idmbfull$ fence. 
Therefore we have to introduce a leading pair of $\idmbld$ and $\idmbst$ fences or a $\idmbfull$ fence for $\movst_\MOat$ mapping. 
 
As a result 
\cref{tab:cxarm} provides more efficient mapping for $\movld_\MOna$ and $\movst_\MOna$ 
accesses, but incurs more cost for $\movst_\MOat$ by introducing 
a leading $\idmbfull$ instead of a $\idmbst$ fence. 
After the mapping we may weaken such a $\idmbfull$ fence whenever appropriate.  

The $\cmm \mapsto \arma \mapsto \arms$ scheme does not introduce fence for mapping non-atomics and 
therefore more efficient than $\arma \mapsto \arms$. 
Note that C11 $\Wlab_{\sqsupseteq\MOrel}$ generates an $\stlr$ in ARMv8 and 
ARMv8 $\str$ is generated only from C11 $\Wlab_{\sqsubseteq\MOrlx}$ 
which does not enforce any such order. 

%
%
%

\subsection{ARMv8 as an intermediate model for mappings between x86 and ARMv7}
\label{sec:armaintermediate}
Now we move to mappings between x86 and ARMv7. 
We do not propose direct mapping schemes, instead we use ARMv8 concurrency as an intermediate concurrency model as $\xarch \mapsto \arms/\armsmca$ and $\arms/\armsmca \mapsto \xarch$ would be same as $\xarch \mapsto \arma \mapsto \arms/\armsmca$ and $\arms/\armsmca \mapsto \arma \mapsto \xarch$ respectively.
%


\paragraph{$\xarch \mapsto \arms$ vs $\xarch \mapsto \arma \mapsto \arms$}
We derive x86 $\mapsto$ ARMv8 $\mapsto$ ARMv7 by combining 
x86 $\mapsto$ ARMv8 (\cref{tab:xarm}) and ARMv8 $\mapsto$ ARMv7 (\cref{tab:armv8v7}) as follows.
\\[1ex]
\begin{tabular}{ll}
$\quad \mfence \mapsto \idmbfull \mapsto \idmb$  \hfill & \hfill $ \qquad \irmw \mapsto \idmbfull;\irmw;\idmbfull \mapsto \idmb;\irmw;\idmb$
\\ $\quad \movld \mapsto \ldr;\idmbld \mapsto \ldr;\idmb$ & $\qquad  \movst \mapsto \idmbst;\str \mapsto \idmb;\str$ 
\end{tabular}
\\[1ex]
The correctness proofs of the x86 to ARMv8 and ARMv8 to ARMv7 mapping schemes in 
\cref{tab:xarm} and \cref{tab:armv8v7} demonstrate the necessity of the introduced fences.  
The introduced fences only allow reordering of an independent store-load access pair on different locations 
which is similar to the allowed reordering restriction of x86. 
Therefore the introduced fences are necessary and sufficient. 

\paragraph{$\arms \mapsto \xarch$ vs $\arms \mapsto \arma \mapsto \xarch$}
We derive ARMv7 $\mapsto$ ARMv8 $\mapsto$ x86  by combining 
ARMv7 $\mapsto$ ARMv8 (\cref{tab:armv7v8}) and ARMv8 to x86 (\cref{tab:armx}) as follows. 
Note that the mapping does not introduce any fence along with the accesses and therefore optimal.
\begin{center}
\begin{tabular}{ll}
$\qquad \qquad\idmb \mapsto \idmbfull \mapsto \mfence \qquad \qquad$  & $\qquad \qquad \irmw \mapsto \irmw \mapsto \irmw$
\\ $\qquad \qquad\ldr \mapsto \ldr \mapsto \movld \qquad \qquad$ &  $\qquad \qquad \str \mapsto \str \mapsto \movst$ 
\end{tabular}
\end{center}
%
%
\begin{center}
\begin{figure}
\begin{subfigure}{0.4\textwidth}
\[
\inparII{
a=X; \codecomment{1}
\\ c = Y[a];
\\ Z=1;
}{
b=Z; \codecomment{1}
\\ V[b] = 1;
\\ X=1;
}
\]
\end{subfigure}
\hfill\vline\hfill
\begin{subfigure}{0.55\textwidth}
\centering
\begin{tikzpicture}[yscale=0.8]
  \node (t11) at (-2.2,0)  {$\Rlab(X,1)$};
  \node (t12) at (-2.2,-1) {$\Rlab(Y[1],1)$};
  \node (t13) at (-2.2,-2)  {$\Wlab(Z,1)$};
  \node (t21) at (2.2,0) {$\Rlab(Z,1)$};
  \node (t22) at (2.2,-1)  {$\Wlab(V[1],1)$};
  \node (t23) at (2.2,-2) {$\Wlab(X,1)$};
%
  \draw[po] (t11) to  node[left]{$\addr$} (t12);
  \draw[po] (t12) to node[left]{$\lpo$} (t13);  
  \draw[po] (t21) to  node[right]{$\addr$} (t22);
  \draw[po] (t22) to  node[right]{$\lpo$} (t23);    
  \draw[rf,bend right=0] (t13.east) to node[below]{$\rfe$} (t21.west);
  \draw[rf,bend left=0] (t23.west) to (t11.east);
\end{tikzpicture}
\end{subfigure}
\caption{Load-store or store-store reorderings introduce $a=b=1$ outcome and are unsound in ARMv8.}
\label{fig:ex:reordering}
\end{figure}
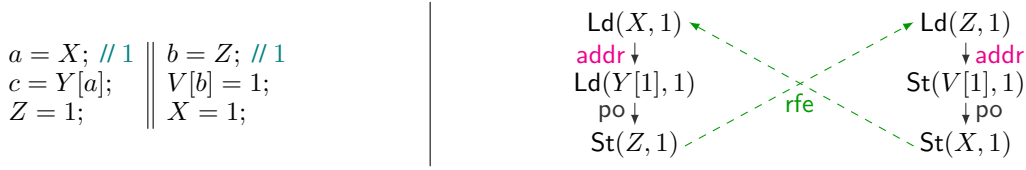
\end{center}
\vspace{-5mm}
\subsection{Common optimizations in ARMv8 concurrency} 
\begin{center}
\begin{figure}
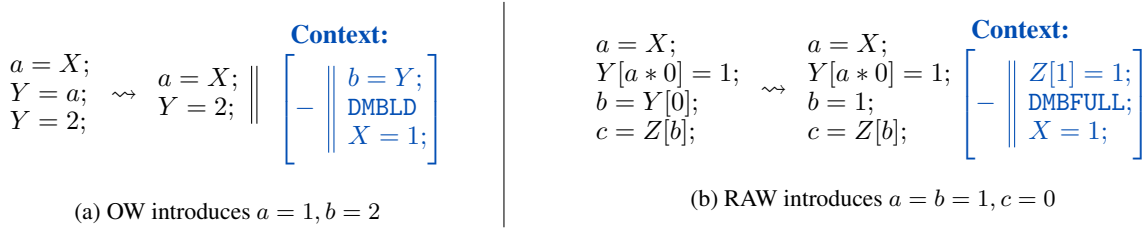

\begin{subfigure}{0.4\textwidth}
\[
\inarr{
a = X; 
\\ Y = a;
\\ Y = 2;
}
\ \ \leadsto\ \
\inparII{
a = X; 
\\ Y = 2;
}
\
\parCtx{
b = Y; 
\\ \idmbld
\\ X = 1;
}
\]
\caption{OW introduces $a=1,b=2$}
\label{fig:ow:elim}
\end{subfigure}
\hfill\vline\hfill
\begin{subfigure}{0.56\textwidth}
\[
\inarr{
a = X; 
\\ Y[a*0] =1;
\\ b = Y[0]; 
\\ c = Z[b]; 
}
\ \ \leadsto\ \
\inarr{
a = X; 
\\ Y[a*0] =1;
\\ b = 1; 
\\ c = Z[b]; 
}
\
\parCtx{
Z[1]=1; 
\\ \idmbfull;
\\ X=1;
}
\]
\caption{RAW introduces $a=b=1,c=0$}
\label{fig:raw:elim}
\end{subfigure}
\caption{Overwritten-write (OW) and Read-after-write elimination (RAW) are unsound in ARMv8.}
\end{figure}
\end{center}

We consider ARMv8 as a concurrency model of an IR and 
find that many common compiler optimizations are unsound in ARMv8.
\begin{itemize}[leftmargin=*]

\item \textit{\emph{ARMv8} does not allow store-store and load-store reorderings}
Consider the program and the execution in \cref{fig:ex:reordering}.
In this execution there are $\addr;[\Rlab];\lpo;[\Wlab]$ and 
$\addr;[\Wlab];\lpo;[\Wlab]$ relations in the first and second threads 
respectively which result in $\dob$ relations and in turn an $\ob$ cycle.
Therefore the execution is not ARMv8 consistent and the outcome $a=b=1$ is  disallowed. 
However, load-store reordering $c=Y[a];Z=1 \leadsto Z=1;c=Y[a]$ or store-store reordering $V[b]=1;Z=1 \leadsto Z=1;V[b]=1$ remove the respective $\dob$ relation(s) and enable $a=b=1$ in the target. 
Thus store-store and load-store reorderings are unsafe in ARMv8.   
\item \textit{Overwritten-write $\mathsf{(OW)}$ is unsound.} 
Consider the program and its outcome $a=1,b=2$ in \cref{fig:ow:elim}. 
In the respective execution the first thread has $\data;\coi \subseteq \dob$ from $\Rlab(X,1)$ to $\Wlab(Y,2)$. 
The other thread has a $\bob$ relation due to $\idmbfull$ fence 
which in turn create an $\ob$ cycle.  
Hence the execution is not ARMv8 consistent and the outcome $a=1,b=2$ is disallowed. 
Overwriting $Y=a$ in the first thread removes the $\dob$ relation and then $a=1,b=2$ becomes possible.
\item \textit{Read-after-write $\mathsf{(RAW)}$ is unsound.} 
We study the RAW elimination in \cref{fig:raw:elim} which is performed based on dependence analysis. 
Before we go to the transformation, we briefly discuss dependence analysis on 
the access sequence $a=X;Y[a*0]=1$. 
In this case there is a false dependence from load of $X$ to store of $Y[a*0]$ as $a*0=0$ always. 
ARMv8 does not allow to remove such a false dependence \cite{Pulte:2018}. 
However, we observe that using a static analysis that distinguishes between true and false dependencies is also wrong in ARMv8. 
In this example we analyze such a false dependency and 
based on that we perform read-after-write elimination on the program, that is, $Y[a*0]=1;b=Y[0] \leadsto Y[a*0]=1;b=1$ . 

The source program does not have any execution $a=1, b=1, c=0$ as 
$\addr;\rfi;\addr \subseteq \dob$ and in the other thread there is a $\bob$ reltion which together  create an $\ob$ cycle. 
In the target execution there is no $\dob$ relation from the load of $X$ to the load of $c=Z[b]$ 
and therefore the outcome $a=1, b=1, c=0$ is possible.   
As a result, the transformation is unsound in ARMv8.
\end{itemize}
\vspace{-7mm}
\subsection{Fence eliminations in ARMv8}
The mapping schemes introduce leading and/or trailing fences for various memory accesses. 
However, some of these fences may be redundant can safely be eliminated. 
Consider the $\xarch \mapsto \arma$ mapping and subsequent redundant fence 
eliminations below.  
\[
\inarr{
\movld;\mfence;\movst \quad \mapsto \quad \ldr;\idmbld;\idmbfull;\idmbst;\str \quad \leadsto \quad \ldr;\idmbld;\str
}
\]
The ARMv8 access sequence generated from x86 to ARMv8 mapping introduces 
three intermediate fences between the load-store pair. 
A $\idmbld$ fence suffices to order a load-store pair and hence the $\idmbfull$ as well as the $\idmbst$ fence are redundant and are safely eliminated.

To perform such fence eliminations, we first identify \emph{non-deletable} fences and then 
delete rest of the fences.
A fence is \emph{non-deletable} if it is placed between a memory access pair in at least one program path so that the access pair may have out-of-order execution without the fence.
Analyzing the ARMv8 sequence above we mark the $\idmbld$ as non-deletable and rest of the fences as redundant.
%
%
\subsection{Analyzing and enforcing robustness}
\label{sec:idearobust}
There are existing approaches  \cite{Lahav:2019,Bouajjani:2013} 
which explores program executions to answer such queries. 
We propose an alternative approach by analyzing memory access sequences. 
In this analysis 
\begin{enumerate}[leftmargin=*]
\item We identify the program components which may run concurrently. 
Currently we consider fork-join parallelism and identify the 
functions which create one or multiple threads.
Our analysis considers that each of such functions creates multiple threads. 
Therefore analyzing these functions $f_1, \ldots f_n$,  
we analyze all programs of the form $f_1 \mid\mid \cdots \mid\mid f_1 \mid\mid f_n \mid\mid \cdots \mid\mid f_n$.

\item Next, we analyze the memory access sequences in $f_1, \ldots f_n$ to check whether the memory access pairs in these functions may create a cycle.  

\item In case a cycle is possible, we check if each access pair on a cycle is ordered by robustness condition. 
If so, then all $K$ consistent executions of these programs are also $M$ consistent. 
\end{enumerate}
Consider the example in \cref{fig:sbrobustness}. 
We analyze the access sequences in thread functions $\sbprg$, $\sbprg'$, and $\sbprg''$ and derive 
a graph by memory access pairs which contains a cycle by the memory access pairs in $\sbprg$ and $\sbprg'$. 
These pairs on the cycle have intermediate $\mfence$ operations which 
enforce interleaving executions only irrespective of the number of threads 
created from $\sbprg$, $\sbprg'$, $\sbprg''$.
Our analysis reports these x86 programs as SC-robust.   
Using this approach we check $M$-robustness against $K$ where $K$ is an weaker models than $M$.
%
\begin{center}
\begin{figure}[t!]
\begin{subfigure}{0.35\textwidth}
\begin{align*}
\sbprg \triangleq & \tup{X = 1; \ \mfence; \ t = Y;} 
\\ \sbprg' \triangleq & \tup{Y = 1; \ \mfence; \ t = X;} 
\\ \sbprg'' \triangleq & \tup{Y = 1; \ t = Z;} 
\end{align*}
\label{fig:sbfunc}
\end{subfigure}
\hfill
\begin{subfigure}{0.6\textwidth}
\centering
\begin{tikzpicture}[yscale=0.8]
  \node (t11) at (-3,0)  {$\Wlab(X,\_)$};
  \node (t12) at (-3,-1) {$\mfence$};
  \node (t13) at (-3,-2) {$\Rlab(Y,\_)$};
  \node (t21) at (0,0)  {$\Wlab(Y,\_)$};
  \node (t22) at (0,-1) {$\mfence$};
  \node (t23) at (0,-2) {$\Rlab(X,\_)$};
  \node (t31) at (3,0) {$\Wlab(Y,\_)$};
  \node (t32) at (3,-2) {$\Rlab(Z,\_)$};
  \draw[po] (t11) to (t12);
  \draw[po] (t12) to (t13);  
  \draw[po] (t21) to (t22);
  \draw[po] (t22) to (t23);
   \draw[po] (t31) to (t32);
  \draw[fr, thick] (t13) to (t21);
  \draw[fr, thick] (t23) to (t11);
  \draw[fr,thick,bend right=15] (t13) to (t31);
\end{tikzpicture}
\label{fig:sbloc}
\end{subfigure}
\caption{
A program of the form $\sbprg \mid\mid \cdots \mid\mid \sbprg \mid\mid \sbprg' \mid\mid \cdots \mid\mid \sbprg' \mid\mid \sbprg'' \mid\mid \cdots \mid\mid \sbprg''$ is SC-robust against x86.}
\label{fig:sbrobustness}
\end{figure}
\end{center}
\vspace{-5mm}
\textit{Enforcing robustness.} If we identify robustness violation for a program 
then we identify memory access pairs which may violate a robustness condition. 
For these access pairs we introduce intermediate fences to 
enforce robustness against a stronger model.  

\paragraph{$\ppo$ does not suffice to enforce robustness in ARMv7}
In addition to fences, $\ppo$ relations also orders a pair of accesses on different locations. 
However, we observe that $\ppo$ relations are not sufficient to ensure robustness 
for ARMv7 model. 
 
Consider the execution in \cref{fig:pponrobust},  
the execution allows the cycle and violates SC robustness. 
Therefore $\ppo$ cannot be used to order $\epo$ relations to preserve robustness.  

\begin{center}
\begin{figure}
\begin{tikzpicture}[yscale=0.8]
  \node (t11) at (-6,-1)  {a:$\Rlab(A,1)$};
  \node (t12) at (-6,-4) {b:$\Wlab(X,2)$};
  \node (t21) at (-4,-2.5)  {c:$\Wlab(X,1)$};
  \node (t31) at (-1,-1)  {d:$\Rlab(X,1)$};
  \node (t32) at (-1,-4) {e:$\Wlab(Y,1)$};
  \node (t41) at (2,-1)  {f:$\Rlab(Y,1)$};
  \node (t42) at (2,-4) {g:$\Wlab(Z,1)$};
  \node (t51) at (4,-2.5) {h:$\Wlab(Z,2)$};
  \node (t61) at (6,-1)  {i:$\Rlab(Z,2)$};
  \node (t62) at (6,-4) {j:$\Wlab(A,1)$};
%
  \draw[po] (t11) to node[left]{$\ppo$} (t12);
  \draw[po] (t31) to node[left]{$\ppo$} (t32);
  \draw[po] (t41) to node[left]{$\fence$} (t42);
  \draw[po] (t61) to node[left]{$\ppo$} (t62);
  \draw[mo,bend left=0] (t12) to node[right] {$\coe$} (t21);
  \draw[rf,bend left=0]  (t21) to node[right] {$\rfe$} (t31);
  \draw[rf,bend left=0]  (t32) to node[right] {$\rfe$} (t41);
  \draw[mo,bend left=0]  (t42) to node[right] {$\coe$} (t51);
  \draw[rf,bend left=0]  (t51) to node[right] {$\rfe$} (t61);
  \draw[rf,bend right=45]  (t62) to node[above] {$\rfe$} (t11);
\end{tikzpicture}
\caption{Execution $\prop(b,g) \land \coe(g,h) \land \ahb(h,b)$ cycle is allowed.}
\label{fig:pponrobust}
\end{figure}
\end{center}

\section{Formal Models}
\label{sec:models}
\paragraph{Syntax} 
Instead of delving into the syntactic notations in  
each instruction set, we use common expressions and commands 
which can be extended in each architecture.
\begin{align*}
E ::= & r \mid  v \mid X \mid E + E \mid E * E \mid E \leq E \mid  \cdots & (Expr)
\\ C ::= & \skips \mid C;C \mid r = E \mid r = X \mid  X = E \mid r = \irmw(X, E, E) \mid r = \irmw(X, E) \mid  \cdots &
\\ &  \mid \br \ label \mid \br \ label \ label  & (Cmd)
\\ P ::= & X = v; \cdots X = v; \set{C \mid \cdots \mid C} & (Program)
\end{align*}

In this notation we use $X \in \Locs$, $r \in \regs$, and $v\in \Val$ where 
$\Locs$, $\regs$, $\Val$ denote finite sets of memory locations, registers, and values 
respectively.
A program $P$ consists of a set of initialization writes followed by a parallel composition of thread commands.  
\paragraph{Semantics}
We follow the per-execution based axiomatic models for these architectures. 
In these models a program's semantics is defined by a set of consistent executions. 
An execution consists of a set of events and relations among the events. 

Given a binary relation $R$ on events, $R^{-1}$, $R^?$, $R^+$, and $R^*$ represent 
inverse, reflexive, transitive, and reflexive-transitive closures of $R$ respectively. 
$\dom(R)$ and $\codom(R)$ denote is its domain and its range respectively. 
Relation $R$ is total on set $S$ when $\total(S,R) \triangleq \forall a,b \in S.~a=b \lor R(a,b) \lor R(b,a)$. 
We compose binary relations $R,S \subseteq \E \times \E$ relationally by $R \mathbin{;}S$. 
$[A]$ denotes an identity relation on a set $A$.
We write $\rloc R$ to denote $R$ related event pairs on same locations, that is, 
$\rloc R \triangleq \set{ (\event,\event')\in R \mid \event.\lloc = \event'.\lloc }$. 
Similarly, $\rnloc R \triangleq R \setminus \rloc R$ is the $R$ related event pairs on different locations.

\begin{definition}
An event is of the form $\langle \lid,\ltid, \llab \rangle$, where $\lid$, $\ltid\in\Nlab$,and $\llab$ are the unique identifier, thread id, and the label of the event based on the respective executed memory access or fence instruction. A label is of the form $\tup{\op,\lloc,\rVal, \wVal}$.
\end{definition}
For an event $\event$, whenever applicable, $\event.\llab$, $\event.\op$, $\event.\lloc$ , $\event.\rVal$, and $\event.\wVal$ to return the label, operation type, location, read value, and written value respectively.
We write $\Rlab$, $\Wlab$, $\Ulab$, and $\Flab$ to represent the set of load, store, update, and fence events. 
Moreover, load or update events represent read events ($\RUlab$) and store or update events are write events ($\WUlab$), that is $\RUlab = \Rlab \cup \Ulab$ and $\WUlab = \Wlab \cup \Ulab$. 
We write $\denot{i}$ to represent the generated event in the respective model from an instruction $i$. 
For example, in x86 $\denot{i} \in \Wlab$ holds when $i$ is a $\movst$ instruction. 
We also overload the notation as $\denot{\mbbp}_M$ to denote the set of execution of program $\mbbp$ 
in model $M$.

In an execution events are related by various types of relations. 
Relation program-order($\lpo$) captures the syntactic order among the events. 
We write $a.b$ to denote that $b$ is immediate $\lpo$-successor of event $a$. 
Reads-from ($\lrf$) associates a write event to a read event that justifies its read value. Relation coherence-order($\co$) is a total-order on same-location writes (stores or updates). 
The from-read ($\fr$) relation relates a pair of same-location read and write events. 
We also categorize the relations as external and internal relations and 
define \emph{extended-coherence-order} ($\eco$).
Relation modification order ($\mo$) is a total-order on writes, updates, and fences such that 
$\mo \subseteq O \times O$ where $O = \Wlab \cup \Ulab \cup \Flab$. 
Note that the $\co$ relation is included in the $\mo$ relation.
The $\mo$ relation is used in x86 model only; the ARM models do not use $\mo$ in their definitions. 
\begin{definition}
An execution is of the form $\ex = \langle \E, \lpo, \lrf, \co, \mo \rangle$ where $\ex.\E$ denotes the set of memory access or fence events and $\ex.\lpo$, $\ex.\lrf$, $\ex.\co$, and $\ex.\mo$ denote the set of program-order, reads-from, coherence order, and modification order relations between the events in $\ex.\E$.
\end{definition} 
\subsection{Concurrency models of x86, ARMv7, ARMv7-mca, and ARMv8}
We now discuss the architectures and follow the axiomatic models of x86 and ARMv7 from \citet{Lahav:2017}, and ARMv8 axiomatic model from \citet{Pulte:2018}. 
We also present ARMv7-mca; a strengthened ARMv7 model with multicopy atomicity (MCA). 

\smallskip
\textbf{x86.} In x86 $\mov$ instruction is used for both loading a value from memory as well as for storing a value to memory.  
To differentiate these two accesses we categorize them as $\movst$ and $\movld$ operations. 
In addition, there are atomic update operations which we denote by $\irmw$.
x86 also provides $\mfence$ which flushes buffers and caches and ensure ordering between the preceding and following memory accesses.

In x86 concurrency $\movst$, $\movld$,  and $\mfence$ generate $\Wlab$, $\Rlab$, and $\Flab$ events respectively.
A successful $\irmw$ generates $\Ulab$ and otherwise an $\Rlab$ event. 
We derive \emph{x86-happens-before} ($\xhb$) relation from program-order and reads-from relations:  
$\xhb \triangleq (\lpo \cup \lrf)^+$.
An x86 execution $\ex$ is consistent when:
\begin{itemize}
\item $\ex.\xhb$ is irreflexive. \hfill (irrHB)
\item $\ex.\mo;\ex.\xhb$ is irreflexive. \hfill (irrMOHB)
\item $\ex.\fr;\ex.\xhb$ is irreflexive. \hfill (irrFRHB)
\item $\ex.\fr;\ex.\mo$ is irreflexive. \hfill (irrFRMO)
\item $\ex.\fr;\ex.\mo;\ex.\rfe;\ex.\lpo$ is irreflexive  \hfill (irrFMRP)
\item $\ex.\fr;\ex.\mo;[\ex.\Ulab \cup \ex.\Flab];\ex.\lpo$ is irreflexive. \hfill (irrUF)
\end{itemize}

\smallskip
\textbf{ARMv7.} It provides $\ldr$ and $\str$ instructions for load and store operations,   
and load-exclusive ($\ldrex$) and store-exclusive($\strex$) instructions to perform atomic update operation $\irmw$ where 
$\irmw \triangleq \mathsf{L:} \ \ldrex; \mathsf{mov};\mathsf{teq \ L'};\strex; \mathsf{teq \ L}; \mathsf{L':}$. 
ARMv7 provides full fence $\idmb$ which orders preceding and following instructions. 
There is also lightweight control fence $\isb$ which is used to construct 
$\cbisb \triangleq \mathsf{cmp}; \mathsf{bc}; \isb$ to order load operations.  

In this model load ($\Rlab$), store ($\Wlab$), $\dmb$ events are generated from the execution of 
$\ldr$ and $\ldxr$, $\str$ and $\stxr$, and $\idmb$ instructions respectively. 
Fence $\isb$ is captured in $\ctrl_\isb$ (similar to $\ctrl_\isync$ in \cite{Lahav:2017}) 
and in turn $\ppo$ relation, but does not create any event in an execution. 
 
ARMv7 defines preserved-program-order ($\ppo$) relation which is a subset of program-order relation. 

We first discuss the primitives of $\ppo$ following \S F.1 in \citet{Lahav:2017}: $\ppo$ is 
based on data ($\subseteq \Rlab \times \Wlab$), control ($ \subseteq \Rlab \times \E$), and address ($ \subseteq  \Rlab \times (\Rlab \cup \Wlab)$) dependencies. 
Moreover, $\isb$ fences along with conditionals introduce $\ctrl_\isb \subseteq \ctrl$ preserved program order. 
Finally,  $\ctrl;\lpo \subseteq \ctrl$ and $\ctrl_\isb;\lpo \subseteq \ctrl_\isb$ holds from definition. 

Based on these primitives ARMv7 define \emph{read-different-writes} ($\rdw$) and \emph{detour} ($\detour$) 
relations as follows.

\begin{center}
\begin{tabular}{cc}
$\rdw \triangleq (\fre;\rfe)\subseteq \lpo$ & $\detour \triangleq (\coe;\rfe) \setminus \lpo$
\end{tabular}
\end{center} 

\emph{read-different-writes} ($\rdw$) relates two reads on same location in a thread which reads from different writes and 
detour captures the scenario where an external write takes place between a pair of same-location write in the same thread, 
and the read reads-from that external write. 

Based on these primitives ARMv7 defines $\ii_0$, $\ci_0$, $\ic_0$, $\cc_0$ components as follows.

$\ii_0 \triangleq \addr \cup \data \cup \rdw \cup \rfi$  \hfill $\ic_0 \triangleq \emptyset$
\hfill $\ci_0 \triangleq \ctrl_\isb \cup \detour$ \hfill $\cc_0 \triangleq \data \cup \ctrl \cup \addr; \lpo^?$
  
Using these components ARMv7 defines $\ii$, $\ic$, $\ci$, $\cc$ relations where each of these relations can be 
derived from the following sequential compositions and the constraints.
\[
\inarr{
xy \triangleq \bigcup_{n \geq 1} x^1 y^1_0; x^2 y^2_0; \cdots x^n y^n_0 
}
\]  
where
\begin{itemize}
\item $x, y, x^1 \cdots x^n, y^1 \cdots y^n \in \set{\mathsf{i}, \mathsf{c}}$.
\item If $x = \mathsf{c}$ then $x^1 = \mathsf{c}$.
\item For every $1 \leq k \leq n-1$, if $y^k = \mathsf{c}$ then $x^{k+1} = \mathsf{c}$.
\item If $y = \mathsf{i}$ then $y^n = \mathsf{i}$.
\end{itemize}
Finally ARMv7 defines $\ppo$ as follows: $\ppo \triangleq [\Rlab];\ii;[\Rlab] \cup [\Rlab];\ii;[\Wlab]$. 
ARMv7 also defines $\fence$, ARM-happens-before ($\ahb$), and propagation ($\prop$) relations as follows.
\[
\inarr{
\fence \triangleq [\Rlab \cup \Wlab];\lpo; [\dmb]; \lpo; [\Rlab \cup \Wlab]
\\ \ahb \triangleq \ppo \cup \fence \cup \rfe
\\ \prop \triangleq \prop_1 \cup \prop_2 \text{ where }
\\ \prop_1 \triangleq [\Wlab]; \rfe^?; \fence; \ahb^*;[\Wlab] \text{ and }
\\  \prop_2 \triangleq (\coe \cup \fre)^?;\rfe^?; (\fence; \ahb^*)^?;\fence; \ahb^*
}
\]
These relations are used to define the consistency constraints of an ARMv7 execution $\ex$ as follows:
\begin{itemize}

\item $\ex.\co$ is total \hfill (total-co)

\item $(\ex.\poloc \cup \ex.\lrf \cup \ex.\fr \cup \ex.\co)$ is acyclic \hfill (sc-per-loc)

\item $\ex.\fre;\ex.\prop;\ex.\ahb^*$ is irreflexive. \hfill (observation)

\item $(\ex.\co \cup \ex.\prop)$ is acyclic. \hfill (propagation) 

\item $[\ex.\rmw];\ex.\fre;\ex.\coe$ is irreflexive \hfill (atomicity)

\item $\ex.\ahb$ is acyclic \hfill (no-thin-air)
\end{itemize}

\smallskip
\textbf{ARMv7-mca.}
We strengthen the ARMv7 model and define ARMv7-mca model 
to support multicopy atomicity. 
To do so, following \citet{Wickerson:2017}, 
we define \emph{write-order} ($\wo$) and impose the additional constraint on ARMv7 
as defined in \cref{fig:consistency}.
\begin{itemize}
\item $\ex.\wo^+ \text{ is acyclic where } \wo = (\rfe;\ppo;\fre)$ \hfill (mca)
\end{itemize}

%
\begin{figure}
\centering
\begin{subfigure}{0.45\textwidth}
\begin{tabular}{l|l}
\quad x86 & \qquad ARMv8
\\ \hline
$\movld$ & $\ldr;\idmbld$  
\\ $\movst$ & $\idmbst;\str$
\\ $\irmw$ & $\idmbfull;\irmw;\idmbfull$ 
\\ $\mfence$ & $\idmbfull$
\end{tabular}
\caption{x86 to ARMv8}
\label{tab:xarm}
\end{subfigure}
\hfill
\begin{subfigure}{0.45\textwidth}
\begin{tabular}{l|l}
\quad C11 to x86 & \qquad ARMv8
\\ \hline
$\movld_\MOna$ & $\ldr$  
\\ $\movst_\MOna$ & $\str$
\\ $\movld_\MOat$ & $\ldr;\idmbld$  
\\ $\movst_\MOat$ & $\idmbfull;\str$
\\ $\irmw$ & $\idmbfull;\irmw;\idmbfull$ 
\\ $\mfence$ & $\idmbfull$
\end{tabular}
\caption{C11 to x86 to ARMv8}
\label{tab:cxarm}
\end{subfigure}
\caption{Mapping schemes from x86 to ARMv8.}
\label{fig:xarm}
\end{figure}

\smallskip
\textbf{ARMv8.} provides load ($\ldr$), store ($\str$) for load and store operations, 
load-exclusive ($\ldxr$) and store-exclusive ($\stxr$) instructions to construct $\irmw$ 
similar to that of ARMv7. 
In addition, ARMv8 provides load-acquire ($\ldar$), store-release ($\stlr$), load-acquire exclusive 
($\ldaxr$), and store-release exclusive ($\stlxr$) instructions which operate as half fences.
In addition to $\idmbfull$ and $\isb$, ARMv8 provides load ($\idmbld$) and store ($\idmbst$) fences. 
A $\idmbld$ fence orders a load with other accesses and a $\idmbst$ orders a pair of store accesses.

%
%
Based on these primitives ARMv8 defines coherence-after ($\mathsf{ca}$), observed-by($\obs$), and atomic-ordered-by ($\aob$) relations on same-location events. 
ARMv8 also defines dependency-ordered-before ($\dob$) and barrier-ordered-by ($\bob$) relations to order a pair of intra-thread events. 
Finally \emph{Ordered-before} ($\ob$) is a transitive closure of $\obs$, $\aob$, $\dob$, and $\bob$ relations. 
\[
\inarr{
\ca \triangleq \fr \cup \co \hspace{10mm}
\obs \triangleq \rfe \cup \fre \cup \coe \hspace{10mm}
\aob \triangleq \rmw \cup [\frange(\rmw)];\rfi; [\Alab]
\\[1.1ex]
\dob \triangleq \addr \cup \data \cup \ctrl;[\Wlab]\cup (\ctrl \cup (\addr; \lpo)); [\isb]; \lpo; [\Rlab] 
\\ \qquad \quad \cup \addr; \lpo; [\Wlab] 
\cup (\ctrl \cup \data); \coi \cup (\addr \cup \data); \rfi
\\[1.1ex]
\bob \triangleq \lpo; [\dmbfull]; \lpo 
 \cup [\Llab]; \lpo; [\Alab]; 
 \cup [\Rlab]; \lpo; [\dmbld]; \lpo
 \cup [\Alab]; \lpo
\\ \qquad \quad \cup [\Wlab]; \lpo; [\dmbst]; \lpo; [\Wlab]
 \cup \lpo; [\Llab]
 \cup \lpo; [\Llab]; \coi
\\[1.1ex]
\hspace{2mm} \ob \triangleq (\obs \cup \dob \cup \aob \cup \bob)^+ 
}
\]
Finally an ARMv8 execution $\ex$ is consistenct when:
\begin{itemize}
\item $\ex.\poloc \cup  \ex.\ca \cup \ex.\lrf$ is irreflexive. \hfill  (internal)

\item $\ex.\ob$ is irreflexive \hfill  (external)

\item $\ex.\rmw \cap (\ex.\fre;\ex.\coe) = \emptyset$ \hfill (atomic)
\end{itemize}

\section{Architecture to Architecture Mappings}
\label{sec:mappings}
We propose correct and efficient mapping schemes between x86 and ARM models.
%
%
These schemes may introduce leading and/or trailing fences 
while mapping memory accesses from one architecture to another.
%
We show that the fences are necessary by examples and prove  
that the fences are sufficient for correctness. 
To prove correctness we show that for each consistent execution of the target program after mapping there exists a corresponding consistent execution of the source program before mapping with same behavior. 

\subsection{x86 to ARMv8 mapping}
\label{sec:x86toarmv8}
The mapping scheme from x86 to ARMv8 is in \cref{tab:xarm}. 
The scheme generates  
a $\idmbfull$ for an $\mfence$. 
While mapping x86 memory accesses to that of ARMv8, the scheme 
introduces a leading $\idmbst$ fence with a store, a trailing $\idmbld$ fence with a load, and 
leading as well as a trailing $\idmbfull$ fences with an update. 
We now discuss why these fences are required.

\begin{center}
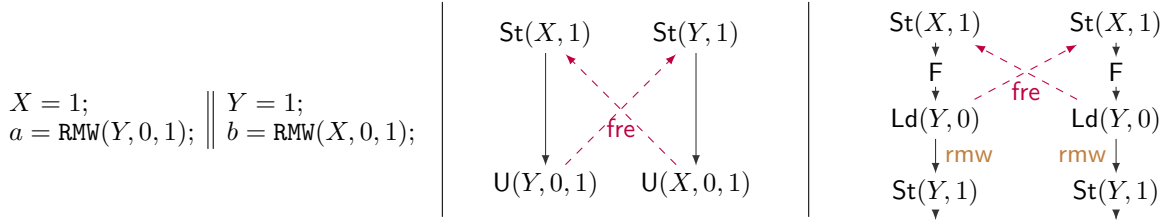
\begin{figure}
\begin{minipage}{0.35\textwidth}
\[
\inparII{
X = 1; 
\\ a = \irmw(Y,0,1);
}{
Y = 1; 
\\ b = \irmw(X,0,1); 
}
\]
\end{minipage}
\hfill\vline\hfill
\begin{minipage}{0.28\textwidth}
\centering
\begin{tikzpicture}[yscale=0.8]
  \node (t11) at (-1,0)  {$\Wlab(X,1)$};
  \node (t12) at (-1,-2.5) {$\Ulab(Y,0,1)$};
  \node (t21) at (1,0) {$\Wlab(Y,1)$};
  \node (t22) at (1,-2.5) {$\Ulab(X,0,1)$};
  \draw[po] (t11) to (t12);
  \draw[po] (t21) to (t22);
  \draw[fr] (t12) to node[below]{$\fre$} (t21);
  \draw[fr] (t22) to (t11);
\end{tikzpicture}
\end{minipage}
\hfill\vline\hfill
\begin{minipage}{0.34\textwidth}
\centering
\begin{tikzpicture}[yscale=0.8]
  \node (t11) at (-1.2,0)  {$\Wlab(X,1)$};
  \node (t12) at (-1.2,-0.8) {$\dmbfull$};
  \node (t13) at (-1.2,-1.6)  {$\Rlab(Y,0)$};
  \node (t14) at (-1.2,-2.8)  {$\Wlab(Y,1)$};
  \node (t21) at (1.2,0)  {$\Wlab(X,1)$};
  \node (t22) at (1.2,-0.8) {$\dmbfull$};
  \node (t23) at (1.2,-1.6)  {$\Rlab(Y,0)$};
  \node (t24) at (1.2,-2.8)  {$\Wlab(Y,1)$};
  \draw[po] (t11) to (t12);
  \draw[po] (t12) to (t13);  
  \draw[po] (t13) to node[right]{$\rmw$} (t14);  
  \draw[po] (t14) to (-1.2,-3.2);  
  \draw[po] (t21) to (t22);
  \draw[po] (t22) to (t23);  
  \draw[po] (t23) to node[left]{$\rmw$} (t24);  
  \draw[po] (t24) to (1.2,-3.2);  
  \draw[fr] (t13) to node[below]{$\fre$} (t21);
  \draw[fr] (t23) to  (t11);
\end{tikzpicture}
\end{minipage}
\caption{In x86 to ARMv8 mapping $\irmw$ requires a leading $\dmbfull$ fence.}
\label{fig:app:xrmwleadf}
\end{figure}
\end{center}
\paragraph{Leading store fence} 
In an x86 execution a pair of stores is ordered unlike that of ARMv8 execution. 
A pair of store events ($\Wlab$) in ARMv8 execution are $\bob$ ordered 
when there is intermediate $\dmbst$ or $\dmbfull$ event, that is $[\Wlab];\lpo;[\dmbst \cup \dmbfull];\lpo;[\Wlab] \subseteq \bob$.
To introduce such a $\bob$ order we require at least an intermediate $\dmbst$ fence event.  
Therefore the scheme generates a leading $\idmbst$ fence 
with a store which ensures store-store order with preceding stores in ARMv8. 
\begin{center}
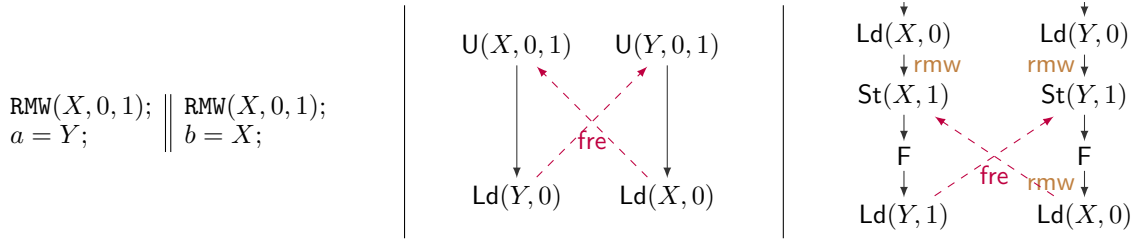
\begin{figure}
\begin{minipage}{0.35\textwidth}
\[
\inparII{
\irmw(X,0,1);
\\ a= Y;
}{
\irmw(X,0,1);
\\ b = X; 
}
\]
\end{minipage}
\hfill\vline\hfill
\begin{minipage}{0.28\textwidth}
\centering
\begin{tikzpicture}[yscale=0.8]
  \node (t11) at (-1,0)  {$\Ulab(X,0,1)$};
  \node (t12) at (-1,-2.5) {$\Rlab(Y,0)$};
  \node (t21) at (1,0) {$\Ulab(Y,0,1)$};
  \node (t22) at (1,-2.5) {$\Rlab(X,0)$};
  \draw[po] (t11) to (t12);
  \draw[po] (t21) to (t22);
  \draw[fr] (t12) to node[below]{$\fre$} (t21);
  \draw[fr] (t22) to (t11);
\end{tikzpicture}
\end{minipage}
\hfill\vline\hfill
\begin{minipage}{0.32\textwidth}
\centering
\begin{tikzpicture}[yscale=0.8]
  \node (t11) at (-1.2,0)  {$\Rlab(X,0)$};
  \node (t12) at (-1.2,-1) {$\Wlab(X,1)$};
  \node (t13) at (-1.2,-2)  {$\dmbfull$};
  \node (t14) at (-1.2,-3)  {$\Rlab(Y,1)$};
  \node (t21) at (1.2,0)  {$\Rlab(Y,0)$};
  \node (t22) at (1.2,-1) {$\Wlab(Y,1)$};
  \node (t23) at (1.2,-2)  {$\dmbfull$};
  \node (t24) at (1.2,-3)  {$\Rlab(X,0)$};
  \draw[po] (t11) to node[right]{$\rmw$} (t12);
  \draw[po] (t12) to  (t13);  
  \draw[po] (t13) to (t14);  
  \draw[po] (-1.2,0.5) to (t11);  
  \draw[po] (t21) to node[left]{$\rmw$} (t22);
  \draw[po] (t22) to (t23);  
  \draw[po] (t23) to node[left]{$\rmw$} (t24);  
  \draw[po] (1.2,0.5) to (t21);  
  \draw[fr] (t14) to node[below]{$\fre$} (t22);
  \draw[fr] (t24) to  (t12);
\end{tikzpicture}
\end{minipage}
\caption{In x86 to ARMv8 mapping $\irmw$ requires a trailing $\dmbfull$ fence.}
\label{fig:app:xrmwtrailf}
\end{figure}
\end{center}
\paragraph{Trailing load fence}
We know a load-store or load-load access pair is ordered in x86. 
To preserve the same access ordering we require a $\dmbld$ 
fence between a load-load or load-store access pair. 
Therefore the scheme generates a trailing $\idmbld$ fence 
with a load which ensures such order. 

\smallskip
\textit{Leading and trailing fence for atomic update} 
Consider the x86 programs and $a=b=0$ outcome. 

No x86 execution would allow $a=b=0$ in the two programs in \cref{fig:app:xrmwleadf,fig:app:xrmwtrailf}. 
However, if we translate these programs without intermediate $\idmbfull$ 
fences between each pair of store and $\irmw$ accesses then $a=b=0$ 
would be possible in these two programs in ARMv8 as shown 
in the corresponding executions. 
As a result, the translations from x86 to ARMv8 would be unsound.
The leading and trailing $\idmbfull$ fences with $\irmw$ accesses 
provide these intermediate 
fences in the respective program to  
disallow $a=b=0$ in both programs. 
%
%

\paragraph{Mapping correctness} 
These fences suffice to preserve mapping correctness as stated in \cref{th:xarm} and proved  in \cref{app:x86toarmv8}.
\begin{restatable}{theorem}{xarm}
\label{th:xarm}
The mappings in \cref{tab:xarm} are correct. 
\end{restatable}

%
\subsection{C11 to x86 to ARMv8 mapping}
\label{sec:cxarm}
In this mapping from x86 to ARMv8 we exploit the C11 semantic rule: data race results in undefined behavior. 
The mapping scheme is in \cref{tab:cxarm}.  In this scheme we categorize the x86 
load and store accesses by whether they are generated from C11 non-atomic or atomic accesses. 
If we know that a load/store 
access is generated from a C11 non-atomic load/store then we do not introduce any trailing or leading fence. 
We prove the correctness of the scheme (\cref{th:cxarm}) in \cref{app:cxarm}.
\begin{restatable}{theorem}{cxarm}
\label{th:cxarm}
The mapping scheme in \cref{tab:cxarm} is correct.
\end{restatable}
In \cref{sec:cmap} we have already demonstrated the tradeoff between the $\xarch \mapsto \arma$ and $\cmm \mapsto \xarch \mapsto \arma$ mapping schemes.
\begin{center}
\begin{figure}[t]
\begin{subfigure}[b]{0.47\textwidth}
\centering
\begin{tabular}{l|l}
ARMv7/ARMv7-mca  &  ARMv8    \\
\hline
$\ldr$ & $\ldr$ \\
$\str$ &  $\str$ \\
$\irmw$ &  $\irmw$ \\
$\idmb$ &  $\idmbfull$  \\
$\isb$ &  $\isb$ \\
\end{tabular}
\caption{ARMv7 or ARMv7-mca to ARMv8}
\label{tab:armv7v8}
\end{subfigure}
\hfill\vline\hfill
\begin{subfigure}[b]{0.47\textwidth}
\centering
\begin{tabular}{l|l}
\qquad ARMv8  & \qquad x86    \\
\hline
$\ldr$ & $\movld$ \\
$\ldar$ & $\movld$ \\
$\str$ & $\movst$ \\
$\stlr$ & $\movst;\mfence$ \\
$\irmw$ & $\irmw$ \\
$\idmbfull$ & $\mfence$  \\
$\idmbld$/$\idmbst$/$\isb$ & $\skips$ \\
\end{tabular}
\caption{ARMv8 to x86}
\label{tab:armx}
\end{subfigure}
\caption{Mapping schemes: ARMv8 to x86 and ARMv7/ARMv7-mca to ARMv8}
\label{fig:xarmas}
\end{figure}
\end{center}
\vspace{-7mm}
\subsection{ARMv8 to x86 mapping}
\label{sec:armv8tox86}
The mapping scheme is in \cref{tab:armx}. 
In this scheme an ARMv8 load or load-acquire is mapped to an x86 load and 
a store is mapped to an x86 store operation. 
The scheme generates a trailing $\mfence$ with a store in x86 
for ARMv8 release-store as $\Llab;\lpo;\Alab \subseteq \bob$
whereas in x86 store-load on different locations are unordered. 
Consider the example below.
%
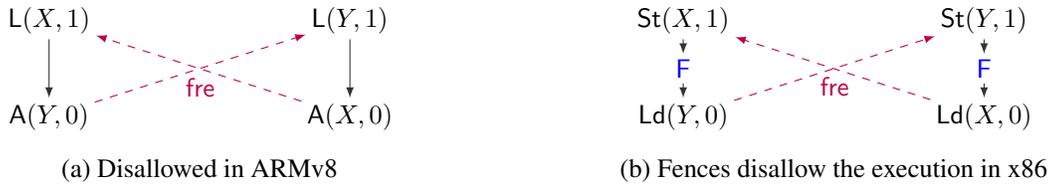
\begin{center}
\begin{minipage}[t]{0.49\linewidth}
\centering
\begin{tikzpicture}[yscale=0.8]
  \node (t11) at (-2,-1)  {$\Llab(X,1)$};
  \node (t12) at (-2,-2.6) {$\Alab(Y,0)$};
  \node (t21) at (2,-1)  {$\Llab(Y,1)$};
  \node (t22) at (2,-2.6) {$\Alab(X,0)$};
  \draw[po] (t11) to (t12);
  \draw[po] (t21) to (t22);
  \draw[fr,bend left=0] (t12) to node[below]{$\fre$} (t21);
  \draw[fr,bend right=0] (t22) to (t11);
  \node at (0,-3.5) {(a) Disallowed in ARMv8};
\end{tikzpicture}
\end{minipage}\hfill
\begin{minipage}[t]{0.49\linewidth}
\centering
\begin{tikzpicture}[yscale=0.8]
  \node (t11) at (-2,-1)  {$\Wlab(X,1)$};
  \node (t12) at (-2,-1.8) {{\color{blue}$\Flab$}};
  \node (t13) at (-2,-2.6) {$\Rlab(Y,0)$};
  \node (t21) at (2,-1)  {$\Wlab(Y,1)$};
  \node (t22) at (2,-1.8) {{\color{blue}$\Flab$}};
  \node (t23) at (2,-2.6) {$\Rlab(X,0)$};
  \draw[po] (t11) to (t12);
  \draw[po] (t12) to (t13);
  \draw[po] (t21) to (t22);
  \draw[po] (t22) to (t23);
  \draw[fr,bend left=0] (t13) to node[below]{$\fre$} (t21);
  \draw[fr,bend right=0] (t23) to (t11);
  \node at (0,-3.5) {(b) Fences disallow the execution in x86};
\end{tikzpicture}
\end{minipage}
\end{center}
The scheme also maps an atomic access pair to an atomic update in x86.  
The $\idmbld$, $\idmbst$, and $\isb$ fences are not mapped to any access. 
\begin{restatable}{theorem}{armx}
\label{th:armx}
The mapping scheme in \cref{tab:armx} is correct. 
\end{restatable}

\paragraph{Proof Strategy}
To prove \cref{th:armx} we first define corresponding ARMv8 execution $\ex_s$ for a given x86 consistent execution $\ex_t$. 
Next we show that $\ex_s$ is ARMv8 consistent. 
To do so, we establish \cref{lem:relcomposex} and then use the same
to establish \cref{lem:noloadx} on x86 consistent execution. 
Next, we define \emph{x86-preserved-program-order} ($\xppo$) and 
then based on $\xppo$ we define \emph{x86-observation} ($\obx$) on an x86 execution and establish \cref{lem:noloadobx}. 
Finally we prove \cref{th:armx} using \cref{lem:noloadx} and \cref{lem:noloadobx}.
The detailed proofs of \cref{lem:relcomposex,lem:noloadx,lem:noloadobx} and \cref{th:armx} 
are discussed in \cref{app:armv8tox86}.
\[
\inarr{
\obx \triangleq \rfe \cup \coe \cup \fre \cup [\Ulab] \cup \xppo \quad \text{ where } \quad
\xppo \triangleq s_1 \cup s_2 \cup s_3  \cup s_4 \cup s_5 \cup s_6 \cup s_7 \cup s_8
}
\]

\vspace{-6mm}

\begin{center}
\begin{minipage}[t]{0.3\textwidth}
\begin{align*}
s_1 \triangleq & [\Rlab];\lpo;[\Rlab \cup \Wlab]
\\ s_2 \triangleq & \lpo;[\Flab];\lpo
\\ s_3 \triangleq & [\Wlab] ; [\Flab]; [\Rlab] 
\end{align*}
\end{minipage}
\hfill
\begin{minipage}[t]{0.3\textwidth}
\begin{align*}
s_4 \triangleq & [\Rlab];\lpo 
\\ s_5 \triangleq & [\Wlab];\lpo; [\Wlab] 
\\ s_6 \triangleq & \lpo;[\Wlab] 
\end{align*}
\end{minipage}
\hfill
\begin{minipage}[t]{0.3\textwidth}
\begin{align*}
s_7 \triangleq & \lpo; [\Wlab];\poloc;[\Wlab]
\\ s_8 \triangleq & [\Ulab];\rfi;[\Rlab]
\end{align*}
\\[1.1ex]
\end{minipage}
\end{center}
\begin{restatable}{lemma}{relcomposex}
\label{lem:relcomposex}
Suppose $\ex$ is an \emph{x86} consistent execution. 
In that case $\ex.\poloc;\ex.\fr \implies \ex.\fr \cup \ex.\co$.
\end{restatable}
%
\begin{restatable}{lemma}{noloadx}
\label{lem:noloadx}
Suppose $\ex=\tup{\E,\lpo,\lrf,\mo}$ is an \emph{x86} consistent execution. 
For each $(\ex.\poloc \cup \ex.\fr \cup \ex.\co \cup \ex.\lrf)^+$ path between 
two events there exists an alternative $(\ex.\xhb \cup \ex.\fr \cup \ex.\co)^+$ 
path between these two events which has no intermediate load event.
\end{restatable}
\begin{restatable}{lemma}{noloadobx}
\label{lem:noloadobx}
Suppose $\ex=\tup{\E,\lpo,\lrf,\mo}$ is an \emph{x86} consistent execution. 
For each $\obx$ path between two events there exists an alternative $\obx$ 
path which has no intermediate load event.
\end{restatable}

\subsection{ARMv7 to ARMv8 mappings}
\label{sec:armv7toarmv8}


The mapping scheme in \cref{tab:armv7v8} from ARMv7 to ARMv8 is straightforward as no fence is introduced along with any memory access. 
\begin{restatable}{theorem}{armvsva}
\label{th:armv7v8}
The mappings in \cref{tab:armv7v8} are correct. 
\end{restatable}
To prove \cref{th:armv7v8} we relate \emph{preserved-program-order} ($\ppo$) in ARMv7 to \emph{Ordered-before} ($\ob$) relation in ARMv8. 
In ARMv7 $\ppo$ relates intra-thread events and in ARMv8 $\dob$, $\bob$, and $\aob$ relates intra-thread event pairs.
Note that ARMv8 $\dob$, $\bob$, and $\aob$ relations together are not enough to capture the ARMv7 $\ppo$ relation as the $\detour$ component of $\ppo$ involves inter-thread relations. 
However, ARMv7 $\detour$ relation implies $\obs$ relation in ARMv8 and therefore 
we can relate $\ppo$ and $\ob$ relations.
Considering these aspects we state the following lemma.
\begin{restatable}{lemma}{ppoob}
\label{lem:ppoob}
Suppose $\ex_s$ is an \emph{ARMv7} consistent execution and $\ex_t$ is corresponding 
\emph{ARMv8} execution. In that case $\ex_s.\ppo \implies \ex_t.\ob$.
\end{restatable}
Based on \cref{lem:ppoob} along with other helper lemmas we prove the mapping soundness \cref{th:armv7v8}.
The detailed proofs of \cref{lem:ppoob}, helper lemmas, and \cref{th:armv7v8} are in \cref{app:armv7toarmv8}.

\subsection{ARMv8 to ARMv7 mapping}
\label{sec:armv8toarmv7}

\begin{figure}
\begin{subfigure}[b]{0.49\textwidth}
\centering
\begin{tabular}{l|l}
ARMv8  & ARMv7/ARMv7-mca    \\
\hline
$\ldr$ & $\ldr;\idmb$ \\
$\str$ & $\str$ \\
$\ldar$ & $\ldr;\idmb$ \\
$\stlr$ & $\idmb;\str;\idmb$ \\
$\irmw$ & $\irmw;\idmb$ \\
$\irmw_\A$ & $\irmw;\idmb$ \\
$\irmw_{\sqsupseteq \Llab}$ & $\idmb;\irmw;\idmb$ \\
$\texttt{DMB(FULL/LD/ST)}$   & $\idmb$  \\
$\isb$ & $\isb$
\end{tabular}
\caption{ARMv8 to ARMv7}
\label{tab:armv8v7}
\end{subfigure}
\hfill\vline\hfill
\begin{subfigure}[b]{0.49\textwidth}
\centering
\begin{tabular}{l|l}
C11 to ARMv8  & ARMv7/ARMv7-mca    \\
\hline
$\ldr_\MOna$ & $\ldr$ \\
$\ldr_\MOat$ & $\ldr;\idmb$ \\
$\str$ & $\str$ \\
$\ldar$ & $\ldr;\idmb$ \\
$\stlr$ & $\idmb;\str;\idmb$ \\
$\irmw$ & $\irmw;\idmb$ \\
$\irmw_\A$ & $\irmw;\idmb$ \\
$\irmw_{\sqsupseteq \Llab}$ & $\idmb;\irmw;\idmb$ \\
$\texttt{DMB(FULL/LD/ST)}$   & $\idmb$  \\
$\isb$ & $\isb$
\end{tabular}
\caption{C11 to ARMv8 to ARMv7}
\label{tab:carmv8v7}
\end{subfigure}
\caption{Mapping schemes: ARMv8 $\mapsto$ ARMv7/ARMv7-mca and C11 $\mapsto$ ARMv8 $\mapsto$ ARMv7/ARMv7-mca.}
\label{fig:carmv8v7}
\end{figure}


The mapping scheme is in \cref{tab:armv8v7}.
Now we show that the fences along with memory accesses are necessary to preserve mapping soundness.
%
%
%
In \cref{sec:case:armv87} we have already shown that $\ldr \mapsto \ldr;\cbisb$ is unsound and therefore $\ldr \mapsto \ldr;\idmb$ is necessary for correctness.
Similarly, $\ldar \mapsto \ldr;\cbisb$ is unsound and $\ldar \mapsto \ldr;\idmb$ is necessary for the same reasons. 

\paragraph{Leading and trailing fences for release-store mapping}

Consider $\lpo;[\Llab] \subseteq \bob$ in ARMv8. 
The $\bob$ relation in the first thread along with other relations 
disallows this behavior. 
Consider the following example.
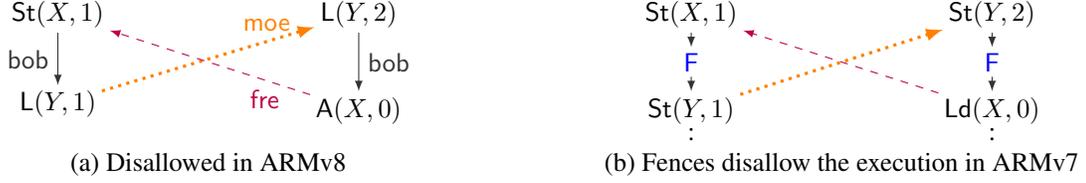
\begin{center}
\begin{minipage}[t]{0.49\linewidth}
\centering
\begin{tikzpicture}[yscale=0.8]
  \node (t11) at (-2,-1)  {$\Wlab(X,1)$};
  \node (t12) at (-2,-2.5) {$\Llab(Y,1)$};
  \node (t21) at (2,-1)  {$\Llab(Y,2)$};
  \node (t22) at (2,-2.6) {$\Alab(X,0)$};
%
  \draw[po] (t11) to node[left]{$\bob$} (t12);
  \draw[po] (t21) to node[right]{$\bob$} (t22);
  \draw[mo,bend left=0] (t12) to node[above,pos=0.8]{$\moe$} (t21);
  \draw[fr,bend right=0] (t22) to node[below,pos=0.2]{$\fre$} (t11);
  \node at (0,-3.5) {(a) Disallowed in ARMv8};
\end{tikzpicture}
\end{minipage}\hfill
\begin{minipage}[t]{0.49\linewidth}
\centering
\begin{tikzpicture}[yscale=0.8]
  \node (t11) at (-2,-1)  {$\Wlab(X,1)$};
  \node (t12) at (-2,-1.8) {{\color{blue}$\dmb$}};
  \node (t13) at (-2,-2.6) {$\Wlab(Y,1)$};
  \node (t14) at (-2,-3) {:};
  \node (t21) at (2,-1)  {$\Wlab(Y,2)$};
  \node (t22) at (2,-1.8) {{\color{blue}$\dmb$}};
  \node (t23) at (2,-2.6) {$\Rlab(X,0)$};
  \node at (2,-3) {:};
%
  \draw[po] (t11) to (t12);
  \draw[po] (t12) to (t13);
  \draw[po] (t21) to (t22);
  \draw[po] (t22) to (t23);
  \draw[mo,bend left=0] (t13) to  (t21);
  \draw[fr,bend right=0] (t23) to (t11);
  \node at (0,-3.5) {(b) Fences disallow the execution in ARMv7};
\end{tikzpicture}
\end{minipage}
\end{center}
Without such an intermediate fence in the first thread 
the ARMv7 execution would be allowed which in turn 
introduce a new outcome in the ARMv7 program and as a result 
the mapping would be incorrect.
Therefore $\stlr$ mapping requires a leading fence to preserve the mapping soundness. 
$\stlr$ mapping requires a trailing 
fence considering the example similar to that of \cref{sec:armv8tox86}.
Considering the mapping, an $\cbisb$ is not required anymore as every load generates a 
trailing $\idmb$ fence. 

In addition to $\irmw$, ARMv8 provides acquire and release or stronger 
$\irmw$ accesses $\irmw_\Alab$ and $\irmw_{\sqsupseteq \Llab}$ respectively. 
Before mapping from ARMv8 we perform the transformations 
$\irmw_\Alab \leadsto \irmw;\idmbld$ and $\irmw_{\sqsupseteq \Llab} \leadsto \idmbfull;\irmw;\idmbfull$.
The trailing $\idmbld$ provides the same ordering as an acquire-exclusive load with following accesses. 
In case of $\irmw_{\sqsupseteq \Llab}$, we introduce leading and trailing $\idmbfull$ fences similar to that of $\stlr$ 
access.

For $\idmbfull$, $\idmbld$, and $\idmbst$ fences in ARMv8 the mapping scheme generates 
$\idmb$ fences so that the $\bob$ orders in ARMv8 executions are preserved in corresponding ARMv7 executions. 
Now we prove the correctness of the mapping as stated in \cref{th:armv8v7}.  
\begin{restatable}{theorem}{armvavs}
\label{th:armv8v7}
The mappings in \cref{tab:armv8v7} are correct. 
\end{restatable}
To prove \cref{th:armv8v7}, we relate ARMv8 and ARMv7 consistent executions in \cref{lem:abdobsppo} and \cref{lem:obprop} as intermediate steps.
\cref{lem:abdobsppo}, \cref{lem:obprop}, and \cref{th:armv8v7} are proved in \cref{app:armv8toarmv7}.
\begin{restatable}{lemma}{abdobsppo}
\label{lem:abdobsppo}
Suppose $\ex_t$ is an \emph{ARMv7} consistent execution and $\ex_s$ is \emph{ARMv8} execution following the mappings in \cref{tab:armv8v7}. 
In this case $\ex_s.\ob \implies (\ex_t.\rfe \cup \ex_t.\coe \cup \ex_t.\fre \cup \ex_t.\rmw \cup \ex_t.\fence)^+$. 
\end{restatable}
\begin{restatable}{lemma}{obprop}
\label{lem:obprop}
Suppose $\ex_t$ is an \emph{ARMv7} consistent execution and $\ex_s$ is \emph{ARMv8} execution following the mappings in \cref{tab:armv8v7}. 
In this case either $\ex_s.\ob \implies (\rloc{(\ex_t.\E \times \ex_t.\E)} \setminus [\E])$ or $\ex_s.\ob \implies (\ex_t.\co;\ex_t.\prop \cup \ex_t.\prop)^+$.
\end{restatable}

\subsection{C11 to ARMv8 to ARMv7 mapping}
\label{sec:carmv8v7}

Similar to $\cmm \mapsto \xarch \mapsto \arma$ we propose C11 to ARMv8 to ARMv7 mapping scheme in \cref{tab:carmv8v7}. 
The proof is discussed in detail in \cref{app:carmv8v7}. 
In \cref{sec:cmap} we already show that this mapping scheme is more efficient than ARMv8 to ARMv7 mapping.
\begin{restatable}{theorem}{carmas}
\label{lem:carmas}
The mapping scheme in \cref{tab:carmv8v7} is correct.
\end{restatable}

\subsection{ARMv7-mca to ARMv8 mapping}
\label{sec:armv7mcaarmv8}

The mapping scheme for ARMv7-mca to ARMv8 is same as the  
ARMv7 to ARMv8 mapping scheme as shown in \cref{tab:armv7v8}.
To prove the mapping soundness we relate an ARMv7 consistent execution to corresponding ARMv8  execution as follows.
\begin{restatable}{lemma}{ppowob}
\label{lem:ppowob}
Suppose $\ex_t$ is an \emph{ARMv8} consistent execution and $\ex_s$ is corresponding 
\emph{ARMv7} consistent execution. In that case $[\ex_s.\Rlab];\ex_s.\ppo;[\ex_s.\Rlab];\ex_s.\poloc;[\ex_s.\Wlab] \implies [\ex_t.\Rlab];\ex_t.\ob;[\ex_t.\Wlab]$
\end{restatable}  
Using \cref{lem:ppowob} we establish the acyclicity of write-order in ARMv7-mca source execution.  
\begin{restatable}{lemma}{acywo}
\label{lem:acywo}
Suppose $\ex_t$ is a target \emph{ARMv8} consistent execution and $\ex_s$ is corresponding \emph{ARMv7} consistent execution. 
In this case $\ex_s.\wo^+$ is acyclic. 
\end{restatable}
The detailed proof of \cref{lem:ppowob} are \cref{lem:acywo} are discussed in \cref{app:sec:armv7mcatoarmv8}. 
The mapping correctness theorem below directly follows from \cref{lem:acywo}. 
\begin{restatable}{theorem}{armsmcaarma}
\label{th:armv7mcav8}
The mappings in \cref{tab:armv7v8} are correct for ARMv7-mca.
\end{restatable}


\subsection{ARMv8 to ARMv7-mca and C11 to ARMv8 to ARMv7-mca mappings}
\label{sec:armv8toarmv7mca}

The mapping schemes, ARMv8 to ARMv7-mca and C11 to ARMv8 to ARMv7-mca, are shown in \cref{fig:carmv8v7}. 
The soundness proofs are same as \cref{th:armv8v7,lem:carmas} respectively. 
We have already discussed in \cref{sec:case:mca} why mapping of a load access requires a trailing $\idmb$ fence to preserve correctness.


\section{ Common Compiler Optimizations in $\mathbf{ARMv8}$}
\label{sec:armaopt}
In this section we study the correctness of independent access reordering, redundant access elimination, and access strengthening in ARMv8 model. 
We prove the correctness of the safe transformations in \cref{app:armaopt}. 

\begin{figure}
\begin{minipage}{0.5\textwidth}
\centering
\begin{tabular}{|c|c|c|c|c|c|c|c|}
\hline
${\downarrow}\,a \;\backslash\; b\,{\rightarrow}$  & $\Wlab$ & $\Rlab$ & $\Llab$ & $\Alab$ & $\dmbfull$ & $\dmbld$ & $\dmbst$ \\ \hline
$\Wlab$        & \no    & \yes  & \no   & \yes  & \no       & \yes    & \no    \\ \hline
$\Rlab$         & \no    & \yes  & \no   & \yes  & \no       & \no    & \yes        \\ \hline
$\Llab$         & \no    &  \yes  & \no   & \no   & \no       & \yes    & \no        \\ \hline
$\Alab$        & \no    &   \no   & \no   & \no   & \yes       & \yes    & \yes        \\ \hline
$\dmbfull$    & \no    &  \no   & \yes   & \no    & =         & \yes    & \yes   \\ \hline
$\dmbld$      & \no    &  \no   & \yes   & \no   & \yes     & =         & \yes   \\ \hline
$\dmbst$      & \no   &  \yes  & \yes & \yes  & \yes     & \yes    & =        \\ \hline
\end{tabular}
\end{minipage}
\hfill
\begin{minipage}{0.45\textwidth}
\yes \quad $\Rlab(X,v') \cdot\Rlab(X,v) \leadsto \Rlab(X,v')$ \hfill (RAR)
\\ \yes \quad $\Alab(X,v') \cdot\Rlab(X,v) \leadsto \Alab(X,v')$ \hfill (RAA)
\\ \yes \quad $\Alab(X,v') \cdot\Alab(X,v) \leadsto \Alab(X,v')$ \hfill (AAA)
\subcaption{$\ldr$ and $\ldar$ eliminations.}
\bigskip
\yes \quad $\Rlab(X,v) \leadsto \Alab(X,v)$ \hfill (R-A)
\\ \yes \quad $\Wlab(X,v) \leadsto \Llab(X,v)$ \hfill (W-L)
\\ \yes \quad $\dmbld/\dmbst \leadsto \dmbfull$ \hfill (F)

\subcaption{Access strengthening.}
\end{minipage}
\caption{Reordering, elimination, and strengthening transformations in ARMv8. 
}
\label{fig:armatrans}
\end{figure}
\textit{Reorderings.}
We show the safe (\yes) and unsafe (\no) reordering transformations of the form $a \cdot b \leadsto b \cdot a$ in \cref{fig:armatrans} 
where $a$ and $b$ represent independent and adjacent shared memory accesses on different locations.
We prove the correctness of the safe reorderings in \cref{app:sreorder}.

In \cref{fig:ex:reordering} we have already shown that we cannot move a store before any load or store in. Same reasoning extends to release-store and acquire-load. 
It is not safe to move a store before any fence as it may violate a $\dob$ relation. 
Similarly a load cannot be moved before an acquire load, $\idmbld$, or $\idmbfull$ operation as it may remove a $\bob$ relation.
However, reordering with a $\idmbst$ is safe as the ordering between them do not affect any component of $\ob$ relation.
A release-store may safely reorder with a preceding fence as it does not eliminate 
any $\bob$ relation.
Similarly moving a load, store, or $\idmbst$ after an acquire-read is allowed 
as it does not eliminate any existing $\bob$ relation. 
We may safely reorder acquire-read 
with $\idmbfull$ as it does not affect the $\bob$ relations among the 
memory accesses. 
A $\idmbld$ between a load and a load or store creates $\bob$ relation. 
Hence moving a load after $\idmbld$ may eliminate a $\bob$ and 
therefore disallowed. 

Finally reorderings fences are safe as it preserves the $\bob$ 
relations between memory accesses.

\smallskip
\textit{Redundant access elimination}
In \cref{sec:armaintermediate} we have shown that overwritten-write and 
read-after-write transformations are unsound. 
However, a read-after-read elimination is safe in ARMv8 as enlisted in \cref{fig:armatrans}. 
We prove the correctness of the transformation in \cref{app:rarelim}.
 
\smallskip
\textit{Access strengthening} 
Strengthening memory accesses and fences may introduce new ordering among events and 
therefore the strengthening transformations enlisted in \cref{fig:armatrans} 
hold trivially.  



\section{Fence Optimizations}
\label{sec:fenceopt}

In this section we prove the correctness of various fence eliminations and  
then propose respective fence elimination algorithms. 
More specifically, 
the proposed mapping schemes in \cref{sec:mappings} 
may introduce fences some of which are redundant in certain scenarios 
and can safely be eliminated. 
To do so, we first check if a fence is \emph{non-eliminable}. 
If not, we delete the fence.  
\subsection{x86 fence elimination}
\label{sec:xfenceelim}
In x86 only a store-load pair on different locations is unordered. 
Therefore if a fence appear between such a pair then it is not safe 
to eliminate the fence. 
Otherwise we may eliminate a fence.
\begin{restatable}{theorem}{xfenceelim}
\label{th:xfenceelim}
An \emph{$\mfence$} in an x86 program thread is non-eliminable if it is the only fence on a program path from a store to a load in the same thread which access different locations.

An \emph{$\mfence$} elimination is safe when it is \emph{not} non-eliminable. 
\end{restatable}
We prove the theorem in \cref{app:xfenceelim}. 
This fence elimination condition is particularly useful after ARMv8 to x86 mapping following the scheme in \cref{tab:armx} as it introduces certain redundant fences. 
For instance, ARMv8 to x86 mapping $\stlr;\str \mapsto \movst;\mfence;\movst$ results 
in an intermediate $\mfence$ which is redudant and can be safely deleted as stores are ordered in x86.
%
%
\subsection{ARMv8 fence elimination (after mapping)}

We identify non-eliminable $\idmbfull$, $\idmbst$, and $\idmbst$ fences and 
then safely eliminate rest of the fences. 
We prove the correctness of these fence eliminations in \cref{app:armv8felim}.

For instance, considering the \cref{tab:xarm} mapping scheme,   
the $\idmbld$ fence after $\movld;\movst \mapsto \ldr;\idmbld;\idmbst;\str$ mapping 
suffices to order the load and store access pair and 
the $\idmbst$ is not required. 
However, we cannot immediately conclude that such a $\idmbst$ fence 
is entirely redundant if we consider a mapping 
$\movst;\movld;\movst \mapsto \idmbst;\str;\ldr;\idmbld;\idmbst;\str$ where the second 
$\idmbst$ orders the two stores and therefore non-eliminable.   
%
%
%
%
\begin{restatable}{theorem}{dmbfullelim}
\label{th:dmbfullelim}
Suppose an ARMv8 program is generated by \emph{$\xarch \mapsto \arma$} mapping (\cref{tab:xarm}). 
A \emph{$\idmbfull$} in a thread of the program is non-eliminable if it is the only fence on a program path from a store to a load in the same thread which access different locations.

A \emph{$\idmbfull$} elimination is safe when it is \emph{not} non-eliminable. 
%
%
%
\end{restatable}
The trailing and leading fences in x86 to ARMv8 mapping ensures that 
a $\idmbfull$ fence can safely be eliminated following \cref{th:dmbfullelim}. 
Otherwise we cannot immediately eliminate a $\idmbfull$; 
rather whenever appropriate, we may weaken such a $\idmbfull$ fence by replacing it with $\idmbst;\idmbld$ fence sequence when a $\idmbfull$ fence is costlier than a pair of $\idmbst$ and $\idmbld$ fences. We define safe fence weakening in \cref{th:dmbfullweaken} below and the detailed proof is in \cref{app:armv8fweaken}.
\begin{restatable}{theorem}{dmbfullweaken}
\label{th:dmbfullweaken}
A \emph{$\idmbfull$} in a program thread is non-eliminable if it is the only fence on a program path from a store to a load in the same thread which access different locations. 

For such a fence \emph{$\idmbfull \leadsto \idmbst;\idmbld$} is safe.
\end{restatable}
While fence weakening can be applied on any ARMv8 program, 
it is especially applicable after ARMv7/ARMv7-mca to ARMv8 mapping. 
ARMv7 has only $\idmb$ fence (except $\isb$) to order any pair of memory 
accesses and these $\idmb$ fences translates to $\idmbfull$ fence in ARMv8.  
In many cases these $\idmbfull$ fences can be weakened and then 
we can eliminate $\idmbld$ and $\idmbst$ fences which are not non-eliminable.
%
\begin{restatable}{theorem}{dmbstelim}
\label{th:dmbstelim}
A \emph{$\idmbst$} in a program thread is non-eliminable if it is placed on a program path between a pair of stores in the same thread which access different locations and there exists no other \emph{$\idmbfull$} or \emph{$\idmbst$} fence on the same path.

A \emph{$\idmbst$} elimination is safe when it is \emph{not} non-eliminable. 
%
%
%
\end{restatable}

\begin{restatable}{theorem}{dmbldelim}
\label{th:dmbldelim}
A \emph{$\idmbld$} in a program thread is non-eliminable if it is placed on a program path from a  load to a store or load access in the same thread which access different locations and there exists no other \emph{$\idmbfull$} or \emph{$\idmbld$} fence on the same path.

A \emph{$\idmbld$} elimination is safe when it is \emph{not} non-eliminable. 
\end{restatable}

\subsection{Fence Elimination in ARMv7}

In ARMv7 we safely eliminate repeated $\idmb$ fences. 
ARMv7 $\idmb$ fence elimination is particularly useful after ARMv8 to ARMv7/ARMv7-mca  mappings. 
For example, $\ldr;\stlr \mapsto \ldr;\idmb;\idmb;\str;\idmb$ generates repeated $\idmb$ fences and one of them can be safely eliminated. 
\begin{restatable}{theorem}{armvsfenceelim}
\label{th:dmbelim}
A \emph{$\idmb$} in a program thread is non-eliminable if it is the only fence on a program path between a pair of memory accesses in the same thread. 

A \emph{$\idmb$} elimination is safe when it is \emph{not} non-eliminable. 
\end{restatable}

\begin{figure}[t]
$\reach(\cfg,i,j) \triangleq (i,j) \in [\cfg.\V];\cfg.\Elab^+;[\cfg.\V]$ \hfill $\opath(\cfg,i,f,j) \triangleq \reach(\cfg,i,f) \land \reach(\cfg,f,j)$
\begin{align*}
\reachwo(\cfg, i, j, F) \triangleq & \reach(\tup{\cfg.\V \setminus F, \cfg.\E \setminus B},i,j) \text{ where } B = (G.\V \times F) \cup (F \times G.\V)
\\ \onelim(\cfg, i,f, j, F) \triangleq & \opath(\tup{\cfg.\V \setminus F, \cfg.\E \setminus B},i,f,j) \text{ where } B = (G.\V \times F) \cup (F \times G.\V)
\\ \mpairs(\cfg, a, b) \triangleq & \set{(i,j) \mid \denot{i} \in a \land \denot{j} \in b \land \reach(\cfg,i,j)}
\\ \rnloc {\mpairs(\cfg, a, b)} \triangleq & \set{(i,j) \mid \mpairs(\cfg,a,b) \land \neg \mustalias(i,j)}
\\ \rloc {\mpairs(\cfg, a, b)} \triangleq & \set{(i,j) \mid \mpairs(\cfg,a,b) \land \mustalias(i,j)}
\\ \texttt{FDelete}(\cfg, F) \triangleq & \tup{\cfg.\V \setminus F, \cfg.\Elab \setminus ((\cfg.\V \times F) \cup (F \times \cfg.\V))} 
\end{align*}
\begin{subfigure}{0.45\textwidth}
\centering
\begin{algorithmic}[1]
\Procedure{getNFS}{$\cfg,PR,F, B$}
\For{$f \in F$}
\For{$(i,j) \in PR$}
\State $\cfg' \leftarrow \texttt{FDelete}(\cfg, B)$
\If{$\opath(\cfg',i,f,j)$}
\State $B \leftarrow B \cup \set{f};$ 
\State $\lbreak; \ \ \codecomment{inner \ loop}$ 
\EndIf
\EndFor
\EndFor
\State \Return $B$
\State \textbf{end procedure}
\EndProcedure
\end{algorithmic}
\end{subfigure}
\begin{subfigure}{0.53\textwidth}
\begin{algorithmic}[1]
\Procedure{FWeaken}{$\cfg,F$}
\For{$f \in F$}
\State $\V_1 \leftarrow \cfg.\V \cup \set{a,b \mid \denot{a} \in \dmbld \land \denot{b} \in \dmbst}$
\State $\Elab_1 \leftarrow \cfg.\E \cup \set{(f,a), (a,b)}$ 
\State $\Elab_2 \leftarrow \Elab_1 \cup \set{(e,a) \mid G.\Elab(e,f)}$ 
\State $\Elab_3 \leftarrow \Elab_2 \cup \set{(b,e) \mid G.\Elab(f,e)}$
\State $\cfg'.\V \leftarrow \V_1 \setminus \set{f}$
\State $\cfg'.\Elab \leftarrow \Elab_3 \setminus ((\cfg'.\V \times \set{f}) \cup (\set{f} \times \cfg'.\V))$
\EndFor
\State \Return $\cfg'$
\State \textbf{end procedure}
\EndProcedure
\end{algorithmic}
\end{subfigure}
\caption{Helpers conditions and functions}
\label{fig:helpers}
\end{figure}

\begin{figure}[t]
\begin{subfigure}[b]{0.43\textwidth}
\centering
\begin{algorithmic}[1]
\Procedure{x86FElim}{$\cfg$}
\State $F = \set{f \mid f \in G.\V \land \denot{f} \in \F}$;
\State $U = \set{f \mid f \in G.\V \land \denot{f} \in \Ulab}$;
\State $SL \leftarrow \rnloc {\mpairs(\cfg,\Wlab,\Rlab)}$
\State $\nelim \leftarrow \texttt{getNFS}(\cfg,SL,F,U)$;
\State \textbf{return } $\texttt{FDelete}(\cfg, F \setminus \nelim)$;
\\ \textbf{end procedure}
\EndProcedure
\bigskip
\Procedure{ARMv7FElim}{$\cfg$}
\State $F = \set{f \mid f \in G.\V \land \denot{f} = \dmb}$;
\State $M \leftarrow \mpairs(\cfg,\E \setminus F,\E \setminus F)$
\State $\nelim \leftarrow \texttt{getNFS}(\cfg,M,F,\emptyset)$;
\State \textbf{return } $\texttt{FDelete}(\cfg, F \setminus \nelim)$;
\State \textbf{end procedure}
\EndProcedure
\end{algorithmic}
\end{subfigure}
\begin{subfigure}[b]{0.54\textwidth}
\begin{algorithmic}[1]
\Procedure{ARMv8FElim}{$\cfg$}
\State $F = \set{f \mid f \in G.\V \land \denot{f} = \idmbfull}$;
\State $SL \leftarrow \rnloc {\mpairs(\cfg,\Wlab,\Rlab)}$
\State $\nelim \leftarrow \texttt{getNFS}(\cfg,SL,F,\emptyset)$;
\If{$\xarch \mapsto \arma$}
 \State $\cfg_1 \leftarrow \texttt{FDelete}(\cfg, F \setminus \nelim)$;
\Else
\State $\cfg_1 \leftarrow \texttt{FWeaken}(\cfg, F \setminus \nelim)$
\EndIf
\State $FS = \set{f \mid f \in \cfg_1.\V \land \denot{f} = \idmbst}$;
\State $SS \leftarrow \rnloc {\mpairs(\cfg_1,\Wlab,\Wlab)}$
\State $FF \leftarrow \texttt{getNFS}(\cfg_1,SS, FS, \nelim)$;
\State $\cfg_2 \leftarrow \texttt{FDelete}(\cfg_1, FS \setminus FF)$;
\State $FL = \set{f \mid f \in \cfg_2.\V \land \denot{f} = \idmbld}$;
\State $LS \leftarrow \rnloc {\mpairs(\cfg_2,\Rlab,\Wlab)}$
\State $LL \leftarrow \rnloc {\mpairs(\cfg_2,\Rlab,\Rlab)}$
\State $FF' \leftarrow \texttt{getNFS}(\cfg_2,LL \cup LS,FL,\nelim)$;
\State \Return $\texttt{FDelete}(\cfg_2, FL \setminus FF')$;
\State \textbf{end procedure}
\EndProcedure
\end{algorithmic}
\end{subfigure}
\caption{Fence elimination algorithms after mappings.}
\label{fig:fenceopt}
\end{figure}

%
%
%
%
We first check if a fence is \emph{non-eliminable} based on the access 
pairs and fence locations on the program paths. 
We perform this analysis on the thread's control-flow-graph $\cfg = \tup{\V, \Elab}$ 
where $\cfg.\V$ denotes the program statements including the accesses and 
$\cfg.\Elab$ represents the set of edges between pair of statements. 
Next, we delete a fence if it is not \emph{non-eliminable}.
 

In \cref{fig:helpers} we define a number of conditions which we use in fence elimination. Condition $\reach(\cfg,i,j)$ holds if there is a path from instruction $i$ to instruction $j$ 
in $\cfg$ and $\opath$ checks if there is any path from $i$ to $j$ through a fence $f$. 
$\mpairs(\cfg,a,b)$ is a set of $(a \times b)$ memory access pairs in $\cfg$. 
We compute 
$\rnloc{\mpairs(\cfg,a,b)}$; the set of memory access pairs on different locations based on 
must-alias analysis. 
$\fdelete$ deletes a set of fences.
Procedure $\getnfs$ updates the set of 
non-eliminable fences considering the positions of 
other fences between the access pairs. 
Given a fence $f$ and an access pair $(i,j)$, 
we check if there is a path from $i$ to $j$ 
through $f$ without passing through 
already identified non-eliminable fences $B$. 
If so, fence $f$ is also non-eliminable.

\textit{Fence elimination in \emph{x86, ARMv7, and ARMv8}.} 
In \cref{fig:fenceopt} we define x86, ARMv8, ARMv7 fence elimination procedures. 
For instance, in $\xfelim$ we first identify store-load access pairs on different locations and 
the $\mfence$ operations in a thread. 
Then we identify the set of non-eliminable fences $\nelim$ using $\texttt{getNFS}$ 
procedure. 
In this case we consider the positions of atomic updates along with fences as 
atomic updates also act as a fence. 
Finally $\fdelete$ eliminates rest of the fences.

Procedure $\armafelim$ works in multiple steps for each of the fences. 
Note that while mapping to ARMv8 we do not use release-write or acquire-load accesses. 
Therefore we use the same $\reachwo$ condition to check if a fence is non-eliminable.
Moreover, in case of x86 to ARMv8 we eliminate $\idmbfull$ fences. 
In this case $\idmbfull$ elimination is safe as it introduces other $\idmbld$ and $\idmbst$ fences. 
However, we do not eliminate $\idmbfull$ when it is generated from ARMv7 as 
it may remove order between a pair of accesses.
In this case or in general we can weaken a $\idmbfull$ fence and then eliminate 
redundant $\idmbst$ and $\idmbld$ fences. 

In ARMv7 a $\dmb$ is redundant when it it appears between a pair of same-location load-load, store-store, store-load, and atomic 
load-store accesses. 
Such redundant fences appear in ARMv7 program after mapping ARMv8 programs to ARMv7/ARMv7-mca following the mapping scheme in \cref{tab:armv8v7}.
For example, a sequence $\ldr;\ldr$ in ARMv8 results in a sequence $\ldr;\dmb;\ldr;\dmb$ in ARMv7 where 
the introduced $\dmb$ instructions are redundant and we eliminate these fences by $\armsfelim$ procedure.

\section{Robustness Analysis}
\label{sec:robustness} 
We first define robustness and then discuss the conditions and its analyses in more details. 
\begin{definition} 
Suppose $M$ and $K$ are concurrency models. A program is $M$-robust against $K$ if all its $K$-consistent executions are also $M$-consistent.
\label{def:robustness}
\end{definition}
We observe that in axiomatic models the axioms are represented in the form irreflexivity 
of a relation or acyclicity of one or a combination of relations. 
When an axiom is violated then it results in a cycle on 
an execution graph. 
Such a cycle consists of a set of internal relations which 
are included in program order ($\lpo$) along with external relations. 
If these involved $\lpo$ relations are appropriately ordered 
then such a cycle would not be possible.
As a result the program would have no weaker behavior and 
would be $M$-robust against a weaker model $K$. 
To capture the idea we define \emph{external-program-order} ($\epo$) relation as follows.
\[
\inarr{\epo \triangleq \lpo \cap \codom(\eco) \times \dom(\eco)
}
\]
Based on this observation we check and enforce $M$-robustness against $K$ considering the relative strength ($\sqsubset$) of the  memory accesses of the memory models: $\SC \sqsubset \xarch \sqsubset \arma \sqsubset \armsmca \sqsubset \arms$. 
In all these cases we define required constraints on the 
\emph{external-program-order} ($\epo$) edges in an execution which preserves robustness. 

\smallskip
\textit{Checking robustness in \emph{x86}.}
A subtle issue in checking SC-robustness against x86 is $\mo$ relation 
may take place between writes on different locations 
and in that case we have to consider a possible through 
different location writes as well.
To avoid this complexity, 
we use the x86A model 
following \cite{Alglave:2014,herd} as shown in \cref{fig:scx86a} for robustness analyses. 
In this model there is no $\mo$ relation and unlike x86 an update operation 
results in $\rmw \subseteq \poloc$ relation instead of an event 
similar to ARM models.
In \cref{fig:scx86a} we also define SC model \cite{Alglave:2014,herd} for robustness analysis.

\begin{figure}
\begin{align*}
\text{(SC-x86A)}  \quad & [\RUlab];\lpo \cup \lpo;[\WUlab] \cup \poloc \cup \fence
\\ \text{(SC-ARMv8)} \quad &  \poloc \cup (\aob \cup \dob \cup \bob)^+ 
\\ \text{(x86A-ARMv8)} \quad & \poloc  \cup (\aob \cup \bob \cup \dob)^+ \cup \awr
\\ \text{(SC-ARMv7)} \quad & \poloc \cup \fence 
\\ \text{(x86A-ARMv7)} \quad & \poloc \cup \fence \cup \awr 
\\\text{(ARMv8-ARMv7)} \quad & \poloc \cup [\Wlab];\lpo \cup \fence
\\ \text{(ARMv7mca-ARMv7)} \quad & [\Wlab];\lpo \cup \lpo;[\Wlab] \cup [\Rlab];(\poloc \cup \fence);[\Rlab]
\end{align*}
\caption{($M\text{-}K$): Condition $R$ for $M$-robust against $K$ analysis.}
\label{fig:robustconditions}
\end{figure}

\begin{figure}
\begin{subfigure}[b]{0.3\textwidth}
\begin{tabular}{l}
(SC) 
\\ $\acy(\lpo \cup \lrf \cup \fr \cup \co)$
\\[1.3ex] (atomicity)  
\\ $\irr([\rmw];\fre;\coe)$
\end{tabular}
\caption{SC}
\end{subfigure}
\begin{subfigure}[b]{0.67\textwidth}
\begin{tabular}{ll}
(sc-per-loc)  & $\acy(\poloc \cup \lrf \cup \fr \cup \co)$
\\ (atomicity)  & $\irr([\rmw];\fre;\coe)$
\\ (GHB)  & $\acy((\lpo \setminus \awr) \cup \fence \cup \rfe \cup \co \cup \fr)$ 
\\ & $\text{ where } \fence = \lpo;[\rmw \cup \F];\lpo$ and  
\\ & $\awr = [\Wlab \setminus \codom(\rmw)];\lpo;[\Rlab \setminus \dom(\rmw)]$ 
\end{tabular}
\caption{x86A}
\end{subfigure}
\caption{SC and x86A model for robustness checking}
\label{fig:scx86a}
\end{figure}

\paragraph{Robustness conditions}
In \cref{fig:robustconditions} we define the conditions which have to be fulfilled by $\epo$ in 
all executions for a given program. 
An x86A execution is SC-robust when all $\epo$ relations are fully ordered as defined in (SC-x86A). 
In ARMv8 model condition (SC-ARMv8) preserves order for all $\epo$ relations. 
Condition (x86A-ARMv8) orders all $\epo$ relations except non-RMW store-load access pairs on different locations similar to x86A. 
ARMv7 model uses $\poloc$ and $\fence$ to order $\epo$ relations fullly to preserve SC robustness. 
We do not use $\ppo$ in these constraints as it violates robustness as shown in the example in \cref{fig:pponrobust}. 
To preserve x86A robustness, ARMv7 orders all $\epo$ relations except non-RMW store-load access pairs on different locations. 
Condition (ARMv8-ARMv7) also does not rely on ordering by dependencies as $\ppo$. 
For example, $\data;\coi \subseteq \dob$ in ARMv8 does not imply $\ppo$ in ARMv7. 
Therefore such a $\dob$ order may disallow an execution to be ARMv8 consistent but 
be allowed in ARMv7 model which would violate ARMv8-robustness. 
Finally, (ARMv7mca-ARMv7) checks if the program may have any MCA behavior.

Now 
we state the robustness theorem based on these constraints and prove the respective robustness results in \cref{app:robustness}.
\begin{restatable}{theorem}{mk}
\label{th:mk}
A program $\mbbp$ is $M$-robust against $K$ if in all its $K$ consistent execution $\ex$, $\ex.\epo \subseteq \ex.R$ holds where $R$ is defined as condition $(M\text{-}K)$ in \cref{fig:robustconditions}.
%
\end{restatable}
\begin{center}
\begin{figure}[t]
\begin{align*}
\reachwo(\cfg, i, j, F) \triangleq & \reach(\tup{\cfg.\V \setminus F, \cfg.\Elab \setminus B},i,j) \text{ where } B = (G.\V \times F) \cup (F \times G.\V)
\\ \ordered(\cfg,(i,j),F) \triangleq & \mustalias(i,j) \lor \neg \reachwo(\cfg,i,j,F)
\\ \oncyc(A) \triangleq & \set{(a,b) \mid (a,b) \in A \land \exists (p,q),(r,s) \in A.~(a,b) \neq (p,q) \land (a,b) \neq (r,s) 
\\ & \land \mayalias(b,p) \land \mayalias(a,s)}
\end{align*}
\\
\begin{subfigure}[b]{0.45\textwidth}
\centering
$\getmf(b) \triangleq  \cfg \text{ where } b \in \cfg.\V$
\\[1.2ex]
\begin{algorithmic}[1]
\Procedure{insertF}{$\mbbp, O$}
\State $H \leftarrow \emptyset$
\For{$(a,b) \in O$}
\If{$b \notin H$}
\State $\cfg \leftarrow \getmf(b)$;
\State $f \leftarrow \textbf{new}(\mfence)$; 
\State $\cfg.\V \leftarrow \cfg.\V \cup \set{f}$; 
\State $P \leftarrow \set{(a,f) \mid (a,e) \in \cfg.\Elab^+}$
\State $Q \leftarrow \set{(f,b) \mid (e,b) \in \cfg.\Elab^+}$
\State $\cfg.\Elab \leftarrow \cfg.\Elab \cup \set{(f,e)} \cup P \cup Q$; 
\State $H \leftarrow H \cup \set{b}$
\EndIf
\EndFor
\\ \textbf{end procedure}
\EndProcedure
\end{algorithmic}
\end{subfigure}
\hfill
\begin{subfigure}[b]{0.53\textwidth}
\centering
\begin{algorithmic}[1]
\Procedure{SCRobustx86}{$\mbbp$, $\N$}
\State $\ell \leftarrow \Wlab \cup \Rlab \cup \Ulab$;
\State $A \leftarrow \bigcup_{i\in \N}\mpairs(\mbbp(i), \ell, \ell)$; 
\State $O \leftarrow \emptyset$;
\For{$(a,b) \in \oncyc(A)$}
\State $B \leftarrow \set{f \mid f \in \cfg.\V \land \denot{f} \in \F \cup \Ulab}$
\If{$\neg \ordered(\getmf(b),(a,b),B)$}
\State $O \leftarrow O \cup \set{(a,b)}$;
\EndIf
\EndFor
\If{$O == \emptyset$} \ \Return $\true$;
\Else  
\State $\textsc{insertF}(\mbbp,O)$;
\State \Return $\false$;
\EndIf
\\ \textbf{end procedure}
\EndProcedure
\end{algorithmic}
\end{subfigure}
\caption{Analysis and enforcement of SC-robustness against x86.}
\label{fig:xrobusproc}
\end{figure}
\end{center}

\begin{figure}[t]
\begin{subfigure}[b]{0.47\textwidth}
\begin{algorithmic}[1]
\Procedure{insertDMBv8}{$\mbbp, O$}
\State $H \leftarrow \emptyset$
\For{$(a,b) \in O$}
\If{$a \notin H$}
\State $\cfg' \leftarrow \getmf(a)$
\If {$\isld(a) \land \neg \isll(a)$} 
\State $f \leftarrow \textbf{new}(\idmbld)$;
\Else 
\State $f \leftarrow \textbf{new}(\idmbfull)$;
\EndIf
\State $\cfg.\V \leftarrow \cfg.\V \cup \set{f}$; 
\State $P \leftarrow \set{(p,f) \mid (p,a) \in \cfg.\Elab^+}$
\State $Q \leftarrow \set{(f,q) \mid (a,q) \in \cfg.\Elab^+}$
\State $\cfg.\Elab \leftarrow \cfg.\Elab \cup \set{(f,a)} \cup P \cup Q$; 
\State $H \leftarrow H \cup \set{a}$
\EndIf
\EndFor
\State \Return $\false$;
\\ \textbf{end procedure}
\EndProcedure
\end{algorithmic}
\caption{ARMv8}
\end{subfigure}
\begin{subfigure}[b]{0.47\textwidth}
\begin{algorithmic}[1]
\Procedure{insertDMBv7}{$\mbbp, O$}
\State $H \leftarrow \emptyset$
\For{$(a,b) \in O$}
\If{$a \notin H \land \neg \isll(a)$}
\State $\cfg' \leftarrow \getmf(a)$
\State $f \leftarrow \textbf{new}(\idmbfull)$;
\State $\cfg.\V \leftarrow \cfg.\V \cup \set{f}$; 
\State $P \leftarrow \set{(p,f) \mid (p,a) \in \cfg.\Elab^+}$
\State $Q \leftarrow \set{(f,q) \mid (a,q) \in \cfg.\Elab^+}$
\State $\cfg.\Elab \leftarrow \cfg.\Elab \cup \set{(f,a)} \cup P \cup Q$; 
\State $H \leftarrow H \cup \set{a}$
\EndIf
\EndFor
\State \Return $\false$;
\\ \textbf{end procedure}
\EndProcedure
\end{algorithmic}
\caption{ARMv7}
\end{subfigure}
\caption{Fence insertion in ARMv8 and ARMv7 for enforcing robustness.}
\end{figure}

\begin{figure}[t]
\begin{subfigure}{0.9\textwidth}
\begin{align*}
\reachwo(\cfg, i, j, F) \triangleq & \reach(\tup{\cfg.\V \setminus F, \cfg.\Elab \setminus B},i,j) \text{ where } B = (G.\V \times F) \cup (F \times G.\V)
\\ RA(i,\rels,\acqs) \triangleq & \set{a \mid \neg \reachwo(\cfg,i,a,\rels) \land a \in \acqs}
\end{align*}
\\
\begin{tabular}{ll}
$\isst(i) \triangleq  \denot{i} \in \Wlab$  
& $\issc(i) \triangleq \denot{i} \in \Wlab \cap \codom(\rmw)$  
\\ $\isld(i) \triangleq \denot{i} \in \Rlab$ &   $\isacq(i) \triangleq \denot{i} \in \Alab$ 
\\ $\isll(i) \triangleq  \denot{i} \in \Rlab \cap \dom(\rmw)$ &  $\isw(i) \triangleq \denot{i} \in \Wlab \cup \Llab$ 
\\ $\isr(i) \triangleq  \denot{i} \in \Rlab \cup \Alab \qquad$ &
\end{tabular}
//
\begin{algorithmic}[1]
\Procedure{Ordered}{$\cfg,i,j$}
\State $FF \leftarrow \set{f \mid f \in G.\V \land \denot{f} \in \dmbfull}$; \quad\qquad $FL \leftarrow \set{f \mid f \in G.\V \land \denot{f} \in \dmbld}$; 
\State $FS \leftarrow \set{f \mid f \in G.\V \land \denot{f} \in \dmbst}$; \quad\qquad $L \leftarrow \set{a \mid a \in G.\V \land \denot{a} \in \Llab}$;
\State $A \leftarrow \set{a \mid a \in G.\V \land \denot{a} \in \Alab}$; \qquad\qquad  $B \leftarrow FF \cup \relacq(i,L,A)$;
\State $\mathbf{Switch} (i,j)$
\State \ \ \ \textbf{Case} $\mustalias(i,j)$:
\State \ \ \ \textbf{Case} $\isst(i) \land \isld(j) \land \neg \reachwo(\cfg, i, j, B)$: 
\State \ \ \ \textbf{Case} $(\isrel(j) \lor \isacq(i)) \lor (\isrel(i) \land \isacq(j))$:
\State \ \ \ \textbf{Case} $\isld(i) \land \isld(j) \land \neg \reachwo(\cfg,i,j,B \cup FL)$:
\State \ \ \ \textbf{Case} $\isld(i) \land \isst(j) \land \neg \reachwo(\cfg,i,j,B \cup FL \cup \srels(\cfg,j))$:
\State \ \ \ \textbf{Case} $\isst(i) \land \isst(j) \land \neg \reachwo(\cfg,i,j,B \cup FS \cup \srels(\cfg,j))$:
\State \ \ \ \textbf{Case} $(\isll(i) \lor \isld(i)) \land (\isst(j) \lor \issc(j))$: \Return $\true;$
\State\Return $\false;$
\EndProcedure
\end{algorithmic}
\caption{Checking order for pairs}
\end{subfigure}
\\[2ex]
\begin{subfigure}{0.47\textwidth}
\begin{algorithmic}[1]

\Procedure{SCRobustARMv8}{$\mbbp, \N$}
\State $\ell \leftarrow \Wlab \cup \Rlab \cup \Llab \cup \Alab$
\State $A \leftarrow \bigcup_{i\in \N}\mpairs(\mbbp(i), \ell, \ell)$; 
\State $O \leftarrow \emptyset$
\For{$(a,b) \in \oncyc(A)$}
\State $B \leftarrow \textsc{getB}(\getmf(b))$
\If{$\neg \ordered(\getmf(b),a,b)$}
\State $O \leftarrow O \cup \set{(a,b)}$;
\EndIf
\EndFor
\If{$O == \emptyset$} \ \Return $\true$;
\Else \ $\textsc{insertDMBv8}(\mbbp, O)$;
\EndIf
\\ \textbf{end procedure}
%
%
%
%
%
\EndProcedure
\end{algorithmic}
\caption{SC-robust against ARMv8}
\end{subfigure}
\hfill
\begin{subfigure}{0.5\textwidth}
\begin{algorithmic}[1]
\Procedure{x86RobustARMv8}{$\mbbp, \N$}
\State $\ell \leftarrow \Wlab \cup \Rlab \cup \Llab \cup \Alab$
\State $A \leftarrow \bigcup_{i\in \N}\mpairs(\mbbp(i), \ell, \ell)$; 
\State $O \leftarrow \emptyset$
\For{$(a,b) \in \oncyc(A)$}
\State $\cfg \leftarrow \getmf(b)$
\State $B \leftarrow \textsc{getB}(\cfg)$
\State $C \leftarrow  \isw(i) \land \isr(j) \land \neg (\issc(i) \land \isll(j))$ 
\If{$\neg (C \lor \ordered(\cfg,a,b))$}
\State $O \leftarrow O \cup \set{(a,b)}$;
\EndIf
\EndFor
\If{$O == \emptyset$} \ \Return $\true$;
\Else \ $\textsc{insertDMBv8}(\mbbp,O)$;
\EndIf
%
\\ \textbf{end procedure}

%
%
%
%
%
%
%
%
%
%
%
%
%
\EndProcedure
\end{algorithmic}
\caption{x86 robust against ARMv8}
\end{subfigure}
\caption{Robustness analysis of ARMv8 programs}
\label{fig:app:armarobusproc}
\end{figure}

\begin{figure}[t]
\begin{align*}
\isawra(i,j) \triangleq &  \isw(i) \land \isr(j) \land \neg (\issc(i) \land \isll(j))
\\ \ordered(\cfg, i, j, F) \triangleq & \mustalias(i,j) \lor \neg \reachwo(\cfg, i, j, F)
\end{align*}
\begin{subfigure}[b]{0.43\textwidth}
\begin{algorithmic}[1]
\Procedure{SCRobustARMv7}{$\mbbp, \N$}
\State $\ell \leftarrow \Wlab \cup \Rlab$
\State $pr \leftarrow \bigcup_{k \in \N} \mpairs(\mbbp(k), \ell, \ell)$;
\State $\uord \leftarrow \emptyset$
\For{$(i,j) \in \oncyc(pr)$}
\State $\cfg \leftarrow \getmf(b)$
\State $F = \set{f \mid f \in G.\V \land \denot{f} \in \F}$;
\If{$\neg \ordered(\cfg, i, j, F)$} 
\State $\uord \leftarrow \uord \cup \set{(i,j)}$
\EndIf
\EndFor
\If{$\uord == \emptyset$} 
\State \Return $\true;$
\Else \  $\textsc{insertDMBv7}(\mbbp,\uord)$;
\EndIf
\\ \textbf{end procedure}
\EndProcedure
\end{algorithmic}
\caption{SC-robustness against ARMv7}
\end{subfigure}
\hfill
\begin{subfigure}[b]{0.53\textwidth}
\begin{algorithmic}[1]
\Procedure{x86RobustARMv7}{$\mbbp, \N$}
\State $\ell \leftarrow \Wlab \cup \Rlab$
\State $pr \leftarrow \bigcup_{k \in \N} \mpairs(\mbbp(k), \ell, \ell)$;
\State $\uord \leftarrow \emptyset$
\For{$(i,j) \in \oncyc(pr)$}
\State $\cfg \leftarrow \getmf(b)$
\State $F = \set{f \mid f \in G.\V \land \denot{f} \in \F}$;
\If{$\neg (\isawra(i,j) \lor \ordered(\cfg, i, j, F))$}  
\State $\uord \leftarrow \uord \cup \set{(i,j)}$
\EndIf
\EndFor
\If{$\uord == \emptyset$} 
\State \Return $\true;$
\Else \  $\textsc{insertDMBv7}(\mbbp,\uord)$;
\EndIf
\\ \textbf{end procedure}
%
%
%
%
\EndProcedure
\end{algorithmic}
\caption{x86-robustness against ARMv7}
\end{subfigure}
\\[2ex]
\begin{subfigure}[b]{0.52\textwidth}
\begin{algorithmic}[1]
\Procedure{ARMv8RobustARMv7}{$\mbbp, \N$}

\State $\ell \leftarrow \Wlab \cup \Rlab$
\State $pr \leftarrow \bigcup_{k \in \N} \mpairs(\mbbp(k), \ell, \ell)$;
\State $\uord \leftarrow \emptyset$
\For{$(i,j) \in \oncyc(pr)$}
\State $\cfg \leftarrow \getmf(b)$
\State $F = \set{f \mid f \in G.\V \land \denot{f} \in \F}$;
\If{$\neg (\isst(i) \lor \ordered(\cfg, i, j, F))$}  
\State $\uord \leftarrow \uord \cup \set{(i,j)}$
\EndIf
\EndFor
\If{$\uord == \emptyset$} 
\State \Return $\true;$
\Else \ $\textsc{insertDMBv7}(\mbbp,\uord)$;
\EndIf
\\ \textbf{end procedure}
%
%
\EndProcedure
\end{algorithmic}
\caption{ARMv8 robust against ARMv7}
\end{subfigure}
\begin{subfigure}[b]{0.45\textwidth}
\begin{algorithmic}[1]
\Procedure{ARMv7mcaRobustARMv7}{$\mbbp, \N$}

\State $pr \leftarrow \bigcup_{k \in \N} \mpairs(\mbbp(k), \Rlab, \Rlab)$;
\State $\uord \leftarrow \emptyset$
\For{$(i,j) \in \oncyc(pr)$}
\State $\cfg \leftarrow \getmf(b)$
\State $F = \set{f \mid f \in G.\V \land \denot{f} \in \F}$;
\If{$\neg \ordered(\cfg, i, j, F)$}  
\State $\uord \leftarrow \uord \cup \set{(i,j)}$
\EndIf
\EndFor
\If{$\uord == \emptyset$} 
\State \Return $\true;$
\Else \  $\textsc{insertDMBv7}(\mbbp,\uord)$;
\EndIf
\\ \textbf{end procedure}
%
%
%
%
%
%
%
%
\EndProcedure
\end{algorithmic}
\caption{ARMv7-mca robust against ARMv7}
\end{subfigure}
\caption{Robustness analysis of ARMv7 programs}
\label{fig:app:armsrobusproc}
\end{figure}

\subsection{Checking and enforcing robustness} 

When an execution is $K$-consistent but violates 
$M$ consistency then it forms a cycle which violates 
certain irreflexivity condition. 
Such a cycle contain events on different locations and therefore 
two or more $\epo$ edges where 
given such an $\epo$ edge $(a,b)$ there exists other 
$\epo$ edge(s) $(p,q)$ and $(r,s)$ such that 
$a$ and $b$ access the same locations as $p$ and $s$ respectively as 
$(b,p), (s,a) \in \eco$.

We lift this semantic notion of 
robustness to program syntax in order to analyze and enforce robustness. 
We first identify the memory access pairs in all threads 
as these are potential $\epo$ edges. 
Next, we conservatively check if the memory access pairs  
would satisfy the robustness conditions in \cref{fig:robustconditions} in 
all its $K$ consistent executions. 
If so, we report the program as $M$-robust against $K$. 
To enforce robustness we insert appropriate fences between the 
memory access pairs.  

We perform such an analysis in \cref{fig:xrobusproc} to check and enforce SC-robustness against in x86 programs by procedure $\textsc{SCRobustx86}$ using a number of helper conditions. 
$\reachwo(\cfg,i,j,F)$ checks if there is a program path from access $i$ to access $j$ without passing through the fences $F$ in $\cfg$. 
$\ordered(\cfg, (i,j),F)$ checks if $(i,j)$ access pair ordered in respective models.
For example, in \cref{fig:xrobusproc} it checks if $i$ and $j$ access same location using $\mustalias$ or 
on all paths from $i$ to $j$ there exists a at least a fence from $F$ by $\reachwo$. 

Finally, given a set of memory access pair $A$, $\oncyc(A) \subseteq A$ 
identifies the set of memory access pairs which may result in $\epo$ edges 
in an execution. 
$\textsc{SCRobustx86}$ checks if all such store-load access pairs appropriately 
ordered which in turn ensure SC-robust for the program $\mbbp$ having $\N$ thread functions. 
If so, we report SC-robustness against x86. 
Otherwise, we insert fences between unordered pairs using $\textsc{insertF}$ procedure to 
enforce robustness. 
Similar to $\textsc{SCRobustx86}$ we also define procedures in \cref{fig:app:armarobusproc} and \cref{fig:app:armsrobusproc} respectively to check and enforce robustness in ARMv8 and ARMv7 programs. 

\section{Experimental Evaluation}
\label{sec:experiment}
Based on the obtained results we have implemented arachitecture to architeture (AA) mapping schemes defined in \cref{fig:xarm,fig:xarmas,fig:carmv8v7}, followed by fence elimination algorithms described in \cref{fig:fenceopt}. 
We have also developed robustness analyses for x86, ARMv8, and ARMv7 programs following the procedures in 
\cref{fig:xrobusproc,fig:app:armarobusproc,fig:app:armsrobusproc}.

We have implemented these mappings, fence eliminations, and robust analyses in LLVM. 
To analyze programs for fence elimination and checking robustness, we leverage the existing control-flow-graph analyses, alias analysis, and memory operand type analysis in LLVM.  
The CFG analyses are used to define $\mpairs$, $\opath$, $\reach$, and $\reachwo$ conditions. 
The $\mayalias$ and $\mustalias$ functions are defined using memory operand type and alias analyses. 

We have experimented these implementations on a number of well-known concurrent algorithms and data structures \cite{Lahav:2019,NorrisDemsky:2013} which use C11 concurrency primitives extensively. 
These programs exihibit fork-join concurrency where the threads are created from 
a set of functions. 
In these programs the memory accesses are \emph{relaxed} accesses in general and 
for wait loops we use \emph{release}/\emph{acquire} accesses.
Some of the programs have release-acquire/TSO/SC versions.
These programs assume the program would run on the respective memory models. 
\begin{center}
\begin{figure}
\begin{subfigure}{0.64\textwidth}
\centering
{\small
\begin{tabular}{|c|c|c|c|c|c|c|c|}
\hline
\multirow{2}{*}{Prog.} & \multirow{2}{*}{Orig} & \multicolumn{2}{c|}{x-v8} & \multicolumn{2}{c|}{C-x-v8} \\ \cline{3-6} 
                          &               & AA  & fd & AA  & fd  \\ \hline
                          
barrier      & 0,6,6  & 5,5,10  & 2,1,6  & 4,0,14   & 2,1,8   \\ \hline
dekker-tso   & 4,7,0  & 5,5,7   & 4,3,4  & 8,0,18   & 4,6,6    \\ \hline
dekker-sc    & 0,7,0  & 5,5,3   & 4,5,0  & 8,0,14   & 4,6,2    \\ \hline
pn-ra  & 4,3,0  & 5,12,7  & 4,7,2  & 4,0,16   & 4,5,6    \\ \hline
pn-ra-b  & 0,9,6  & 5,10,7  & 4,7,2  & 4,0,14   & 4,5,4    \\ \hline
pn-ra-d  & 0,5,4  & 5,10,7  & 4,5,4  & 4,0,14   & 4,5,4    \\ \hline
pn-tso & 2,3,0  & 5,12,7  & 4,7,2  & 4,0,14   & 4,5,4    \\ \hline
pn-sc  & 0,3,0  & 5,12,3  & 4,9,0  & 4,0,12   & 4,5,2    \\ \hline
lamport-ra   & 4,3,7  & 7,5,5   & 5,4,4  & 10,0,12  & 4,2,8    \\ \hline
lamport-tso  & 2,3,5  & 7,5,3   & 5,4,2  & 8,0,10   & 4,2,6    \\ \hline
lamport-sc   & 0,3,5  & 7,5,1   & 5,4,0  & 8,0,8    & 4,2,4    \\ \hline
spinlock     & 0,8,6  & 5,7,10  & 2,6,0  & 2,0,14   & 2,10,0  \\ \hline
spinlock4    & 0,14,12 & 9,11,18 & 4,10,0 & 4,0,24  & 4,18,0   \\ \hline
tlock   & 0,8,4   & 7,8,8   & 4,5,2  & 4,0,16  & 2,5,4    \\ \hline
tlock4  & 0,12,8  & 13,12,12 & 8,7,4 & 8,0,24  & 4,7,8   \\ \hline
seqlock      & 0,6,4 & 6,4,12    & 5,3,2  & 5,0,16  & 5,3,2    \\ \hline
nbw          & 0,3,4 & 10,8,12   & 6,6,1  & 7,0,18  & 6,7,6   \\ \hline
rcu          & 0,2,10 & 12,15,2 & 3,12,0   & 12,0,11 & 2,4,4   \\ \hline
rcu-ofl  & 4,16,8 & 17,18,24 & 12,6,9 & 15,0,51 & 11,2,42 \\ \hline
cilk-tso     & 2,7,4 & 15,15,15 & 13,4,11 & 13,0,29 & 9,4,14 \\ \hline
cilk-sc      & 0,7,4 & 15,15,13 & 13,6,9 & 13,0,27 & 9,6,12   \\ \hline
cldq-ra  & 3,4,0  & 7,5,9 & 6,2,1 & 6,0,14 & 6,2,2   \\ \hline
cldq-tso & 1,4,0  & 9,5,7 & 6,2,1 & 6,0,12 & 6,2,2 \\ \hline
cldq-sc  & 0,4,0 & 7,5,6 & 6,2,1 & 6,0,11  & 6,2,1 \\ \hline
\end{tabular}
}
\caption{x86 to ARMv8}
\label{fig:mappingxv8}
\end{subfigure}
\hfill
\begin{subfigure}{0.34\textwidth}
\centering
{\small
\begin{tabular}{|c|c|c|c|c|}
\hline
\multirow{2}{*}{Prog.} & \multirow{2}{*}{Orig} & \multicolumn{2}{c|}{v8-x} \\ \cline{3-4} 
                          &                           & AA  & fd   \\ \hline
barrier     & 6,0  & 4,2  & 4,1  \\ \hline
dekker-tso  & 3,4  & 0,11 & 0,6    \\ \hline
dekker-sc   & 3,0  & 0,7  & 0,3   \\ \hline
pn-ra       & 3,4  & 0,7  & 0,3   \\ \hline
pn-ra-b       & 5,0  & 2,7  & 2,3   \\ \hline
pn-ra-d       & 5,0  & 2,3  & 2,1   \\ \hline
pn-tso      & 3,2  & 0,5  & 0,5    \\ \hline
pn-sc       & 3,0  & 0,3  & 0,1    \\ \hline
lamport-ra  & 1,4  & 0,7  & 0,5    \\ \hline
lamport-tso & 1,2  & 0,5  & 0,3    \\ \hline
lamport-sc  & 1,0  & 0,3  & 0,1    \\ \hline
spinlock    & 4,0  & 2,4  & 2,1    \\ \hline
spinlock4   & 6,0  & 4,6  & 4,2    \\ \hline
tlock       & 6,0  & 2,6  & 2,3    \\ \hline
tlock4      & 8,0  & 4,8  & 4,2    \\ \hline
seqlock     & 6,0 & 4,4 & 4,1   \\ \hline
nbw         & 4,0  & 2,3  & 2,2    \\ \hline
rcu         & 2,0  & 0,2  & 0,1    \\ \hline
rcu-ofl     & 16,4 & 1,20 & 1,12    \\ \hline
cilk-tso    & 5,2  & 2,9  & 2,2     \\ \hline
cilk-sc     & 5,0  & 2,7  & 2,1     \\ \hline
cldq-ra     & 4,3  & 2,5  & 2,2    \\ \hline
cldq-tso    & 4,1  & 2,3  & 2,2    \\ \hline
cldq-sc     & 4,0  & 2,2  & 2,2   \\ \hline
\end{tabular}
}
\caption{ARMv8 to x86}
\label{fig:mappingv8x}
\end{subfigure}
\caption{Mappings between x86 and ARMv8. 
In x86 to ARMv8: \#(ish, stl, lda) in original and \#(ishld, ishst, ish) after mapping. In ARMv8 to x86: \#(RMW,mfence) in original and generated programs.}
\end{figure}
\end{center}
\vspace{-7mm}
\subsection{Mapping Schemes}
We have modified the x86, ARMv7, and ARMv8 code generation phases in LLVM to capture the effect of mapping schemes on C11 programs. 
For example, in original LLVM mapping a non-atomic store ($\Wlab_\MOna$) results in $\movst$ and $\str$ 
accesses in x86 and ARMv8 respectively. 
Following the AA-mapping in \cref{tab:xarm}, $\movst$ results in $\idmbst;\str$ in ARMv8. 
Therefore to capture the effect of x86 to ARMv8 translation we generate $\idmbst;\str$ in ARMv8 instead of a $\str$ for a C11 non-atomic store access.
We modify the code lowering phase in LLVM to generate the required leading and trailing fences along with the memory accesses.  
The AA-mapping schemes introduce additional fences compared to original mapping in all mapping schemes which is evident in \cref{fig:mappingv8v7,fig:mappingv7v8,fig:mappingv8x,fig:mappingxv8} in `Orig' and `AA' columns respectively.

\smallskip
\textit{\emph{x86} to \emph{ARMv8} mappings (\cref{tab:xarm}).} 
In \cref{fig:mappingxv8} we show the numbers of different fences 
resulted from $\cmm \mapsto \arma$ (Orig), $\xarch \mapsto \arma$ (AA in x-v8), 
and $\cmm \mapsto \xarch \mapsto \arma$ (AA in C-x-v8).
Both $\xarch \mapsto \arma$ and $\cmm \mapsto \xarch \mapsto \arma$ mapping schemes generate more fences compared to the original $\cmm \mapsto \arma$ mapping. 
$\xarch \mapsto \arma$ (x-v8) generates more $\idmbld$ fences compared to $\cmm \mapsto \xarch \mapsto \arma$ (C-x-v8) as the earlier scheme generates trailing $\idmbld$ fence for non-atomic loads. 
However, the number of $\idmbfull$ fences are more in C-x-v8 compared to x-v8 as atomic stores introduce leading $\idmbfull$ fences instead of $\idmbst$. 
For the same reason there is no $\idmbst$ in C-x-v8 column.
%

\smallskip

\textit{\emph{ARMv8} to \emph{x86} mappings (\cref{tab:armx}).} 
As shown in \cref{fig:mappingv8x}, 
the number of atomic updates and fence operations in AA-mapping 
varies from Orig due to the mapping of C11 $\Wlab_\MOsc$ and $\Wlab_\MOrel$ accesses. 
In original mapping $\Wlab_\MOsc \mapsto \irmw$ and $\Wlab_\MOrel \mapsto \movst$
whereas in AA-mapping 
$\Wlab_{(\MOrel \mid \MOsc)} \mapsto \stlr \mapsto \movst;\mfence$. 
As a result, the number of atomic updates are less and the number of fences are more in AA-mapping compared to the original x86 mapping in LLVM. 
We can observe the tradeoff between $\xarch \!\mapsto\! \arma$ and $\cmm \!\mapsto\!\xarch \!\mapsto\! \arma$ 
considering the number of generated $\idmbld$ and $\idmbfull$ fences. 
For example, in \emph{Barrier} program $\xarch \!\mapsto\! \arma$ generates more $\idmbld$ than $\cmm \!\mapsto\!\xarch \!\mapsto\! \arma$ as it generates $\idmbld$ fences for non-atomic loads. 
On the other hand, $\cmm \!\mapsto\!\xarch \!\mapsto\! \arma$ generates $\idmbfull$ fences for relaxed atomic stores instead of $\idmbst$ fences.    

\smallskip

\textit{\emph{ARMv8} to \emph{ARMv7} mappings (\cref{tab:armv8v7})} 
We show the number of $\idmb$ fences in \cref{fig:mappingv8v7} due to $\cmm\!\mapsto\!\arma$ (Orig), $\arma \!\mapsto\!\arms$ (AA in v8-v7),  $\cmm\!\mapsto\!\arma\!\mapsto\!\arms$ (AA in C-v8-v7) 
mappings. 
Both $\arma\!\mapsto\!\arms$ and $\cmm\!\mapsto\!\arma\!\mapsto\!\arms$ generate more fences than $\cmm \!\mapsto\!\arma$ mapping. 
Moreover, $\cmm\!\mapsto\!\arma\!\mapsto\!\arms$ generates less number of fences than 
$\arma\!\mapsto\!\arms$ as we do not generate trailing $\idmb$ fences for non-atomic loads.

\smallskip
\textit{\emph{ARMv7} to \emph{ARMv8} mappings (\cref{tab:armv7v8}).} 
The result is in  \cref{fig:mappingv7v8} where 
The original $\cmm \mapsto \arma$ mapping generates $\idmbfull$, release-store, and 
acquire-load operations for these programs whereas the AA-mapping generates 
respective $\idmbfull$ fences only as ARMv7 does not have release-store, and 
acquire-load operations.

\begin{center}
\begin{figure}
\begin{subfigure}{0.52\textwidth}
\centering
{\small
\begin{tabular}{|c|c|c|c|c|}
\hline
\multirow{2}{*}{Programs} & \multirow{2}{*}{Orig} & \multicolumn{2}{c|}{v7-v8} \\ \cline{3-4} 
                          &                           & AA  & fd   \\ \hline
barrier      & 0,6,6   & 0,0,12    & 1,1,8   \\ \hline
dekker-tso   & 4,7,0   & 0,0,11    & 2,5,4    \\ \hline
dekker-sc    & 0,7,0   & 0,0,7     & 2,5,0    \\ \hline
pn-ra  & 4,3,0   & 0,0,7     & 0,2,4    \\ \hline
pn-ra-b  & 0,9,6   & 0,0,15     & 0,4,6    \\ \hline
pn-ra-d  & 0,5,4   & 0,0,9     & 0,2,6    \\ \hline
pn-tso & 2,3,0   & 0,0,5     & 0,2,2    \\ \hline
pn-sc  & 0,3,0   & 0,0,3     & 0,2,0    \\ \hline

lamport-ra   & 4,3,7   & 0,0,14    & 1,2,11    \\ \hline
lamport-tso  & 2,3,5   & 0,0,10    & 1,2,7    \\ \hline
lamport-sc   & 0,3,5   & 0,0,8     & 1,2,5    \\ \hline
spinlock     & 0,8,6   & 0,0,14    & 2,11,0    \\ \hline
spinlock4    & 0,14,12 & 0,0,26    & 4,21,0   \\ \hline
tlock   & 0,8,4   & 0,0,12    & 2,5,4   \\ \hline
tlock4  & 0,12,8  & 0,0,20    & 4,7,8   \\ \hline
seqlock      & 0,4,4   & 0,0,8     & 4,3,2    \\ \hline
nbw          & 0,3,4   & 0,0,9     & 1,3,4   \\ \hline
rcu          & 0,2,10  & 0,0,12    & 0,1,10  \\ \hline
rcu-ofl  & 4,16,8  & 0,0,29    & 1,1,22  \\ \hline
cilk-tso     & 2,7,4 & 0,0,15    & 4,4,8   \\ \hline
cilk-sc      & 0,7,4 & 0,0,13    & 4,6,6   \\ \hline
cldq-ra  & 3,4,0 & 0,0,7 & 3,1,6  \\ \hline
cldq-tso & 1,4,0  & 0,0,5 & 1,1,2  \\ \hline
cldq-sc  & 0,4,0  & 0,0,4 & 1,1,1  \\ \hline
\end{tabular}
}
\caption{ARMv7 to ARMv8}
\label{fig:mappingv7v8}
\end{subfigure}
\hfill
\begin{subfigure}{0.45\textwidth}
\centering
{\small
\begin{tabular}{|c|c|c|c|c|c|c|c|}
\hline
\multirow{2}{*}{Prog.} & \multirow{2}{*}{Orig} & \multicolumn{2}{c|}{v8-v7} & \multicolumn{2}{c|}{C-v8-v7} \\ \cline{3-6} 
                          &                           & AA  & fd  & AA   & fd  \\ \hline

barrier     & 13     & 19        & 16    & 13     & 12          \\ \hline
 dekker-tso  & 12     & 25        & 23    & 22     & 19         \\ \hline
 dekker-sc   & 8      & 21        & 20    & 18     & 15         \\ \hline
 pn-ra       & 8      & 17        & 16    & 12     & 11         \\ \hline
 pn-ra-b     & 14     & 19        & 16    & 18     & 13         \\ \hline
 pn-ra-d     & 10     & 17        & 16    & 12     & 11         \\ \hline
 pn-tso      & 6      & 15        & 14    & 10     & 9         \\ \hline
 pn-sc       & 4      & 13        & 12    & 8      & 7        \\ \hline
 lamport-ra  & 15     & 21        & 20    & 18     & 17       \\ \hline
 lamport-tso & 11     & 18        & 17    & 15     & 14         \\ \hline
 lamport-sc  & 9      & 16        & 15    & 13     & 12          \\ \hline
 spinlock    & 13     & 18        & 17    & 15     & 12        \\ \hline
 spinlock4   & 23     & 32        & 31    & 27     & 22         \\ \hline
 tlock       & 13     & 20        & 16    & 17     & 12         \\ \hline
 tlock4      & 21     & 34        & 29    & 29     & 20        \\ \hline
 seqlock     & 11     & 19        & 15    & 12     & 9         \\ \hline
 nbw         & 7      & 23        & 21    & 15     & 13         \\ \hline
 rcu         & 13     & 36        & 32    & 15     & 14         \\ \hline
 rcu-ofl     & 30     & 55        & 49    & 39     & 36         \\ \hline
 cilk-tso    & 13     & 34        & 31    & 30     & 22         \\ \hline
 cilk-sc     & 11     & 32        & 31    & 28     & 23          \\ \hline
 cldq-ra     & 8      & 19        & 18    & 12     & 12        \\ \hline
 cldq-tso    & 6      & 18        & 17    & 12     & 11         \\ \hline
 cldq-sc     & 5      & 18        & 17    & 11     & 11        \\ \hline
\end{tabular}
}
\caption{ARMv8 to ARMv7}
\label{fig:mappingv8v7}
\end{subfigure}
\caption{Mappings between ARMv7 and ARMv8. Original mapping to ARMv8 is ($\idmbfull$, release-store, acquire-load). In ARMv7-ARMv8 mapping the numbers are of ($\idmbld$, $\idmbst$, $\idmbfull$).}
\end{figure}
\end{center}
\vspace{-7mm}
\subsection{Fence elimination} 
The fence optimization passes remove significant number of fences as shown in the `fd' columns in \cref{fig:mappingv8v7,fig:mappingv7v8,fig:mappingv8x,fig:mappingxv8}. 
We have implemented the fence elimination algorithms as LLVM passes and run the pass after AA-mappings to eliminate redundant fences. 
The pass extends LLVM \emph{MachineFunctionPass} and run on each machine function of the 
program. The precision of our analyses depend upon underlying LLVM functions which we have used. 
For example, we apply alias analysis and memory operand analysis to identify the memory location accessed by a particular access. 
Consider a scenario where we have identified an $\mfence$ between a store-load pair. 
If we precisely identify that the store-load pair access same-location then we can eliminate the fence. 
Otherwise we conservatively mark the fence as non-eliinable. 

\smallskip
\textit{Fence elimination after x86 to ARMv8 mapping.} 
The fence elimination algorithms have eliminated a number of redundant fences 
after the mapping. 
In some scenarios original C11 to ARMv8 mapping 
is too restrictive as it generates release-store and acquire-load 
accesses for C11  release-store and acquire-load accesses 
respectively. 
In our scheme we prefer to generate fences separately and 
fence elimination eliminates those extra fences.

\textit{Fence elimination after C11 to x86 to ARMv8 mapping.} 
In this case we first weaken the $\idmbfull$ fences to a pair of $\idmbst$ and $\idmbld$ fences whenever appropriate  
and then perform the fence elimination. 
Therefore it introduces some $\idmbst$ fences in the 'fd' column in C-x-v8. 

\smallskip
\textit{Fence elimination after ARMv8 to x86 mapping.} 
The mapping generates $\mfence$ for release-store mapping and the 
fence elimination safely eliminate these fences whenever possible.  

\smallskip
\textit{Fence elimination after ARMv7 to ARMv8 mapping.} 
In this case the mapping introduce $\idmbfull$ fences in ARMv8 
from ARMv7 $\idmb$ fences. 
We eliminate the repeated fences if any and then 
weaken the $\idmbfull$ fences to 
$\idmbst$ and $\idmbld$ fences, and 
further eliminate redundant fences.

\smallskip
\textit{Fence elimination after ARMv8 to ARMv7 mapping.} 
ARMv8 to ARMv7 mapping generates extra fences 
in certain scenarios such as $\ldr;\stlr \mapsto \ldr;\idmb;\idmb;\str;\idmb$ 
where we can safely remove a repeated $\idmb$ fence. 
Similar scenario takes place for $\ldr_\MOat;\stlr$ mapping in C11 to ARMv8 to ARMv7 
mapping.

\begin{center}
\begin{figure}
\centering
{\small
\begin{tabular}{|c|c|c|c|c|c|c|c|c|c|}
\hline
\multirow{2}{*}{Prog.} & \begin{tabular}[c]{@{}c@{}}x86A\end{tabular} & \multicolumn{2}{c|}{ARMv8} & \multicolumn{4}{c|}{ARMv7} & \begin{tabular}[c]{@{}c@{}}Rocker\\ (RA)\end{tabular} & \begin{tabular}[c]{@{}c@{}}Trencher\\ (TSO)\end{tabular} \\ \cline{2-10} 
                       & SC                                                  & SC          & x86A          & SC   & x86A   & v8   & mca  & SC                                                    & SC                                                       \\ \hline
barrier    &8$\mid$0\no 1   & 12$\mid$6\no 6 & \no 5 &  12$\mid$10\no 1&\yes&\yes&\yes& \yes(\#2) & \no(\#2) \\ \hline
dekker-tso   &20$\mid$ 4\yes & 20$\mid$8 \no 6 & \no 6  &  20$\mid$ 8 \no 8  &  \no 8   &  \no 8  & \no 4  &   \yes(\#2)   &  \yes(\#2)   \\ \hline
dekker-sc   &20$\mid$ 0 \no 10 & 20 $\mid$ 4 \no 12 & \no 9& 20$\mid$ 4 \no 12  & \no 8  & \no 8   & \no 4 & \no(\#2)   &  \no(\#2)   \\ \hline
pn-ra   &12$\mid$ 4 \yes & 12 $\mid$ 4 \no 7 & \no 7 & 12$\mid$ 4 \no 8   &  \no 8   &  \no 6  & \no 4  &   \yes(\#2)   &  \yes(\#2)   \\ \hline
pn-ra-b   &10 $\mid$ 0 \no 2 & 12 $\mid$ 10 \no 2 & \no 2 & 12 $\mid$ 12 \no 2   &  \yes   &  \yes  & \yes  &   \no(\#2)   &  \no(\#2)   \\ \hline
pn-ra-d   &10$\mid$ 0 \yes & 12 $\mid$ 4 \no 8 & \no 8 & 12 $\mid$ 6 \no 8   &  \no 8   &  \no 4  & \no 2  &   \yes(\#2)   &  \yes(\#2)   \\ \hline
pn-tso  &12$\mid$ 2 \yes& 12$\mid$ 2 \no 9 & \no 9 & 12 $\mid$ 2 \no 10   & \no 10    &  \no 6  & \no 4  &   \no(\#2)   &  \yes(\#2)   \\ \hline
pn-sc    &12 $\mid$ 0 \no 4 &12 $\mid$ 0 \no 11 & \no 11 & 12$\mid$ 0 \no 10 &\no 10 &\no 6 &\no 4 &   \no(\#2)   &  \no(\#2)   \\ \hline
lmprt-ra   &19$\mid$ 4\no 8 & 18 $\mid$ 13 \no 7 & \no 4 & 19 $\mid$ 13 \no 6  &  \no 4   & \no 3   & \yes 0 & \yes (\#2/3)   &  \yes(\#2)   \\ \hline
lmprt-tso  & 17$\mid$ 2 \no 6 & 16$\mid$ 9 \no 11 & \no 10  & 17 $\mid$ 9 \no 8   &  \no 7  & \no 6   & \no 1   &   \no (\#2)   &  \yes(\#2)   \\ \hline
lmprt-sc   &17$\mid$ 0 \no 8 & 16$\mid$ 7 \no 14 & \no 13 &  17$\mid$ 7 \no 10  &  \no 9 & \no 8 & \no 3  & \no (\#2)   &  \no(\#2)   \\ \hline
spinlock   &8$\mid$ 0 \yes & 10 $\mid$ 8 \yes & \yes &  12$\mid$ 12 \yes  &  \yes   &  \yes  & \yes  &   \yes (\#2)   &  \yes(\#2)   \\ \hline
spinlock4   &16$\mid$0\yes&20$\mid$16\yes & \yes  &  24$\mid$24\yes & \yes & \yes  & \yes &  \yes (\#4)   &  \yes(\#4)   \\ \hline
tlock    &10$\mid$0\yes& 10$\mid$6\yes & \yes & 12$\mid$8\yes  &  \yes  & \yes  & \yes  &   \yes (\#2)   &  \yes(\#2)   \\ \hline
tlock4   &20$\mid$0\yes& 20$\mid$\yes & \yes & 24$\mid$16\yes  &  \yes  & \yes  & \yes  &   \yes (\#2)   &  \yes(\#2)   \\ \hline
seqlock  &7$\mid$0\yes& 11$\mid$8\no 3 & \no 3 &  11$\mid$8\no 1 &  \no 1   &  \no 1  &  \no 1 &   \yes (\#2)   &  \yes(\#2)   \\ \hline
nbw    &15$\mid$0\yes& 18$\mid$2 \no 12 & \no 12 &  20$\mid$ 7 \no 10  & \no 10  & \no 9  & \no 8  &   \yes (\#4)   &  \yes(\#4)   \\ \hline
rcu     &27$\mid$0\no 10& 25$\mid$ 10 \no 16 & \no 12 &  27$\mid$ 10 \no 18  &  \no 18   &  \no 9  & \no 7  &   \yes (\#4)   &  \no(\#4)   \\ \hline
rcu-ofl   &30$\mid$ 4 \no 7 & 33$\mid$ 14 \no 27 & \no 25 &  36$\mid$ 19 \no 17 &  \no 16  & \no 15  & \no 6  &   \yes (\#3)   &  \no(\#3)   \\ \hline
cilk-tso  &11$\mid$2\yes& 28$\mid$ 6\no 8 & \no 8 &  29$\mid$ 10 \no 7  &  \no 7  & \no 7  & \no 7   &   \yes (\#2)   &  \yes(\#2)   \\ \hline
cilk-sc  &11$\mid$0\yes& 28$\mid$ 4 \no 9 & \no 9 & 29 $\mid$ 8 \no 8 & \no 8  & \no 8  & \no 8  &   \no (\#2)   &  \no(\#2)   \\ \hline
cldq-ra    &9$\mid$3\yes& 11$\mid$5\no 3 & \no 3 & 11$\mid$ 5 \no 3   &  \no 3   & \no 3   &  \no 1 &   \yes (\#3)   &  \yes(\#3)   \\ \hline
cldq-tso    &9$\mid$1\yes& 11$\mid$ 3 \no 5 & \no 5 &  11$\mid$ 3 \no 5  &  \no 5   &  \no 5  & \no 3  &   \no (\#3)   &  \yes(\#3)   \\ \hline
cldq-sc    &9$\mid$ 0 \no 1& 11$\mid$ 2\no 6 & \no 7 & 11$\mid$ 2 \no 6   &  \no 6   & \no 6   & \no.4  &   \no (\#3)   &  \no(\#3)   \\ \hline
\end{tabular}
}
\caption{Robustness analyses. 
Entry (a$\mid$b\yes/\no c) where a: \# fences inserted by naive scheme excluding the existing fences,  
b: \#existing fences, \yes/\no: program is robust or not, [c] \#fences inserted to enforce robustness.  
Rocker and Trencher robustness results (for \#k number of threads) are taken from \cite{Lahav:2019}.
Our SC-robustness against x86A analysis matches Trencher in a number of cases. 
ARMv8 and ARMv7 is weaker than RA and therefore we report non-robustness in these programs.  
}
\label{fig:tabrobustness}
\end{figure}
\end{center}

 \vspace{-7mm}
\subsection{Robustness analysis}

We implement the robustness analysis as LLVM passes following the procedures in   \cref{fig:xrobusproc} as well as following \cref{fig:app:armarobusproc,fig:app:armsrobusproc} in the appendix 
after instruction lowering in x86, ARMv8, and ARMv7. 
We report the analyses results on the concurrent programs in \cref{fig:tabrobustness}. 
In these results we mark both robustness checking and robustness enforcement results. 
We have also included the results from \cite{Lahav:2019} about two other robustness checker: Trencher \cite{Bouajjani:2013} and Rocker \cite{Lahav:2019}.

Now we discuss the robustness results of the benchmarks programs which are marked by \yes or \no. 
Among these programs \emph{spinlock}, \emph{spinlock4}, \emph{seqlock}, \emph{ticketlock} (tlock), 
and \emph{ticketlock4} (tlock4) provide robustness in all models. 
These results also match the results from both Trencher and Rocker; both SC-robustness checkers. 
In rest of the programs we observe robustness violations due to various unordered accesses sequences. 
For example, (St-Ld) violates SC-robustness in all architectures, (SC-St/Ld) violate x86A robustness in ARMv8 and ARMv7, and (Ld-Ld) violate all robustness in ARMv8 and ARMv7 models.
%

\smallskip
\textit{Robustness of x86 programs.}
We first focus on SC-robustness against x86A and compare the result with Trencher. 
Our analysis precisely analyze robustness and agrees to Trencher in all cases except \emph{lamport-ra} (lmprt-ra),  \emph{lamport-tso} (lmprt-tso), and cilk-sc. 
Both \emph{lamport-ra} and \emph{lamport-tso} has (St-Ld) sequence in different thread functions. 
As a result, our analysis reports SC-robustness violation which is a false positive as  
in actual executions these access pairs never execute in concurrence. 
In \emph{cilk-sc} we report SC-robustness as the program has store-load sequences of the form 
$a = \Rlab_\MOrlx(T);\Wlab_\MOrlx(T,a\!-\!1);\Rlab_\MOrlx(H)$. 
In this case the $\Wlab_\MOrlx(T,a\!-\!1);\Rlab_\MOrlx(H)$ may yield non-SC behavior during an execution 
which is reported by Trencher and Rocker.
However, LLVM combines the load and store of $T$ into an atomic fetch-and-sub ($\mathsf{fsub}$) operation, that is, $a = \Rlab_\MOrlx(T);\Wlab_\MOrlx(T,a\!-\!1) \leadsto a = \mathsf{fsub}(T,1)$. 
As a result the program turns into SC-robust against x86 in LLVM as reported by our analysis.

\smallskip 
\textit{Robustness of ARMv8 programs.}
Next, we study SC-robustness and x86A-robustness against ARMv8 for the benchmark programs. 
ARMv8 allows out-of-order executions of memory accesses on different locations which do not affect dependencies. 
Therefore many of these programs in ARMv8 are not SC or x86A robust. 
Also our robustness analyzer do not rely on $\dob$ ordering as it performs the analysis before the ARMv8 machine code is generated during the code lowering phase. 
Therefore LLVM may perform optimizations after the analysis which may remove 
certain dependencies and in that case our analysis would be unsound and may report false negative.    

As ARMv8 is weaker than x86A, the program which are not SC-robust in x86A are also not SC-robust in ARMv8. 
Programs like \emph{barrier}, \emph{peterson-ra-Bartosz} (pn-ra-b), \emph{peterson-sc} (pn-sc), \emph{lamport-ra/tso/sc}, \emph{rcu}, \emph{rcu-offline} (rcu-ofl), and \emph{chase-lev-dequeue-tso/sc} (cldq-tso/sc) are in this category. 
There are programs which are SC-robust in x86 but not in ARMv8 such as \emph{dekker-tso} and so on. 
These programs violate both SC and x86A robustness due to unordered (Ld-Ld) or (SC-St/Ld) pairs.   
%

\smallskip 
\textit{Robustness of ARMv7 programs.}
Now we move to the robustness analysis in ARMv7. 
Except \emph{spinlock}, \emph{spinlock4}, and \emph{seqlock} programs,  all other programs violate SC-robustness due to the similar pattern as discussed in ARMv8 robustness. 
Among these programs SC-robustness is violated in \emph{barrier} due to (St-Ld) unordered sequence. 
This access pattern is allowed in x86A, ARMv8, and ARMv7-mca and therefore these 
ARMv7 programs are robust in these models.
Program \emph{rcu} has unordered (St-St) pairs which violates SC and x86A robustness. 
However, these pairs does not violate ARMv8 and ARMv7-mca robustness. 
Rest of the programs exihibit certain (Ld-Ld) pairs which result 
in x86, ARMv8, and ARMv7-mca robustness violations. 

\subsubsection{Enforcing robustness}

Whenever we identify a program as non-robust we insert appropriate fences 
to enforce respective robustness. 
For example in \cref{fig:xrobusproc} we identify the different-location 
store-load access pairs which may violate robustness. 
We introduce leading $\mfence$ operations for the load operation in the pair as required. 
   
A naive scheme does not use robustness information. 
It first eliminates existing fences in concurrent threads and then 
insert fences after each memory accesses except atomic update in x86 and 
load-exclusive accesses in ARM models to restrict program behavior.
In both naive scheme and our approach we do not insert fences 
for atomic update.
In ARMv8 we insert $\idmbld$ and $\idmbfull$ trailing fences for load, and 
store and store-exclusive respectively when they are unordered with a successor. 
In ARMv7 we insert $\idmbfull$ trailing fences for load,  
store, and store-exclusive when they are unordered with a successor. 
%
%
%

In \cref{fig:tabrobustness} we report the number of fences 
required in the naive scheme, robustness analyses results 
in our proposed approach along with the number of introduce fences 
to enforce robustness. 
We compare our result to the naive schemes as explained in \cref{fig:tabrobustness} 
and find that our approach insert less number of fences in major instances. 
However, our fence insertion is not optimal; we leave the optimal fence insertion
for enforcing robustness for future investigation.

\section{Related Work}
\label{sec:related}

\textit{Architecture to architecture mapping} 
There are a number of dynamic binary translators \cite{Ding:2011,Wang:2011,Hong:2012,Lustig:2015,Cota:2017} 
emulate mutithreaded program. 
Among these earlier translators such as PQEMU\cite{Ding:2011}, COREMU\cite{Wang:2011}, HQEMU \cite{Hong:2012} and so on do not address the memory consistency model mismatches. 
ArMOR \cite{Lustig:2015} proposes a specification format to define the ordering requirements for different 
memory models which is used in translating between architectural concurrency models in dynamic translation. 
The specification format is used in specifying TSO and Power architectures. 
\cite{Cota:2017} uses the rules from ArMOR in Pico dynamic translator for QEMU.
Our mapping schemes provide the ordering rules which can be used to populate the 
ordering tables for x86 and ARM models. 
Moreover the ARMv8 reordering table in \cref{fig:armatrans}  
demonstrates that reordering 
certain independent access pairs are not safe if they are part of 
certain dependency based ordering.
In addition to the QEMU based translators, LLVM based decompilers \cite{dagger,mcsema,Yadavalli:2019,retdec,Shen:2012} raise binary code to LLVM IR and then compiles to another architecture. 
These decompilers do not support relaxed memory concurrency. 

\textit{Fence optimization} Redundant fence elimination is addressed by  \cite{Vafeiadis:2011,Elhorst:2014,Morisset:2017}. \citet{Vafeiadis:2011} performs safe fence elimination in x86,  \citet{Elhorst:2014} eliminate adjacent fences in ARMv7, and \citet{Morisset:2017} perform efficient fence elimination in x86, Power, and ARMv7. 
However, none of these approaches perform ARMv8 fence elimination. 
%
%
%

\textit{Robustness analysis.} 
Sequential consistency robustness has been explored against TSO \cite{Bouajjani:2013}, POWER \cite{Derevenetc:2014}, and Release-Acquire \cite{Lahav:2019} models by exploring executions using model checking tools. 
\citet{Alglave:2017} proposed fence insertion in POWER to strengthen a program to release/acquire semantics 
which has same \emph{preserved-program-order} constraints between memory aceesses as TSO. 
On the contrary, we identify robustness checking conditions in ARMv7 and ARMv8 where 
we show that \emph{preserved-program-order} is not sufficient to recover sequential consistency 
in ARMv7 models. 
Identifying minimal set of fences is NP-hard \cite{Lee:2001} and a number of approaches such as \cite{Shasha-Snir,Bouajjani:2013,Lee:2001,Alglave:2017} proposed fence insertion to recover stonger order, particularly sequential consistency.  
Similar to \cite{Lee:2001} our approach is based on analyzing control flow graphs without exploring the 
possible executions by model checkers. 
Though in certain scenarios we report false positives, our approach precisely identifies 
robustness for a number of well-known programs.


\section{Conclusion and Future Work}
In this paper we propose correct and efficient mapping schemes 
between x86, ARMv8, and ARMv7 concurrency models.  
We have shown that ARMv8 can indeed serve as an intermediate model for 
mapping between x86 and ARMv7. 
We have also shown that removing non-multicopy atomicity from ARMv7 
does not affect the mapping schemes. 
We also show that ARMv8 model cannot serve as an IR in a decompiler 
as it does not support all common compiler optimizations.
Next,we propose fence elimination algorithms to remove additional fences 
generated by the mapping schemes. 
We also propose robustness analyses and enforcement techniques based on 
memory access sequence analysis for x86 and ARM programs. 

Going forward we want to extend these schemes and analyses  
to other architectures as well. 
We believe these results would play a crucial role in a number of 
translator, decompilers, and state-of-the-art systems. 
Therefore integrating these results to these systems 
is another direction we would like to pursue in future.


\bibliographystyle{abbrvnat}
\bibliography{paper}

\begin{thebibliography}{37}
\providecommand{\natexlab}[1]{#1}
\providecommand{\url}[1]{\texttt{#1}}
\expandafter\ifx\csname urlstyle\endcsname\relax
  \providecommand{\doi}[1]{doi: #1}\else
  \providecommand{\doi}{doi: \begingroup \urlstyle{rm}\Url}\fi

\bibitem[map()]{mappings}
{C/C++11} mappings to processors.
\newblock \url{https://www.cl.cam.ac.uk/~pes20/cpp/cpp0xmappings.html}.

\bibitem[Alglave and Maranget()]{herd}
J.~Alglave and L.~Maranget.
\newblock herd7 consistency model simulator.
\newblock \url{http://diy.inria.fr/www/}.

\bibitem[Alglave et~al.(2014)Alglave, Maranget, and Tautschnig]{Alglave:2014}
J.~Alglave, L.~Maranget, and M.~Tautschnig.
\newblock Herding cats: modelling, simulation, testing, and data-mining for
  weak memory.
\newblock \emph{{ACM} Trans. Program. Lang. Syst.}, 36\penalty0 (2):\penalty0
  7:1--7:74, 2014.
\newblock \doi{10.1145/2627752}.

\bibitem[Alglave et~al.(2017)Alglave, Kroening, Nimal, and
  Poetzl]{Alglave:2017}
J.~Alglave, D.~Kroening, V.~Nimal, and D.~Poetzl.
\newblock Don't sit on the fence: {A} static analysis approach to automatic
  fence insertion.
\newblock \emph{{ACM} Trans. Program. Lang. Syst.}, 39\penalty0 (2):\penalty0
  6:1--6:38, 2017.

\bibitem[Android-x86()]{androidx}
Android-x86.
\newblock \url{https://www.android-x86.org/}.

\bibitem[Arm()]{arm57}
Arm.
\newblock Migrating a software application from armv5 to armv7-a/r application.
\newblock
  \url{http://infocenter.arm.com/help/index.jsp?topic=/com.arm.doc.dai0425/chapter1intendreader.html}.

\bibitem[avast()]{retdec}
avast.
\newblock A retargetable machine-code decompiler based on llvm.
\newblock \url{https://github.com/avast/retdec}.

\bibitem[Barbalace et~al.(2017)Barbalace, Lyerly, Jelesnianski, Carno, Chuang,
  Legout, and Ravindran]{Barbalace:2017}
A.~Barbalace, R.~Lyerly, C.~Jelesnianski, A.~Carno, H.~Chuang, V.~Legout, and
  B.~Ravindran.
\newblock Breaking the boundaries in heterogeneous-isa datacenters.
\newblock In \emph{{ASPLOS} 2017}, pages 645--659, 2017.
\newblock \doi{10.1145/3037697.3037738}.

\bibitem[Barbalace et~al.(2020)Barbalace, Karaoui, Wang, Xing, Olivier, and
  Ravindran]{Barbalace:2020}
A.~Barbalace, M.~L. Karaoui, W.~Wang, T.~Xing, P.~Olivier, and B.~Ravindran.
\newblock Edge computing: the case for heterogeneous-isa container migration.
\newblock In \emph{{VEE}'20}, pages 73--87, 2020.
\newblock \doi{10.1145/3381052.3381321}.

\bibitem[Bits()]{mcsema}
L.~Bits.
\newblock Framework for lifting x86, amd64, and aarch64 program binaries to
  llvm bitcode.
\newblock \url{https://github.com/lifting-bits/mcsema}.

\bibitem[Bouajjani et~al.(2013)Bouajjani, Derevenetc, and
  Meyer]{Bouajjani:2013}
A.~Bouajjani, E.~Derevenetc, and R.~Meyer.
\newblock Checking and enforcing robustness against {TSO}.
\newblock In \emph{ESOP 2013}, pages 533--553, 2013.
\newblock \doi{10.1007/978-3-642-37036-6\_29}.

\bibitem[Bougacha()]{dagger}
A.~Bougacha.
\newblock Binary translator to llvm ir.
\newblock \url{https://github.com/repzret/dagger}.

\bibitem[{Chernoff} et~al.(1998){Chernoff}, {Herdeg}, {Hookway}, {Reeve},
  {Rubin}, {Tye}, {Bharadwaj Yadavalli}, and {Yates}]{Chernoff:1998}
A.~{Chernoff}, M.~{Herdeg}, R.~{Hookway}, C.~{Reeve}, N.~{Rubin}, T.~{Tye},
  S.~{Bharadwaj Yadavalli}, and J.~{Yates}.
\newblock Fx\!32 a profile-directed binary translator.
\newblock \emph{IEEE Micro}, 18\penalty0 (2):\penalty0 56--64, 1998.

\bibitem[Cota et~al.(2017)Cota, Bonzini, Benn\'{e}e, and Carloni]{Cota:2017}
E.~G. Cota, P.~Bonzini, A.~Benn\'{e}e, and L.~P. Carloni.
\newblock Cross-isa machine emulation for multicores.
\newblock In \emph{CGO'2017}, page 210–220. IEEE Press, 2017.

\bibitem[Derevenetc and Meyer(2014)]{Derevenetc:2014}
E.~Derevenetc and R.~Meyer.
\newblock Robustness against power is pspace-complete.
\newblock In \emph{ICALP'14}, volume 8573 of \emph{LNCS}, pages 158--170, 2014.
\newblock \doi{10.1007/978-3-662-43951-7\_14}.

\bibitem[Ding et~al.(2011)Ding, Chang, Hsu, and Chung]{Ding:2011}
J.~Ding, P.~Chang, W.~Hsu, and Y.~Chung.
\newblock {PQEMU:} {A} parallel system emulator based on {QEMU}.
\newblock In \emph{ICPADS'11}, pages 276--283, 2011.
\newblock \doi{10.1109/ICPADS.2011.102}.

\bibitem[Docs()]{msxa}
M.~Docs.
\newblock How x86 emulation works on arm.
\newblock
  \url{https://docs.microsoft.com/en-us/windows/uwp/porting/apps-on-arm-x86-emulation}.

\bibitem[Elhorst(2014)]{Elhorst:2014}
R.~Elhorst.
\newblock Lowering {C11} atomics for {ARM} in {LLVM}.
\newblock In \emph{European {LLVM} Conference}, 2014.

\bibitem[Hong et~al.(2012)Hong, Hsu, Yew, Wu, Hsu, Liu, Wang, and
  Chung]{Hong:2012}
D.-Y. Hong, C.-C. Hsu, P.-C. Yew, J.-J. Wu, W.-C. Hsu, P.~Liu, C.-M. Wang, and
  Y.-C. Chung.
\newblock Hqemu: A multi-threaded and retargetable dynamic binary translator on
  multicores.
\newblock In \emph{CGO'12}, page 104–113, 2012.
\newblock \doi{10.1145/2259016.2259030}.

\bibitem[{ISO/IEC 14882}(2011)]{cppstandard}
{ISO/IEC 14882}.
\newblock Programming language {C++}, 2011.

\bibitem[{ISO/IEC 9899}(2011)]{cstandard}
{ISO/IEC 9899}.
\newblock Programming language {C}, 2011.

\bibitem[Lahav and Margalit(2019)]{Lahav:2019}
O.~Lahav and R.~Margalit.
\newblock Robustness against release/acquire semantics.
\newblock In \emph{PLDI 2019}, pages 126--141, 2019.
\newblock \doi{10.1145/3314221.3314604}.

\bibitem[Lahav and Vafeiadis(2016)]{Lahav:fm16}
O.~Lahav and V.~Vafeiadis.
\newblock Explaining relaxed memory models with program transformations.
\newblock In \emph{FM'16}, pages 479--495, 2016.
\newblock \doi{10.1007/978-3-319-48989-6_29}.

\bibitem[Lahav et~al.(2017)Lahav, Vafeiadis, Kang, Hur, and Dreyer]{Lahav:2017}
O.~Lahav, V.~Vafeiadis, J.~Kang, C.-K. Hur, and D.~Dreyer.
\newblock Repairing sequential consistency in {C/C++11}.
\newblock In \emph{PLDI 2017}, pages 618--632, 2017.
\newblock \doi{10.1145/3062341.3062352}.
\newblock Technical Appendix Available at
  \url{https://plv.mpi-sws.org/scfix/full.pdf}.

\bibitem[{Lee} and {Padua}(2001)]{Lee:2001}
J.~{Lee} and D.~A. {Padua}.
\newblock Hiding relaxed memory consistency with a compiler.
\newblock \emph{IEEE Transactions on Computers}, 50\penalty0 (8):\penalty0
  824--833, 2001.

\bibitem[Lustig et~al.(2015)Lustig, Trippel, Pellauer, and
  Martonosi]{Lustig:2015}
D.~Lustig, C.~Trippel, M.~Pellauer, and M.~Martonosi.
\newblock Armor: Defending against memory consistency model mismatches in
  heterogeneous architectures.
\newblock In \emph{ISCA'15}, page 388–400, 2015.
\newblock \doi{10.1145/2749469.2750378}.

\bibitem[Morisset and Nardelli(2017)]{Morisset:2017}
R.~Morisset and F.~Z. Nardelli.
\newblock Partially redundant fence elimination for x86, arm, and power
  processors.
\newblock In \emph{CC'17}, pages 1--10, 2017.

\bibitem[Norris and Demsky(2013)]{NorrisDemsky:2013}
B.~Norris and B.~Demsky.
\newblock {CDSChecker}: Checking concurrent data structures written with
  {C/C++} atomics.
\newblock In \emph{OOPSLA'13}, 2013.

\bibitem[notaz(2014)]{notaz:2014}
notaz.
\newblock Starcraft.
\newblock \url{http://repo.openpandora.org/}, 2014.

\bibitem[Pulte et~al.(2018)Pulte, Flur, Deacon, French, Sarkar, and
  Sewell]{Pulte:2018}
C.~Pulte, S.~Flur, W.~Deacon, J.~French, S.~Sarkar, and P.~Sewell.
\newblock Simplifying {ARM} concurrency: multicopy-atomic axiomatic and
  operational models for {ARMv8}.
\newblock \emph{{PACMPL}}, 2\penalty0 ({POPL}):\penalty0 19:1--19:29, 2018.
\newblock \doi{10.1145/3158107}.

\bibitem[QEMU()]{qemu}
QEMU.
\newblock the fast! processor emulator.
\newblock \url{https://www.qemu.org/}.

\bibitem[Shasha and Snir(1988)]{Shasha-Snir}
D.~E. Shasha and M.~Snir.
\newblock Efficient and correct execution of parallel programs that share
  memory.
\newblock \emph{{ACM} Trans. Program. Lang. Syst.}, 10\penalty0 (2):\penalty0
  282--312, 1988.
\newblock \doi{10.1145/42190.42277}.

\bibitem[Shen et~al.(2012)Shen, Chen, Hsu, and Yang]{Shen:2012}
B.-Y. Shen, J.-Y. Chen, W.-C. Hsu, and W.~Yang.
\newblock Llbt: An llvm-based static binary translator.
\newblock In \emph{CASES 2012}, page 51–60, 2012.
\newblock \doi{10.1145/2380403.2380419}.

\bibitem[Vafeiadis and Zappa~Nardelli(2011)]{Vafeiadis:2011}
V.~Vafeiadis and F.~Zappa~Nardelli.
\newblock Verifying fence elimination optimisations.
\newblock In \emph{SAS'11}, volume 6887 of \emph{LNCS}, pages 146--162.
  Springer, 2011.
\newblock \doi{10.1007/978-3-642-23702-7_14}.

\bibitem[Wang et~al.(2011)Wang, Liu, Chen, Wu, Chen, Zhang, and
  Zang]{Wang:2011}
Z.~Wang, R.~Liu, Y.~Chen, X.~Wu, H.~Chen, W.~Zhang, and B.~Zang.
\newblock {COREMU:} a scalable and portable parallel full-system emulator.
\newblock In C.~Cascaval and P.~Yew, editors, \emph{PPOPP'11}, pages 213--222,
  2011.
\newblock \doi{10.1145/1941553.1941583}.

\bibitem[Wickerson et~al.(2017)Wickerson, Batty, Sorensen, and
  Constantinides]{Wickerson:2017}
J.~Wickerson, M.~Batty, T.~Sorensen, and G.~A. Constantinides.
\newblock Automatically comparing memory consistency models.
\newblock In \emph{POPL'17}, pages 190--204. ACM, 2017.
\newblock \doi{10.1145/3009837.3009838}.

\bibitem[Yadavalli and Smith(2019)]{Yadavalli:2019}
S.~B. Yadavalli and A.~Smith.
\newblock Raising binaries to llvm ir with mctoll (wip paper).
\newblock In \emph{LCTES 2019}, page 213–218, 2019.
\newblock \doi{10.1145/3316482.3326354}.

\end{thebibliography}

\appendix



%
%
%

%
%

\clearpage

%

\section{Proofs of Mapping Schemes}
\label{app:mappings}

\subsection{x86 to ARMv8 Mappings}
\label{app:x86toarmv8}

We first restate \cref{th:xarm}.

\xarm*

To prove \cref{th:xarm}, we prove the following formal statement.
\[
\inarr{
\xp \leadsto \armp \implies \forall \ex_t \in \denot{\armp}.~\exists \ex_s \in \denot{\xp}.~\behav(\ex_t) = \behav(\ex_s)
}
\]

Given an ARM execution $\ex_t$ we define correxponding x86 execution $\ex_s$ where 
%

\begin{enumerate}

\item $[\ex_t.\Wlab \cup \ex_t.\dmbfull];\ex_t.\ob;[\ex_t.\Wlab \cup \ex_t.\dmbfull] \implies \ex_s.\mo$

\item $[\ex_t.\Wlab \cup \ex_t.\dmbfull];\ex_t.\lpo;[\ex_t.\Wlab \cup \ex_t.\dmbfull] \implies \ex_s.\mo$

\item $ [\ex_t.\dmbfull];\ex_t.\lpo;\ex_t.\fr \implies \ex_s.\mo$

\item $\ex_t.\co \implies \rloc {\ex_s.\mo}$
\end{enumerate}

We know that $\ex_t$ is ARMv8 consistent. 
Now we show that $\ex_s$ is x86 consistent. 

\begin{proof}

We prove by contradiction. 

%
%
%
%
%
%
%
%

\bigskip

(irrHB)

Assume $\ex_s$ has an $\ex_s.\xhb$ cycle. 

It implies a $(\ex_s.\lpo \cup \ex_s.\rfe)^+$ cycle. 

Considering the possible cases of $\ex_s.\lpo$ edges on the cycle:

\Case{$[\ex_s.\Rlab];\ex_s.\lpo;[\ex_s.\WUlab]$}

$\implies$ $[\ex_t.\Rlab];\ex_t.\lpo;[\ex_t.\dmbld];\ex_t.\lpo;[\ex_t.\WUlab]$. 

$\implies$ $[\ex_t.\Rlab];\ex_t.\bob;[\ex_s.\WUlab]$

$\implies$ $[\ex_t.\Rlab];\ex_t.\ob;[\ex_s.\WUlab]$

\smallskip

\Case{$[\ex_s.\Ulab];\ex_s.\lpo;[\ex_s.\WUlab]$}

$\implies$ $[\ex_t.\Rlab];\ex_t.\rmw;\ex_t.\lpo;[\ex_t.\dmbfull];\ex_t.\lpo;\ex_t.\rmw;[\ex_t.\Wlab]$

$\implies$ $[\ex_t.\Rlab];\ex_t.\aob;\ex_t.\bob;\ex_t.\aob;[\ex_t.\Wlab]$

$\implies$ $[\ex_t.\Rlab];\ex_t.\ob;[\ex_t.\Wlab]$ 

Thus in both cases $\ex_s.\xhb \implies (\ex_t.\ob \cup \ex_t.\rfe)^+ \subseteq \ex_t.\ob$. 
However, $\ex_t$ is ARM consistent and $\ex_t.\ob$ is irreflexive. 
Hence a contradiction and $\ex_s.\xhb$ is irreflexive.

\bigskip

(irrMOHB)

Assume $\ex_s$ has a $\ex_s.\mo;\ex_s.\xhb$ cycle. 

However, from definition, 
$[\ex_s.\WUlab \cup \ex_s.\F];\ex_s.\xhb;[\ex_s.\WUlab \cup \ex_s.\F]$ 

Considering the $\lpo$ and $\rfe$ from $\xhb$:

\medskip

\Case{$[\ex_s.\WUlab \cup \ex_s.\F];\ex_s.\lpo;[\ex_s.\WUlab \cup \ex_s.\F]$}

We know,

$[\ex_s.\WUlab \cup \ex_s.\F];\ex_s.\lpo;[\ex_s.\WUlab \cup \ex_s.\F]$

%
%
%
%
%
%
%
%
%

Considering the subcases:

\Subcase{$[\ex_s.\Wlab \cup \ex_s.\F];\ex_s.\lpo;[\ex_s.\Wlab \cup \ex_s.\F]$}

It implies $[\ex_t.\Wlab \cup \ex_t.\dmbfull];\ex_t.\lpo;[\ex_t.\Wlab \cup \ex_t.\dmbfull]$.

From definitions, $[\ex_t.\Wlab \cup \ex_t.\dmbfull];\ex_t.\lpo;[\ex_t.\Wlab \cup \ex_t.\dmbfull] \implies \ex_s.\mo \land \neg \ex_s.\mo^{-1}$.

\medskip

\Subcase{Otherwise}

Possible scenarios are $[\ex_s.\Ulab];\ex_s.\lpo;[\ex_s.\WUlab \cup \ex_s.\F]$ or $[\ex_s.\WUlab \cup \ex_s.\F];\ex_t.\lpo;[\ex_t.\Ulab]$.

Now,

$[\ex_s.\Ulab];\ex_s.\lpo;[\ex_s.\WUlab \cup \ex_s.\F]$

$\implies \ex_t.\rmw;\ex_t.\lpo;[\ex_t.\dmbfull];\ex_t.\lpo;[\ex_t.\WUlab \cup \ex_t.\dmbfull]$ 

$\implies \ex_t.\bob$

$\implies \ex_t.\ob$ 

Similarly, 

$[\ex_s.\WUlab \cup \ex_s.\F];\ex_t.\lpo;[\ex_t.\Ulab]$

$\implies \ex_t.\lpo;[\ex_t.\dmbfull];\ex_t.\lpo;[\ex_s.\WUlab]$

$\implies \ex_t.\bob$

$\implies \ex_t.\ob$ 

\smallskip 

From definitions, $[\ex_t.\Wlab \ex_t.\dmbfull];\ex_t.\ob;[\ex_t.\Wlab \ex_t.\dmbfull] \implies \ex_s.\mo \land \neg \ex_s.\mo^{-1}$. 

\medskip

\Case{$[\ex_s.\WUlab \cup \ex_s.\F];\ex_s.\rfe;[\ex_s.\WUlab \cup \ex_s.\F]$}

It implies $[\ex_s.\WUlab];\ex_s.\rfe;[\ex_s.\Ulab]$

$\implies ([\ex_t.\Rlab];\ex_t.\rmw)^?;[\ex_t.\Wlab];\ex_t.\rfe;[\ex_t.\Rlab];\ex_t.\rmw;[\ex_t.\Wlab]$ following the mappings.

$\implies ([\ex_t.\Rlab];\ex_t.\aob)^?;[\ex_t.\Wlab];\ex_t.\obs;[\ex_t.\Rlab];\ex_t.\aob;[\ex_t.\Wlab]$ 

$\implies ([\ex_t.\Rlab];\ex_t.\aob)^?;[\ex_t.\Wlab];\ex_t.\ob;[\ex_t.\Wlab]$ 

From definitions we know that $[\ex_t.\Wlab];\ex_t.\ob;[\ex_t.\Wlab]  \implies \ex_s.\mo \land \neg \ex_s.\mo^{-1}$.

Therefore $\ex_s.\xhb \implies \ex_s.\mo$ and hence $\ex_s.\mo;\ex_s.\xhb$ is acyclic and  $\ex_s$ satisfies (irrMOHB).

\bigskip

(irrFRHB)

Assume $\ex_s$ has a $\ex_s.\fr;\ex_s.\xhb$ cycle. 

We already know that $\ex_s.\xhb \implies \ex_t.\ob$ holds. 

Considering the cases of $\ex_s.\fr$:

\Case{$\ex_s.\fre$}

In this case $\ex_s.\fre \implies \ex_t.\fre \implies \ex_t.\obs$. 

In this case there exists a  $\ex_t.\obs;\ex_t.\ob$ cycle which violates (external) in $\ex_t$. 

Hence a contradiction and $\ex_s$ satisfies (irrFRHB).

\Case{$\ex_s.\fri$} 

Following the mappings $\ex_s.\fri \implies \ex_t.\bob$. 

In this case there exists a  $\ex_t.\bob;\ex_t.\ob$ cycle which violates (external) in $\ex_t$. 

Hence a contradiction and $\ex_s$ satisfies (irrFRHB).

\bigskip

(irrFRMO) 

Assume $\ex_s$ has a $\ex_s.\fr;\ex_s.\mo$ cycle. 

It implies a $\ex_s.\fr;\ex_s.\co$ cycle and in consequence a $\ex_t.\fr;\ex_t.\co$ cycle which violates (internal) in $\ex_t$. 

Hence a contradiction and $\ex_s$ satisfies (irrFRMO).

\bigskip

(irrFMRP)

Assume $\ex_s$ has a $\ex_s.\fr;\ex_s.\mo;\ex_s.\rfe;\ex_s.\lpo$ cycle. 

It implies a $\ex_s.\rfe;\ex_s.\lpo;\ex_s.\fr;\ex_s.\mo$ cycle. 

Now we consider a $\ex_s.\rfe;\ex_s.\lpo;\ex_s.\fr$ path.
%
%
%
%
%

Thus 

$[\ex_s.\WUlab];\ex_s.\rfe;\ex_s.\lpo;\ex_s.\fr;[\ex_s.\WUlab]$

\begin{flalign*}
\implies & [\ex_s.\WUlab];\ex_s.\rfe;\ex_s.\lpo;[\ex_s.\RUlab];\ex_s.\fre;[\ex_s.\WUlab] & \\
& \cup [\ex_s.\WUlab];\ex_s.\rfe;\ex_s.\lpo;[\ex_s.\RUlab];\ex_s.\fri;[\ex_s.\WUlab]       &
\end{flalign*}


\begin{flalign*}
\implies & [\ex_t.\Wlab];\ex_t.\rfe;[\ex_t.\Rlab];\ex_t.\lpo;[\ex_t.\dmbld \cup \ex_t.\dmbfull];\ex_t.\lpo;[\ex_t.\Rlab];\ex_t.\fre;[\ex_t.\Wlab] & \\
& \cup  [\ex_t.\Wlab];\ex_t.\rfe;[\ex_t.\Rlab];\ex_t.\lpo;[\ex_t.\dmbld \cup \ex_t.\dmbfull];\ex_t.\lpo;[\ex_t.\Rlab];\ex_t.\fri;[\ex_t.\Wlab]  &
\end{flalign*}
\begin{flalign*}
\implies & [\ex_t.\Wlab];\ex_t.\obs;[\ex_t.\Rlab];\ex_t.\bob;[\ex_t.\Rlab];\ex_t.\obs;[\ex_t.\Wlab] & \\
& \cup  [\ex_t.\Wlab];\ex_t.\obs;[\ex_t.\Rlab];\ex_t.\bob;[\ex_t.\Rlab];\ex_t.\bob;[\ex_t.\Wlab]  &
\end{flalign*}
\begin{flalign*}
\implies & [\ex_t.\Wlab];\ex_t.\ob;[\ex_t.\Wlab] \cup  [\ex_t.\Wlab];\ex_t.\ob;[\ex_t.\Wlab] &
\end{flalign*}
\begin{flalign*}
\implies & [\Wlab];\ex_t.\ob;[\ex_t.\Wlab] &
\end{flalign*}

\medskip

However, we know $[\ex_t.\Wlab];\ex_t.\ob;[\Wlab] \implies [\ex_s.\WUlab];\ex_s.\mo;[\ex_s.\WUlab]$.

Thus $[\ex_s.\WUlab];\ex_s.\rfe;\ex_s.\lpo;\ex_s.\fr;[\ex_s.\WUlab] \implies \ex_s.\mo \land \neg \ex_s.\mo^{-1}$. 

Hence a contradiction and thus $\ex_s$ satisfies (irrFMRP).

\bigskip

(irrUF)

Assume $\ex_s$ has a $\ex_s.\fr;\ex_s.\mo;[\ex_s.\Ulab \cup \ex_s.\F];\ex_s.\lpo$ cycle. 

It implies $[\ex_s.\Ulab \cup \ex_s.\F];\ex_s.\lpo;[\ex_s.\RUlab];\ex_s.\fr;[\ex_s.\WUlab];\ex_s.\mo$ cycle.

Now, we consider a $[\ex_s.\Ulab \cup \ex_s.\F];\ex_s.\lpo;[\ex_s.\RUlab];\ex_s.\fr;[\ex_s.\WUlab]$ path.

Considering possible cases:

\medskip

\Case{$[\ex_s.\Ulab];\ex_s.\lpo;[\ex_s.\RUlab];\ex_s.\fr;[\ex_s.\WUlab]$}

$\implies [\ex_t.\Rlab];\ex_t.\lpo;[\ex_t.\dmbfull];\ex_t.\lpo;[\ex_t.\Rlab];(\ex_t.\fre \cup \ex_t.\fri);[\ex_t.\Wlab]$

\begin{flalign*}
\implies & [\ex_t.\Rlab];\ex_t.\lpo;[\ex_t.\dmbfull];\ex_t.\lpo;[\ex_t.\Rlab];\ex_t.\fre;[\ex_t.\Wlab] & \\
& \cup [\ex_t.\Rlab];\ex_t.\lpo;[\ex_t.\dmbfull];\ex_t.\lpo;[\ex_t.\Rlab];\ex_t.\fri;[\ex_t.\Wlab] &
\end{flalign*}

$\implies [\ex_t.\Rlab];\ex_t.\bob;\ex_t.\obs;[\ex_t.\Wlab] 
\cup [\ex_t.\Rlab];\ex_t.\bob;[\ex_t.\Wlab]$

$\implies [\ex_t.\Rlab];\ex_t.\ob;[\ex_t.\Wlab]$

$\implies \ex_s.\mo$ following the definition.

\medskip

\Case{$[\ex_s.\F];\ex_s.\lpo;\ex_s.\fr;[\ex_s.\WUlab]$}

$\implies [\ex_t.\dmbfull];\ex_s.\lpo;\ex_t.\fr;[\ex_t.\Wlab]$ following the mappings.

$\implies \ex_s.\mo \land \neg \ex_s.\mo^{-1}$ following the definition.

Therefore $[\ex_s.\Ulab \cup \ex_s.\F];\ex_s.\lpo;\ex_s.\fr;\ex_s.\mo$ does not have a cycle.

Hence a contradiction and $\ex_s$ satisfies (irrUF).

\bigskip

From definition we know $\ex_t.\co \implies \rloc {\ex_s.\mo}$ and therefore $\behav(\ex_s) = \behav(\ex_t)$ holds.

\end{proof}

\subsection{Correctness of C11 to x86 to ARMv8 Mapping}
\label{app:cxarm}

We restate the theorem and then prove the same.

\cxarm*

\begin{proof}
The mapping can be represented as a combination of following transformation steps.
\begin{enumerate}
\item $\mbbp_{\cmm} \mapsto \mbbp_{\xarch}$ mapping from \citet{mappings}.
\item $\mbbp_{\xarch} \mapsto \mbbp_{\arma}$ mappings from \cref{tab:xarm}. 
\item Fence strengthening $\idmbst;\str \leadsto \idmbfull;\str$ in $\mbbp_{\arma}$.
\item Elimination of leading $\idmbfull$ and trailing $\idmbld$ fences in following cases.
\begin{enumerate}
\item $\idmbfull;\str \leadsto \str$ where $\movst_\MOna \mapsto \str$.
\item $\ldr;\idmbld \leadsto \ldr$ where $\movld_\MOna \mapsto \ldr$.
\end{enumerate}
\end{enumerate}

We know (1), (2), (3) are sound and therefore it suffices to show that transformation (4) is sound.

Let $\ex_a$ and $\ex'_a$ be the consistent execution of $\mbbp_{\arma}$ 
before and after the transformation (3).
Let $\ex$ be correspnding C11 execution $\mbbp_{\cmm}$.
 and 
we know $\mbbp_{\cmm}$ is race-free. 
Therefore for all non-atomic event $a$ in $\ex$ if there exist 
another same-location event $b$ then $\ex.\lhb^=(a,b)$ holds. 

Now we consider x86 to ARMv8 mapping scheme in \cref{tab:xarm}.

Considering the $\lhb$ definition following are the possibilities:
 
\Case{$[\E_\MOna];\ex.\lpo;[\WUlab_{\sqsupseteq \MOrlx}]$}

$\implies [\E];\ex_a.\lpo;[\dmbld];\ex_a.\lpo;[\dmbfull];\ex_a.\lpo;[\E]$

$\implies [\E];\ex'_a.\lpo;[\dmbfull];\ex'_a.\lpo;[\E]$

$\implies [\E];\ex'_a.\bob;[\E]$

\Case{$[\RUlab_{\sqsupseteq \MOrlx}];\ex.\lpo;[\E_\MOna]$} 

$\implies [\Rlab];(\ex_a.\rmw;\ex_a.\dmbfull \cup \ex_a.\lpo;[\dmbld]);\ex_a.\lpo;[\E]$

$\implies [\E];\ex'_a.\bob;[\E]$

Hence $\ex'_a.\bob = \ex_a.\bob$ and the transfmation is sound for x86 to ARMv8 mapping. 

As a result, the mapping scheme in \cref{tab:cxarm} is sound.
\end{proof}


\subsection{ARMv8 to x86 Mappings}
\label{app:armv8tox86}

%

We restate \cref{lem:relcomposex}.

\relcomposex*

%
%
%

\begin{proof}


We consider two cases in $\ex$:

\medskip

\Case{$[\ex.\Rlab];\ex.\poloc;\ex.\fr;[\ex.\WUlab]$}

Let $(r,e) \in [\ex.\Rlab];\ex.\poloc;[\ex.\RUlab]$, $(e,w') \in [\ex.\RUlab];\ex.\fr;[\ex.\WUlab]$ holds. 

Also consider $\ex.\lrf(w_e,e)$ and $\ex.\lrf(w,r)$ holds.

We show by contradiction that $\ex.\co(w,w')$ and in consequence $\ex.\fr(r,w')$ holds. 

Assume $\ex.\co(w_e,w)$ holds. 
Therefore $\ex.\fr(e,w)$ holds. 
However, from definition, $\ex.\xhb(w,e)$ holds. 
It is not possible in a x86 consistent execution as it violates $\irreflexive(\ex.\fr;\ex.\xhb)$ condition. 
Hence a contradiction and $\ex.\co(w,w_e)$ holds. 

We also know that $\ex.\co(w_e,w')$ holds as from definition $\ex.\lrf(w_e,e) \land \ex.\fr(e,w')$. 

As a result $\ex.\co(w,w')$ holds. 

Therefore $\ex.\fr(r,w')$ holds. 

Thus $[\ex.\Rlab];\ex.\poloc;\ex.\fr;[\ex.\WUlab] \implies \ex.\fr$.

\medskip

\Case{$[\ex.\WUlab];\ex.\poloc;\ex.\fr;[\ex.\WUlab]$} 

Let $(w,w') \in [\ex.\WUlab];\ex.\poloc;\ex.\fr;[\ex.\WUlab]$ and 

$(w,r) \in [\ex.\WUlab];\ex.\poloc;[\ex.\RUlab] \land (r,w') \in [\ex.\RUlab];\ex.\fr;[\ex.\WUlab]$ holds. 

Two subcases:

\Subcase{$\ex.\lrf(w,r)$} 

In this case $\ex.\co(w,w')$ holds by definition. 

\Subcase{$\ex.\rfe(w_r,r)$} 

In this case $w \neq w_r$. 

We show $\ex.\co(w,w_r)$ holds by contradiction.

Assume $\ex.\co(w_r,w)$ holds. In that case $\ex.\fr(r,w)$ holds. 
This violates $\irreflexive(\ex.\fr;\ex.\xhb)$ constraint and hence a contradiction. 

Therefore, $\ex.\co(w,w_r)$ holds and in consequence $\co(w,w')$ holds.

Thus $[\ex.\WUlab];\ex.\poloc;\ex.\fr;[\ex.\WUlab] \implies \ex.\co$.     
\end{proof}

We restate \cref{lem:noloadx}.

\noloadx*


\begin{proof}

Consider a load event $r$ on $(\ex.\poloc \cup \ex.\fr \cup \ex.\co \cup \ex.\lrf)^+$ path. 
Considering the path, the possible incoming edges to $r$ are $\ex.\lrf$, $\ex.\poloc$, and 
the outgoing edges are $\ex.\fr$, $\ex.\poloc$.

Let $a$ and $b$ be the source and destination of the incoming and outgoing edges on the path. 

Possible cases:

\Case{$\ex.\lrf(a,r) \land \ex.\fr(r,b)$} 

From definition $\ex.\co(a,b)$ holds.

\Case{$\ex.\lrf(a,r) \land \ex.\poloc(r,b)$} 
 
From definition, $\ex.\xhb(a,b)$.

\Case{$\ex.\poloc(a,r) \land \ex.\fr(r,b)$}
 
From \cref{lem:relcomposex}, $\ex.\fr(a,b) \lor \ex.\co(a,b)$ holds.
 
\Case{$\ex.\poloc(a,r) \land \ex.\poloc(r,b)$} 
 
From definition, $\ex.\xhb(a,b)$ holds.

\end{proof}

We restate \cref{lem:noloadobx}.

\noloadobx*


\begin{proof}

Consider a load event $r$ on $\ex.\obx$ path. 
Considering the path, the possible incoming edges to $r$ are $\ex.\lrf$, $\ex.\xppo$, and 
the outgoing edges are $\ex.\fr$, $\ex.\xppo$.

Let $a$ and $b$ be the source and destination of the incoming and outgoing edges on the path. 

Possible cases: 

\Case{$\ex.\lrf(a,r) \land \ex.\fr(r,b)$} 

From definition $\ex.\mo(a,b)$ holds.

\Case{$\ex.\lrf(a,r) \land \ex.\xppo(r,b)$} 

From definition $\ex.\xhb(a,b)$ holds as $\xppo \subseteq \lpo$. 

\Case{$\ex.\lpo(a,r) \land \ex.\lpo(r,b)$} 

From definition $\ex.\xhb(a,b)$ holds as $\xppo \subseteq \lpo$. 

\Case{$\ex.\xppo(a,r) \land \ex.\fr(r,b)$} 

Considering the subcases of $a$:

\Subcase{$a \in (\WUlab \cup \F)$}

We show $\ex.\mo(a,b)$ holds.

In this case following the definition of $\xppo$ we know $(a,r) \in [\Wlab];\lpo;[\F];\lpo;[\Rlab]$. 
Let $c \in \F$ such that $\ex.\lpo(a,c) \land \ex.\lpo(c,r)$ holds. 

We show $\ex.\mo(c,b)$ holds by contradiction.

Assume $\ex.\mo(b,c)$ holds. 

In this case $\ex.\fr(r,b) \land \ex.\mo(b,c) \land c \in \F \land \ex.\lpo(c,r)$ creates a cycle. 
Hence a contradiction as $\ex$ is x86 consistent. 

Therefore $\ex.\mo(c,b)$ holds.

We also know that $\ex.\mo(a,c)$ holds as $\ex.\mo(c,a)$ would lead to a $\ex.\mo;\ex.\xhb$ cycle which is a contradiction. 

As a result, $\ex.\mo(a,c) \land \ex.\mo(c,b)$ implies that $\ex.\mo(a,b)$ holds.

\Subcase{$a \in \Rlab$}

Let $\ex.\lrf(w,a)$. 
We consider two scenarios based on whether there is an intermediate fence:

\Subsubcase{$(a,r) \in [\Rlab];\ex.\lpo;[\F];\ex.\lpo;[\Rlab]$} 

Let $c \in \F$ be the intermediate fence event.

It implies $(a,c) \in \ex.\xppo$ following $s_6$ and $\ex.\mo(c,b)$ holds. \hfill (see earlier subcase)

Thus there is a $\ex.\obx$ path from $a$ to $b$ without passing through $r$.

\Subsubcase{Otherwise}

In this case $(a,r) \in [\Rlab];\ex.\lpo;[\Rlab] \land \nexists e.~\ex.\lpo(a,e) \land \ex.\lpo(e,r)$. 

Let $c$ be the event such that $(c,r) \in \ex.\lpo \cap \ex.\obx$ and there is no such $c'$ in between $c$ and $r$.

The scenarios are as follows:

\begin{itemize}

\item $c \in \Ulab \cup \F$.

In this case $\ex.\mo(c,b)$ holds as otherwise $\ex.\mo(b,c)$ creates a $\ex.\fr;\ex.\mo;[\Ulab \cup \F];\ex.\lpo$ cycle which results in a contradiction.

Thus $\ex.\obx$ path between the same events does not pass through $r$.

\item $c \in \Wlab$. 

Following the definition of $\xppo$, there is an intermediate fence event $d \in \F$ such that $\ex.\lpo(c,d) \land \ex.\lpo(d,r)$ holds. 
In this case $\ex.\mo(d,b)$ holds and also $\ex.\mo(c,d)$ holds. 
Hence $\ex.\mo(c,b)$ also holds.

Thus $\ex.\obx$ path between the same events does not pass through $r$.

\item $c \in \Rlab$. 

Let $w \in \WUlab$ be the event on the $\ex.\obx$ path and $\ex.\rfe(w,c)$ holds. 

In this case we show by contradiction that $\ex.\mo(w,b)$ holds. 

Assume $\ex.\mo(b,w)$ holds. 

In that case $\ex.\fr(r,b) \land \ex.\mo(b,w) \land \ex.\rfe(w,c) \land \ex.\lpo(c,r)$ creates a cycle which violates x86 consistency for $\ex$. 
Hence a contradiction and $\ex.\mo(w,b)$ holds.

Thus $\ex.\obx$ path between the same events does not pass through $r$.

\end{itemize}

\end{proof}


We restate the theorem.

\armx*

To prove \cref{th:armx}, we prove the following formal statement.
\[
\inarr{
\armp \leadsto \xp \implies \forall \ex_t \in \denot{\xp}.~\exists \ex_s \in \denot{\armp}.~\behav(\ex_t) = \behav(\ex_s)
}
\]

%
%
%

\begin{proof}

Given an x86 execution $\ex_t$ we define the correxponding ARM execution $\ex_s$. 
%
%
%
%
%
%
%

We know that $\ex_t$ is x86 consistent. 
Now we show that $\ex_s$ is ARM consistent. 
We prove by contradiction. 

\bigskip

(internal)

Assume $\ex_s$ contains $\ex_s.\poloc \cup  \ex_s.\ca \cup \ex_s.\lrf$ cycle. 

It implies a $\ex_t.\poloc \cup  \ex_t.\ca \cup \ex_t.\lrf$ cycle following the mappings.

In that case we can derive a $(\ex_t.\xhb \cup \ex_t.\fr \cup \ex_t.\co)^+$ cycle with no load event in $\ex_t$ following \cref{lem:noloadx}. 

Thus the cycle contains only same-location write events. 

In that case $\ex_t.\fr \implies \ex_t.\co$ and $\rloc{([\ex_t.\WUlab];\ex_t.\xhb;[\ex_t.\WUlab])} \implies \ex_t.\co$ which 
implies a $\ex_t.\mo$ cycle as $\ex_t.\co \subseteq \ex_t.\mo$ 
However, we know $\ex_t.\mo$ is has no cycle and hence a contradiction. 

Therefore the source execution $\ex_s$ in ARMv8 satisfies (internal).

\medskip

(external)

We prove this by contradiction. 
Assume $\ex_s$ contains a $\ob$ cycle. 
In that case $\ex_t$ contains a $\obx$ cycle. 
In that case, from \cref{lem:noloadobx}, we know that there exists a $\ex_t.\obx$ cycle 
which has no load event. 
Therefore the cycle contains only $\WUlab \cup \F$ events and thus there is a $\ex_t.\mo$ cycle. 
However, $\ex_t$ is x86 consistent and hence there is no $\ex_t.\mo$ cycle. 
Thus a contradiction and $\ex_s$ has no $\ob$ cycle.
Therefore the source execution $\ex_s$ in ARMv8 satisfies (external).

\medskip

(atomic)

We prove this by contradiction. 
Assume $[\ex_s.\rmw] \cap ;\ex.\fre;\ex_s.\coe \neq \emptyset$. 

In that case there exists $u \in \ex_t.\Ulab$, $w \in \ex_t.\WUlab$ in x86 consistent execution $\ex_t$ such that 
$\ex_t.\fre(u,w)$, $\ex_t.\coe(w,u)$ hold. 

It implies there is a $\ex_t.\fr;\ex_t.\mo$ cycle as $\fre \subseteq \fr$ and $\coe \subseteq \mo$ hold.

However, $\ex_t.\fr;\ex_t.\mo$ cycle is not possible as $\ex_t$ is consistent. 
Hence a contradiction and therefore $[\ex_s.\rmw] \cap ;\ex.\fre;\ex_s.\coe = \emptyset$.

\bigskip

Thus $\ex_s$ is ARMv8 consistent as it satisfies (internal), (external), and (atomic) constraints.

\end{proof}


\subsection{ARMv7-mca to ARMv8 Mappings}
\label{app:sec:armv7mcatoarmv8}

In \cref{app:armv7toarmv8} we have already 
shown all the relevant consistency constraints. 
It remains to show that (mca) holds for ARMv7-mca to ARMv8 mappings.

We restate \cref{lem:ppowob} and then prove the same.

\ppowob*


\begin{proof}

We start with

$[\ex_s.\Rlab];\ex_s.\ppo;[\ex_s.\Rlab];\ex_s.\poloc;[\ex_s.\Wlab]$

Considering the final incoming edge to $[\ex_s.\Rlab]$, we consider following cases:

\Case{$[\ex_s.\Rlab];\ex_s.\ppo^?;[\ex_s.\E];\ex_s.\addr;[\ex_s.\Rlab];\ex_s.\poloc;[\ex_s.\Wlab]$}

It implies $[\ex_s.\Rlab];\ex_s.\ppo^?;[\ex_s.\E];\ex_s.\addr;\ex_s.\po;[\ex_s.\Wlab]$

$\implies [\ex_t.\Rlab];\ex_t.\ob^?;[\ex_t.\E];\ex_t.\dob;[\ex_t.\Wlab]$ from \cref{lem:ppoob}.

$\implies [\ex_t.\Rlab];\ex_t.\ob;[\ex_t.\Wlab]$

\smallskip

\Case{$[\ex_s.\Rlab];\ex_s.\ppo^?;[\ex_s.\E];\ex_s.\rdw;[\ex_s.\Rlab];\ex_s.\poloc;[\ex_s.\Wlab]$}

It implies $[\ex_s.\Rlab];\ex_s.\ppo^?;[\ex_s.\E];\ex_s.\coe;\ex_s.\rfe;[\ex_s.\Rlab];\ex_s.\poloc;[\ex_s.\Wlab]$

$\implies [\ex_s.\Rlab];\ex_s.\ppo^?;[\ex_s.\E];\ex_s.\coe;\ex_s.\coe;[\ex_s.\Wlab]$

$\implies [\ex_t.\Rlab];\ex_t.\ob^?;[\ex_t.\E];\ex_t.\obs;\ex_t.\obs;[\ex_t.\Wlab]$ from \cref{lem:ppoob}.

$\implies [\ex_t.\Rlab];\ex_t.\ob;[\ex_t.\Wlab]$

\smallskip

\Case{$[\ex_s.\Rlab];\ex_s.\ppo;[\ex_s.\Wlab];\ex_s.\rfi;[\ex_s.\Rlab];\ex_s.\poloc;[\ex_s.\Wlab]$}

It implies $[\ex_s.\Rlab];\ex_s.\ppo^?;[\ex_s.\Rlab];(\ex_s.\ctrl \cup \ex_s.\data \cup \ex_s.\addr);[\ex_s.\Wlab];\ex_s.\coi;[\ex_s.\Wlab]$ as $\ex_s$ satisfies (sc-per-loc).

\begin{flalign*}
\implies & [\ex_s.\Rlab];\ex_s.\ppo^?;[\ex_s.\Rlab];(\ex_s.\ctrl \cup \ex_s.\data);[\ex_s.\Wlab];\ex_s.\coi; [\ex_s.\Wlab] & \\
& \cup [\ex_s.\Rlab];\ex_s.\ppo^?;[\ex_s.\Rlab];\ex_s.\addr;\ex_s.\po;[\ex_s.\Wlab] &
\end{flalign*}

$\implies [\ex_t.\Rlab];\ex_t.\ob^?;[\ex_t.\Rlab];\ex_t.\dob; [\ex_t.\Wlab] 
\cup [\ex_t.\Rlab];\ex_t.\ob^?;[\ex_t.\Rlab];\ex_t.\dob;[\ex_t.\Wlab]$ from \cref{lem:ppoob}.

$\implies [\ex_t.\Rlab];\ex_t.\ob;[\ex_t.\Wlab]$

\smallskip

\Case{$[\ex_s.\Rlab];\ex_s.\ppo^?;[\ex_s.\Rlab];\ex_s.\ctrl_\isb;[\ex_s.\Rlab];\ex_s.\poloc;[\ex_s.\Wlab]$}

It implies $[\ex_s.\Rlab];\ex_s.\ppo^?;[\ex_s.\Rlab];\ex_s.\ctrl;[\ex_s.\Wlab]$ as $\ctrl_\isb;\lpo \subseteq \ctrl_\isb$ and $\ctrl_\isb \subseteq \ctrl$.

$\implies  [\ex_t.\Rlab];\ex_t.\ob^?;[\ex_t.\Rlab];\ex_t.\dob; [\ex_t.\Wlab]$ from \cref{lem:ppoob}.

$\implies [\ex_t.\Rlab];\ex_t.\ob;[\ex_t.\Wlab]$

\smallskip

\Case{$[\ex_s.\Rlab];\ex_s.\ppo;[\ex_s.\Wlab];\ex_s.\detour;[\ex_s.\Rlab];\ex_s.\poloc;[\ex_s.\Wlab]$}

It implies $[\ex_s.\Rlab];\ex_s.\ppo;[\ex_s.\Wlab];\ex_s.\coe;[\ex_s.\Wlab];\ex_s.\rfe;[\ex_s.\Rlab];\ex_s.\poloc;[\ex_s.\Wlab]$ from the definition of $\detour$.

$\implies [\ex_s.\Rlab];\ex_s.\ppo;[\ex_s.\Wlab];\ex_s.\coe;[\ex_s.\Wlab];\ex_s.\coe;[\ex_s.\Wlab]$

$\implies [\ex_t.\Rlab];\ex_t.\ob;[\ex_t.\Wlab];\ex_t.\obs;[\ex_t.\Wlab];\ex_t.\obs;[\ex_t.\Wlab]$

$\implies [\ex_t.\Rlab];\ex_t.\ob;[\ex_t.\Wlab]$

\smallskip

\Case{$[\ex_s.\Rlab];\ex_s.\ppo^?;[\ex_s.\Rlab];\ex_s.\ctrl;[\ex_s.\Rlab];\ex_s.\poloc;[\ex_s.\Wlab]$}

It implies $[\ex_s.\Rlab];\ex_s.\ppo^?;[\ex_s.\Rlab];\ex_s.\ctrl;[\ex_s.\Wlab]$ as $\ctrl;\lpo \subseteq \ctrl$.

$\implies \implies [\ex_t.\Rlab];\ex_t.\ob^?;\ex_t.\dob;[\ex_t.\Wlab]$

$\implies [\ex_t.\Rlab];\ex_t.\ob;[\ex_t.\Wlab]$

\smallskip

\Case{$[\ex_s.\Rlab];\ex_s.\ppo^?;[\ex_s.\Rlab];\ex_s.\addr;\ex_s.\po^?;[\ex_s.\Rlab];\ex_s.\poloc;[\ex_s.\Wlab]$}

It implies $[\ex_s.\Rlab];\ex_s.\ppo^?;[\ex_s.\Rlab];\ex_s.\addr;\ex_s.\po^?;[\ex_s.\Wlab]$ 

$\implies \implies [\ex_t.\Rlab];\ex_t.\ob^?;\ex_t.\dob;[\ex_t.\Wlab]$

$\implies [\ex_t.\Rlab];\ex_t.\ob;[\ex_t.\Wlab]$

\end{proof}

Now we show that $\ex_s$ satisfies (mca).
We restate \cref{lem:acywo} and then prove the same.

\acywo*


\begin{proof}

Following the definition of $\ex_s.\wo$:

$\ex_s.\wo \triangleq ((\ex_s.\rfe; \ex_s.\ppo; \ex_s.\rfe^{-1}) \setminus [\ex_s.\E]); \ex_s.\co$

It implies 

$\ex_s.\rfe; \ex_s.\ppo;[\ex_s.\Rlab];\ex_s.\fri;[\ex_s.\Wlab] \cup \ex_s.\rfe; \ex_s.\ppo; [\ex_s.\Rlab];\ex_s.\fre; [\ex_s.\Wlab]$ 

$\implies \ex_s.\rfe; [\ex_s.\Rlab];\ex_s.\ppo;[\ex_s.\Rlab]; \ex_s.\poloc; [\ex_s.\Wlab] \cup \ex_s.\rfe;\ex_s.\ppo; \ex_s.\fre$ from definitions.

$\implies \ex_t.\rfe; [\ex_t.\Rlab];\ex_t.\ob;[\ex_s.\Wlab] \cup  \ex_t.\rfe; \ex_t.\ob; \ex_t.\fre$ from \cref{lem:ppowob}.
 
$\implies \ex_t.\obs; [\ex_t.\Rlab];\ex_t.\ob;[\ex_s.\Wlab] \cup  \ex_t.\obs; \ex_t.\ob; \ex_t.\obs$ from \cref{lem:frerfeobs}.

$\implies \ex_t.\ob$.

Thus $\ex_s.\wo^+ \implies \ex_t.\ob \cup \ex_t.\ob \implies \ex_t.\ob$. 

We know $\ex_t.\ob$ is acyclic.

Therefore $\ex_s.\wo^+$ is acyclic.

\end{proof}

%
%
%
%
%
%
%
%
%
%
%
%
%
%
%
%
%
%

We restate \cref{th:armv7mcav8} and then prove the same.

\armsmcaarma*

We formally show
\[
\inarr{
\oarmpmca \leadsto \armp \implies \forall \ex_t \in \denot{\armp}.~\exists \ex_s \in \denot{\oarmpmca}.~\behav(\ex_t) = \behav(\ex_s)
}
\]

\begin{proof}
Follows from \cref{th:armv7v8} and \cref{lem:acywo}.
Moreoveover, $\ex_s.\co \iff \ex_t.\co$ holds. 
Therefore $\behav(\ex_t) = \behav(\ex_s)$ also holds.
\end{proof}

\subsection{ARMv7 to ARMv8 Mappings}
\label{app:armv7toarmv8}


We restate \cref{th:armv7v8}.

\armvsva*


We prove the following formal statement.
\[
\inarr{
\oarmp \leadsto \armp \implies \forall \ex_t \in \denot{\armp}.~\exists \ex_s \in \denot{\oarmp}.~\behav(\ex_t) = \behav(\ex_s)
}
\]

Given an ARMv8 execution $\ex_t$ we define the correxponding ARMv7 execution $\ex_s$ 
such that $\ex_t.\lpo \iff \ex_s.\lpo$, $\ex_t.\lrf \iff \ex_s.\lrf$, and $\ex_t.\co \iff \ex_s.\co$ hold.

%
%
%
%
%
%
%
%

We know that $\ex_t$ is ARMv8 consistent. 
We will show that $\ex_s$ is ARMv7 consistent. 

First we relate the $\ex_s$ and $\ex_t$ relations. 

\begin{lemma}
\label{lem:frerfeobs}
Suppose $\ex_s$ is an ARMv7 consistent execution and $\ex_t$ is corresponding 
ARMv8 execution. In that case $\ex_s.\fre \implies \ex_t.\obs$ and $\ex_s.\rfe \implies \ex_t.\obs$.
\end{lemma}

\begin{proof}
Follows from definition.
\end{proof}

\begin{lemma}
\label{lem:fencebob}
Suppose $\ex_s$ is an ARMv7 consistent execution and $\ex_t$ is corresponding 
ARMv8 execution. In that case $\ex_s.\fence \implies \ex_t.\bob$.
\end{lemma}
\begin{proof}
$\ex_s.\fence \implies \ex_t.\lpo;[\ex_s.\dmbfull];\ex_t.\lpo \implies \ex_t.\bob$.
\end{proof}


\begin{restatable}{lemma}{norfi}
\label{lem:norfi}
$(\ii_0 \cup \ci_0 \cup \cc_0);[\Wlab];\rfi \implies \ob$
\end{restatable}

\begin{proof}
We know $\rfi \subseteq \ii_0$ and $\ppo$ does not have $\cc_0;\ii_0$ subsequence following the constraint. 
Therefore we show $(\ii_0 \cup \ci_0);[\Wlab];\rfi \implies\ob$. 
It implies $(\addr \cup \data \cup \ctrl_\isb);[\Wlab];\rfi$

$\implies (\addr \cup \data);\rfi \cup \ctrl;[\Wlab] \implies \dob \cup \dob \implies \ob$
\hfill $\qed$
\end{proof}

Let $\dobcc_0 = \data \cup \ctrl;[\Wlab] \cup \addr \cup \addr;\lpo;[\Wlab]$ and 
$\ndobcc_0 = \ctrl;[\Rlab] \cup \addr;\lpo;[\Rlab]$. 
Therefore $\cc_0 = \dobcc_0 \cup \ndobcc_0$.

\begin{restatable}{lemma}{lemdobcc}
\label{lem:dobcc}
$\cc_0^+ = \dobcc_0 \cup \ndobcc_0$
\end{restatable}

\begin{proof}
From definition $\cc_0^+ = (\dobcc_0 \cup \ndobcc_0)^+$

Consider the following cases:
\begin{itemize}[leftmargin=*]

\item $\dobcc_0;\dobcc_0$ 

$\implies \addr;\addr \implies \addr;\lpo;[\Wlab] \cup \addr;\lpo;[\Rlab] \implies \dobcc_0 \cup \ndobcc_0$ 

\smallskip

\item $\dobcc_0;\ndobcc_0 \implies \addr;(\ctrl;[\Rlab] \cup \addr;\lpo[\Rlab]) \implies\addr;\lpo;[\Rlab] \implies \ndobcc_0$ 

\smallskip

\item $\ndobcc_0;\dobcc_0 \implies (\ctrl;[\Rlab] \cup \addr;\lpo;[\Rlab]);\dobcc_0;[\Rlab \cup \Wlab]$ 

$\implies \ctrl;[\Wlab] \cup \addr;\lpo;[\Wlab] \cup \ctrl;[\Rlab] \cup \addr;\lpo;[\Rlab] \implies \dobcc_0 \cup \ndobcc_0$

\smallskip

\item $\ndobcc_0;\ndobcc_0 \implies (\ctrl;[\Rlab] \cup \addr;\lpo;[\Rlab]);\ndobcc_0;[\Rlab] \implies \ndobcc_0$ 
\end{itemize}
Therefore $\cc_0^+ = \dobcc_0 \cup \ndobcc_0$.
\end{proof}

Now we restate \cref{lem:ppoob} and then prove the same.

\ppoob*

\begin{proof}
From the definition of $\ppo$ and \cref{lem:dobcc}:

$[\Rlab];\tgtx.\ppo \implies [\Rlab];(\tgtx.\ii_0 \cup \tgtx.\ci_0 \cup \tgtx.\dobcc_0;\ci_0^? \cup \tgtx.\ndobcc_0;\tgtx.\ci_0)^+$

$\implies [\Rlab];(\tgtx.\ii_0 \cup \tgtx.\ci_0 \cup \tgtx.\dobcc_0;\tgtx.\ci_0 \cup \tgtx.\ndobcc_0;\tgtx.\ci_0)^+$

$\implies [\Rlab];(\srcx.\addr \cup \srcx.\data \cup \srcx.\rdw \cup \srcx.\ob \cup \srcx.\ctrl_\isb \cup \srcx.\detour \cup \srcx.\dobcc_0;\ci_0^? \cup \srcx.\ndobcc_0;\ci_0)^+$ by reducing the $\rfi$ edges following \cref{lem:norfi}.

$\implies [\Rlab];(\srcx.\ob \cup \srcx.\ndobcc_0;\ci_0)^+$ as 

\begin{itemize}

\item $\srcx.\addr \cup \srcx.\data \cup \srcx.\ctrl_\isb \subseteq \srcx.\dob \subseteq \srcx.\ob$

\item $\srcx.\rdw = \srcx.\fre;\srcx.\rfe \subseteq \srcx.\obs;\srcx.\obs \subseteq \srcx.\ob$

\item $\srcx.\detour = \srcx.\coe;\srcx.\rfe \subseteq \srcx.\obs;\srcx.\obs \subseteq \srcx.\ob$

\item $\srcx.\dobcc_0 = (\srcx.\data \cup \srcx.\ctrl;[\Wlab] \cup \srcx.\addr \cup \srcx.\addr;\srcx.\lpo;[\Wlab]) \subseteq \srcx.\ob$

\end{itemize}

Now, $\srcx.\ndobcc_0;\srcx.\ci_0 = (\srcx.\ctrl;[\Rlab] \cup \srcx.\addr;\srcx.\lpo;[\Rlab]);(\srcx.\ctrl_\isb \cup \srcx.\detour)$ from definition.

$ \implies (\srcx.\ctrl;[\Rlab];\srcx.\ctrl_\isb \cup \srcx.\addr;\srcx.\lpo;[\Rlab];\srcx.\ctrl_\isb)$ as $\dom(\srcx.\detour) \not\subseteq \Rlab$.

$\implies (\srcx.\dob \cup \srcx.\dob)$ as $\dom(\srcx.\detour) \not\subseteq \Rlab \implies \srcx.\ob$

$\implies [\Rlab];(\srcx.\ob \cup \srcx.\ndobcc_0;\srcx.\ci_0)^+ \implies \srcx.\ob$.

Therefore $\tgtx.\ppo \implies \srcx.\ob$.

\end{proof}

\begin{lemma}
\label{lem:hbpropob}
Suppose $\ex_s$ is an ARMv7 consistent execution and $\ex_t$ is corresponding 
ARMv8 execution. In that case (i) $\ex_s.\hbppc \implies \ex_t.\ob$ and (ii) $\ex_s.\prop \implies \ex_t.\ob$
\end{lemma}

\begin{proof}

(i) $\ex_s.\hbppc \implies \ex_s.\ppo \cup \ex_s.\fence \cup \ex_s.\rfe \implies \ex_t.\ob \cup \ex_t.\bob \cup \ex_t.\obs$ 
from \cref{lem:frerfeobs}, \cref{lem:fencebob}, and \cref{lem:ppoob}.

(ii) We know $\ex_s.\prop = \ex_s.\prop_1 \cup \ex_s.\prop_2$ from definition.

Now,

$\ex_s.\prop_1$ 

$\implies [\ex_t.\WUlab];\ex_t.\rfe^?;\ex_t.\fence;\ex_t.\hbppc^*;[\ex_t.\WUlab]$ 

$\implies \ex_t.\obs; \ex_t.\bob ; ( \ex_t.\dob \cup \ex_t.\bob \cup \ex_t.\obs);[\ex_t.\WUlab]$

$\implies \ex_t.\ob$

\medskip

Also $\ex_s.\prop_2$

$\implies ((\ex_t.\co \cup \ex_t.\fr) \setminus \ex_t.\lpo)^?;\ex_t.\rfe^?;(\ex_t.\fence;\ex_t.\hbppc^*)^?;\ex_t.\fence;\ex_t.\hbppc^*$.

$\implies (\ex_t.\coi \cup \ex_t.\coe \cup \ex_t.\fri \cup \ex_t.\fre) \setminus \ex_t.\lpo)^?;\ex_t.\rfe^?;(\ex_t.\fence;\ex_t.\hbppc^*)^?;\ex_t.\fence;\ex_t.\hbppc^*$ 

$\implies (\ex_t.\coe \cup \ex_t.\fre) \setminus \ex_t.\lpo)^?;\ex_t.\rfe^?;(\ex_t.\fence;\ex_t.\hbppc^*)^?;\ex_t.\fence;\ex_t.\hbppc^*$ 

$\implies \ex_t.\obs;(\ex_t.\fence;\ex_t.\hbppc^*)^?;\ex_t.\fence;\ex_t.\hbppc^*$ 

$\implies \ex_t.\obs;(\ex_t.\bob;(\ex_t.\dob \cup \ex_t.\bob \cup \ex_t.\obs)^*)^?;\ex_t.\bob;(\ex_t.\dob \cup \ex_t.\bob \cup \ex_t.\obs)^*$ 

$\implies \ex_t.\ob$ 

\medskip

Hence $\ex_s.\prop \implies \ex_t.\ob$.

\end{proof}

Now we prove \cref{th:armv7v8}.

\begin{proof}

We show $\ex_s$ is ARMv7 consistent by contradiction. 

\medskip

(total-mo), (sc-per-loc), (atomicity) hold on $\ex_s$ as they hold on $\ex_t$. 
It remains to show that (observation) and (propagation) hold on $\ex_s$.

\smallskip

(observation)

Assume there is a $\ex_s.\fre;\ex_s.\prop;\ex_s.\hbppc^*$ cycle.

Considering the relations above, 

$\ex_s.\fre;\ex_s.\prop;\ex_s.\hbppc^* \implies \ex_t.\obs;\ex_t.\ob;( \ex_t.\dob \cup \ex_t.\bob \cup \ex_t.\obs)^* \implies \ex_t.\ob$. 

However, we know that $\ex_t.\ob$ is irreflexive and hence a contradiction.

Therefore, $\ex_s.\fre;\ex_s.\prop;\ex_s.\hbppc^*$ is irreflexive and $\ex_s$ satisfies (observation).

\medskip

(propagation)

Assume there is a $\ex_s.\co \cup \ex_s.\prop$ cycle. 

It implies a $\ex_t.\co \cup \ex_t.\ob$ cycle. 

We know $\ex_t.\co;\ex_t\co \implies \ex_t.\co$ and $\ex_t.\prop;\ex_t.\prop \implies \ex_t.\prop$. 

Thus a $\ex_t.\co \cup \ex_t.\ob$ cycle can be reduced to a $\ex_t.\co \cup \ex_t.\ob$ cycle where $\ex_t.\co$ and $\ex_t.\prop$ 
take place alternatively. 

In this case each of $\ex_t.\prop \subseteq \rloc {(\ex_t.\WUlab \times \ex_t.\WUlab)} \subseteq \ex_t.\co$. 
 
It implies there is a $\ex_t.\co$ cycle which is a contradiction.

Hence $\ex_s.\co \cup \ex_s.\prop$ is acyclic and $\ex_s$ satisfies (propagation). 

Therefore $\ex_s$ is ARMv7 consistent.

Moreover, $\behav(\ex_t) = \behav(\ex_s)$ holds as $\ex_t.\co \iff \ex_s.\co$.

\end{proof}

\subsection{ARMv8 to ARMv7 Mappings}
\label{app:armv8toarmv7}

We restate \cref{lem:abdobsppo} and then prove the same.

\abdobsppo*

%

\begin{proof}

(1) $\ex_s.\obs \implies \ex_s.\rfe \cup \ex_s.\coe \cup \ex_s.\fre$ 

$\implies \ex_t.\rfe \cup \ex_t.\coe \cup \ex_t.\fre$ from definition.

\bigskip

(2)  We know

$\ex_s.\dob \subseteq [\ex_s.\Rlab];\ex_s.\lpo;[\ex_s.\E]$.

$\implies  [\ex_t.\Rlab]; \ex_t.\lpo; [\ex_t.\dmb];\ex_t.\lpo;[\ex_s.\E]$ following the mappings in \cref{tab:armv8v7}.

$\implies [\ex_t.\Rlab];\ex_t.\fence;[\ex_t.\E]$ from the definition.

\bigskip

(3) 

We know $\aob \triangleq \rmw \cup [\frange(\rmw)];\rfi; [\Alab]$

Hence $\ex_s.\rmw \cup [\frange(\ex_s.\rmw)];\ex_s.\rfi;[\ex_s.\Alab \cup \ex_s.\Qlab]$

$\implies \ex_t.\rmw \cup [\frange(\ex_t.\rmw)];\ex_t.\rfi;[\ex_t.\Rlab];\ex_t.\lpo;[\ex_t.\dmb]$

\bigskip

(4) 

Following the definition of $\ex_s.\bob$, we consider its components:

\begin{itemize}

\item $\ex_s.\lpo;[\ex_s.\dmbfull];\ex_s.\lpo$ 

$\implies \ex_t.\lpo;[\ex_t.\dmb];\ex_t.\lpo$ 

$\implies \ex_t.\fence$

\item $[\ex_s.\stlr];\ex_s.\lpo;[\ex_s.\ldar]$ 

$\implies [\ex_t;\dmb];\ex_t.\lpo;[\ex_t.\Wlab];\ex_t.\lpo;[\ex_t;\dmb];\ex_t.\lpo;[\ex_t.\Rlab]$ 

$\implies \ex_t.\fence$

\item $[\ex_s;\Rlab];\ex_s.\lpo;[\ex_s.\dmb];\ex_s.\lpo$ 

$\implies \ex_t.\fence$

\item $[\ex_s.\ldar];\ex_s.\lpo$ 

$\implies [\ex_t;\Rlab];\ex_t.\lpo;[\ex_t.\dmb];\ex_t.\lpo$ 

$\implies \ex_t.\fence$ 

\item $[\ex_s.\Wlab];\ex_s.\lpo;[\ex_s.\dmbst];\ex_s.\lpo;[\ex_s.\Wlab]$ 

$\implies [\ex_t.\Wlab];\ex_t.\lpo;[\ex_t.\dmb];\ex_t.\lpo;[\ex_t.\Wlab]$ 

$\implies \ex_t.\fence$

\item $\ex_s.\lpo;[\ex_s.\stlr]$ 

$\implies \ex_t.\lpo;[\ex_t.\dmb];\ex_t.\lpo;[\ex_t.\Wlab]$ 

$\implies \ex_t.\fence$

\item $\ex_s.\lpo;[\ex_s.\stlr];\ex_s.\coi$ 

$\implies \ex_t.\lpo;[\ex_t.\dmb];\ex_t.\lpo;[\ex_t.\Wlab];\ex_t.\lpo$ 

$\implies \ex_t.\fence$
\end{itemize}

Thus $\ex_s.\bob \implies \ex_t.\fence$ .

Therefore $\ex_s.\ob \implies$

$(\ex_t.\rfe \cup \ex_t.\coe \cup \ex_t.\fre \cup \ex_t.\fence \cup \ex_t.\rmw \cup [\frange(\ex_t.\rmw)];\ex_t.\rfi;[\ex_t.\Rlab];\ex_t.\lpo;[\ex_t.\dmb])^+$

Considering the outgoing edges from $\Rlab$ event in $[\frange(\ex_t.\rmw)];\ex_t.\rfi;[\ex_t.\Rlab];\ex_t.\lpo;[\ex_t.\dmb]$ 
we consider two cases:

\textbf{case} $[\frange(\ex_t.\rmw)];\ex_t.\rfi;[\ex_t.\Rlab];\ex_t.\lpo;[\ex_t.\dmb];\ex_t.\lpo$

$\implies \ex_t.\fence$

\textbf{case} $[\frange(\ex_t.\rmw)];\ex_t.\rfi;[\ex_t.\Rlab];\ex_t.\fre$

$\implies [\frange(\ex_t.\rmw)];\ex_t.\coe$ by definition of $\fre$.

\bigskip

Therefore 
$\ex_s.\ob \implies (\ex_t.\rfe \cup \ex_t.\coe \cup \ex_t.\fre \cup \ex_t.\fence \cup \ex_t.\rmw)^+$.

\end{proof}

We restate the \cref{lem:obprop} and then prove the same.

\obprop*
%

\begin{proof}

We know $\ex_s.\ob \implies (\ex_t.\rfe \cup \ex_t.\coe \cup \ex_t.\fre \cup \ex_t.\rmw \cup \ex_t.\fence)^+$ from \cref{lem:abdobsppo}.

\textbf{Scenario (1):} $(\ex_t.\rfe \cup \ex_t.\coe \cup \ex_t.\fre \cup \ex_t.\rmw \cup \ex_t.\fence)^+$ has no $\ex_t.\fence$.

In this case $(\ex_t.\rfe \cup \ex_t.\coe \cup \ex_t.\fre \cup \ex_t.\rmw)^+ \implies \rloc{(\ex_t.\E \times \ex_t.\E)} \setminus [\Elab]$ from the definitions.

\bigskip

\textbf{Scenario (2): Otherwise}

\medskip


In this case 
$\ex_s.\ob \implies ((\ex_t.\rfe \cup \ex_t.\coe \cup \ex_t.\fre \cup \ex_t.\rmw)^*;\ex_t.\fence)^+$

Now we consider following cases:

\begin{enumerate}
\item[(RR)] $[\ex_t.\Rlab];(\ex_t.\rfe \cup \ex_t.\coe \cup \ex_t.\fre \cup \ex_t.\rmw)^+;[\ex_t.\Rlab];\ex_t.\fence$

\item[(RW)] $[\ex_t.\Rlab];(\ex_t.\rfe \cup \ex_t.\coe \cup \ex_t.\fre \cup \ex_t.\rmw)^+;[\ex_t.\Wlab];\ex_t.\fence$

\item[(WR)] $[\ex_t.\Wlab];(\ex_t.\rfe \cup \ex_t.\coe \cup \ex_t.\fre \cup \ex_t.\rmw)^+;[\ex_t.\Rlab];\ex_t.\fence$

\item[(WW)] $[\ex_t.\Wlab];(\ex_t.\rfe \cup \ex_t.\coe \cup \ex_t.\fre \cup \ex_t.\rmw)^+;[\ex_t.\Wlab];\ex_t.\fence$
\end{enumerate}

\Case{(RR)}

$[\ex_t.\Rlab];(\ex_t.\rfe \cup \ex_t.\coe \cup \ex_t.\fre \cup \ex_t.\rmw)^+;[\ex_t.\Rlab];\ex_t.\fence$

$\implies [\ex_t.\Rlab];(\ex_t.\rfe \cup \ex_t.\coe \cup \ex_t.\fre \cup \ex_t.\rmw)^+;[\ex_t.\Wlab];\ex_t.\rfe;[\ex_t.\Rlab];\ex_t.\fence$

$\implies [\ex_t.\Rlab];\ex_t.\fr;[\ex_t.\Wlab];\ex_t.\rfe;[\ex_t.\Rlab];\ex_t.\fence$ as $\ex_t$ satisfies (sc-per-loc).


$\implies [\ex_t.\Rlab];\ex_t.\fri;[\ex_t.\Wlab];\ex_t.\rfe;[\ex_t.\Rlab];\ex_t.\fence \cup [\ex_t.\Rlab];\ex_t.\fre;[\ex_t.\Wlab];\ex_t.\rfe;[\ex_t.\Rlab];\ex_t.\fence$


$\implies [\ex_t.\Rlab];(\ex_t.\rmw \cup \ex_t.\fence);[\ex_t.\Wlab];\ex_t.\rfe;[\ex_t.\Rlab];\ex_t.\fence \cup \ex_t.\prop_2$

following the mapping of \cref{tab:armv8v7} and definition of $\prop_2$.

\begin{flalign*}
\implies & [\ex_t.\Rlab];\ex_t.\rmw;[\ex_t.\Wlab];\ex_t.\rfe;[\ex_t.\Rlab];\ex_t.\fence & \\
& \cup [\ex_t.\Rlab];\ex_t.\fence;[\ex_t.\Wlab];\ex_t.\rfe;[\ex_t.\Rlab];\ex_t.\fence \cup \ex_t.\prop_2 &
\end{flalign*}

\begin{flalign*}
\implies & [\ex_t.\Rlab];\ex_t.\ppo;[\ex_t.\Wlab];\ex_t.\rfe;[\ex_t.\Rlab];\ex_t.\fence & \\
& \cup [\ex_t.\Rlab];\ex_t.\fence;[\ex_t.\Wlab];\ex_t.\rfe;[\ex_t.\Rlab];\ex_t.\fence \cup \ex_t.\prop_2 & 
\end{flalign*}
as $\ex_t.\rmw \implies \ex_t.\ppo$.

\begin{flalign*}
\implies & [\ex_t.\Rlab];\ex_t.\hbppc;\ex_t.\fence & \\ 
& \cup [\ex_t.\Rlab];\ex_t.\fence;[\ex_t.\Wlab];\ex_t.\hbppc;[\ex_t.\Rlab];\ex_t.\fence \cup \ex_t.\prop_2 &
\end{flalign*}
from definition of $\prop_2$.

$\implies \ex_t.\prop_2 \cup \prop_2 \cup \prop_2$
 
$\implies \ex_t.\prop$

\medskip

\Case{(RW)}

$[\ex_t.\Rlab];(\ex_t.\rfe \cup \ex_t.\coe \cup \ex_t.\fre \cup \ex_t.\rmw)^+;[\ex_t.\Wlab];\ex_t.\fence$

$\implies [\ex_t.\Rlab];\ex_t.\fr;\ex_t.\fence$

$\implies [\ex_t.\Rlab];\ex_t.\fri;\ex_t.\fence \cup [\ex_t.\Rlab];\ex_t.\fre;\ex_t.\fence$

$\implies [\ex_t.\Rlab];\ex_t.\fence \cup [\ex_t.\Rlab];\ex_t.\fre;\ex_t.\fence$

$\implies \ex_t.\prop_2 \cup \ex_t.\prop_2$ from definition of $\prop_2$.

$\implies \ex_t.\prop$

\medskip

\Case{(WR)}

$[\ex_t.\Wlab];(\ex_t.\rfe \cup \ex_t.\coe \cup \ex_t.\fre \cup \ex_t.\rmw)^+;[\ex_t.\Rlab];\ex_t.\fence$

$\implies [\ex_t.\Wlab];(\ex_t.\rfe \cup \ex_t.\coe \cup \ex_t.\fre \cup \ex_t.\rmw)^*;[\ex_t.\Wlab];\ex_t.\rfe;[\ex_t.\Rlab];\ex_t.\fence$

$\implies [\ex_t.\Wlab];\ex_t.\co;[\ex_t.\Wlab];\ex_t.\rfe;[\ex_t.\Rlab];\ex_t.\fence$ as $\ex_t$ satisfies (sc-per-loc).

$\implies \ex_t.\co;\ex_t.\prop_1$ from definition.

$\implies \ex_t.\co;\ex_t.\prop$ as $\prop_1 \subseteq \prop$

$\implies [\ex_t.\Wlab];\ex_t.\coi;[\ex_t.\Wlab];\ex_t.\rfe;[\ex_t.\Rlab];\ex_t.\fence \cup [\ex_t.\Wlab];\ex_t.\coe;[\ex_t.\Wlab];\ex_t.\rfe;[\ex_t.\Rlab];\ex_t.\fence$
 

$\implies [\ex_t.\Wlab];\ex_t.\coi;[\ex_t.\Wlab];\ex_t.\rfe;[\ex_t.\Rlab];\ex_t.\fence \cup \ex_t.\prop_2$ from definitions

\Case{(WW)}

$[\ex_t.\Wlab];(\ex_t.\rfe \cup \ex_t.\coe \cup \ex_t.\fre \cup \ex_t.\rmw);[\ex_t.\Wlab];\ex_t.\fence$

$\implies [\ex_t.\Wlab];\ex_t.\co;[\ex_t.\Wlab];\ex_t.\fence$

$\implies [\ex_t.\Wlab];\ex_t.\coi;[\ex_t.\Wlab];\ex_t.\fence \cup [\ex_t.\Wlab];\ex_t.\coe;[\ex_t.\Wlab];\ex_t.\fence$

$\implies \ex_t.\fence \cup [\ex_t.\Wlab];\ex_t.\coe;[\ex_t.\Wlab];\ex_t.\fence$

$\implies \ex_t.\prop_2 \cup \ex_t.\prop_2$ from definition of $\prop_2$.

$\implies \ex_t.\prop$

Thus (in {\bf Scenario-II}) $\ex_s.\ob \implies (\ex_t.\co;\ex_t.\prop \cup \ex_t.\prop)^+$. 

\end{proof}

Finally we restate \cref{th:armv8v7} and then prove the same.

\armvavs*


To prove \cref{th:armv8v7}, we prove the following formal statement.
\[
\inarr{
\armp \leadsto \oarmp \implies \forall \ex_t \in \denot{\oarmp}.~\exists \ex_s \in \denot{\armp}.~\behav(\ex_t) = \behav(\ex_s)
}
\]

\begin{proof}
We know that $\ex_t$ is ARMv7 consistent. 
Now we show that $\ex_s$ is ARMv8 consistent. 
We prove by contradiction. 

\bigskip

\Case{(internal) }
We know that (sc-per-loc) holds in $\ex_t$. Hence (internal) trivially holds in $\ex_s$.

\Case{(external)}
Assume there is a $\ex_s.\ob$ cycle.

From \cref{lem:obprop} we know that $\ex_s.\ob \implies (\rloc{(\ex_t.\E \times \ex_t.\E)} \setminus [\Elab]) \cup (\ex_t.\co;\ex_t.\prop \cup \ex_t.\prop)^+$.

We know both $ (\rloc{(\ex_t.\E \times \ex_t.\E)} \setminus [\Elab])$ is acyclic as $\ex_t$ satisfies (sc-per-loc) and 
$(\ex_t.\co;\ex_t.\prop \cup \ex_t.\prop)^+$ is acyclic as $\ex_t$ satisfies (propagation).

\Case{(atomic)}

We know that (atomic) holds in $\ex_t$. Hence (atomic) trivially holds in $\ex_s$.

\bigskip

Therefore $\ex_s$ is consistent. 
Moreover, as $\ex_s.\co \iff \ex_t.\co$ holds, $\behav(\ex_s) = \behav(\ex_t)$ also holds.

\end{proof}

\subsection{Proff of correctness: C11 to ARMv8 to ARMv7}
\label{app:carmv8v7} 

We restate the theorem and then prove the correctness.

\carmas*

\begin{proof}
The mapping can be represented as a combination of following transformation steps.
\begin{enumerate}
\item $\mbbp_{\cmm} \mapsto \mbbp_{\arma}$ mapping from \citet{mappings}.
\item $\mbbp_{\arma} \mapsto \mbbp_{\arms}$ mappings from \cref{tab:armv8v7}. 
\item Elimination of leading $\idmb$ fences for $\ldr_\MOna \mapsto \ldr$ mapping, that is, 
$\ldr_\MOna \mapsto \ldr;\idmb \leadsto \ldr$.
\end{enumerate}

We know (1), (2) are sound and therefore it suffices to show that transformation (4) is sound.

Let $\ex_a$ and $\ex'_a$ be the consistent execution of $\mbbp_\arma$ 
before and after the transformation (4).
Let $\ex$ be correspnding C11 execution $\mbbp_{\cmm}$.
 and 
we know $\mbbp_{\cmm}$ is race-free. 
Therefore for all non-atomic event $a$ in $\ex$ if there exist 
another same-location event $b$ then $\ex.\lhb^=(a,b)$ holds.

Now we consider ARMv8 to ARMv7 mapping scheme.

Considering the $\lhb$ definition following are the possibilities:
 
\Case{$[\E_\MOna];\ex.\lpo;[\Flab_{\sqsupseteq \MOrel}];\ex.\lpo;[\WUlab_{\MOrlx}] \cup [\E_\MOna];\ex.\lpo;[\WUlab_{\sqsupseteq \MOrel}]$}

$\implies [\E];\ex_a.\lpo;[\dmbfull];\ex_a.\lpo;[\Wlab \cup \rmw]$

$\implies [\E];\ex'_a.\lpo;[\dmb];\ex'_a.\lpo;[\E]$

$\implies [\E];\ex'_a.\fence;[\E]$

\Case{$[\RUlab_{\sqsupseteq \MOrlx}];\ex.\lpo;[\E_\MOna] \cup [\RUlab_{\MOrlx}];\ex_\lpo;[\Flab_{\sqsupseteq \MOacq}];\ex.\lpo;[\E_\MOna]$} 

$\implies [\Rlab];\ex_a.\lpo;[\dmbfull];\ex_a.\lpo$

$\implies [\E];\ex'_a.\fence;[\E]$

Therefore $\ex'_a.\fence = \ex_a.\fence$ and the transfmation is sound for ARMv8 to ARMv7 mapping.
\end{proof}

\clearpage


\section{Proofs and counter-examples for Optimizations in $\arma$}
\label{app:armaopt}

\subsection{Proofs of Safe reorderings}
\label{app:sreorder}

We prove the following theorem for safe reorderings in \cref{fig:armatrans}.
\[
\inarr{
\srcp \leadsto \tgtp \implies \forall \ex_t \in \denot{\tgtp}~\exists \ex_s \in \denot{\srcp}.~\behav(\ex_t) = \behav(\ex_s) 
}
\]

\begin{proof}

We know $\ex_t$ is ARMv8 consistent. 
We define $\ex_s$ where $a\cdot b \leadsto b \cdot a$.

$\ex_s.\E = \ex_t.\E$

$\ex_s.\lpo = (\ex_t.\lpo \setminus \set{(b,a)} \cup \set{(a,b)})^+$

$\ex_s.\lrf = \ex_t.\lrf$

$\ex_s.\co = \ex_t.\co$

We show that $\ex_s$ is ARMv8 consistent.

(internal)

\smallskip

We know that $\ex_t.\poloc = \ex_s.\poloc$, $\ex_s.\lrf = \ex_t.\lrf$, $\ex_s.\fr = \ex_t.\fr$, $\ex_s.\co = \ex_t.\co$ hold.
We also know that $\ex_t$ satisfies (internal). 
Therefore $\ex_s$ also satisfies (internal).

\bigskip

(external)

\smallskip

We relate the $\ob$ relations between memory accesses in $\ex_t$ and $\ex_s$. 
Let $M = \Wlab \cup \Rlab \cup \Llab \cup \Alab$.

\begin{itemize}

\item $\Wlab(x)/\Llab(x)\cdot\Rlab(y) \leadsto \Rlab(y)\cdot\Wlab(x)/\Llab(x)$. 
In this case $\ex_s.\aob = \ex_t.\aob$, $\ex_s.\bob \subseteq \ex_t.\bob$, and 
$\ex_s.\dob = \ex_t.\dob$ hold.

%
\item $\Rlab(x)\cdot\Rlab(y) \leadsto \Rlab(y)\cdot\Rlab(x)$
In this case $\ex_s.\aob = \ex_t.\aob$, $\ex_s.\bob = \ex_t.\bob$, and 
$\ex_s.\dob = \ex_t.\dob$ hold.
%
%
%

\item $\dmbst \cdot \Rlab(y) \leadsto \Rlab(y) \cdot\dmbst$. 
In this case $\ex_s.\aob = \ex_t.\aob$, $\ex_s.\bob = \ex_t.\bob$, and 
$\ex_s.\dob = \ex_t.\dob$ hold.

\item $\Wlab(x)/\Rlab(x)/\dmbst\cdot\Alab(y) \leadsto \Alab(y)\cdot\Wlab(x)/\Rlab(x)/\dmbst$
In this case $\ex_s.\aob = \ex_t.\aob$, $\ex_s.\bob \subseteq \ex_t.\bob$, and 
$\ex_s.\dob = \ex_t.\dob$ hold.

\item $\dmbld/\dmbst/\dmbfull \cdot\Llab(y) \leadsto \dmbld/\dmbst/\dmbfull \cdot \Llab(y)$. 
In this case $\ex_s.\aob = \ex_t.\aob$, and 
$\ex_s.\dob = \ex_t.\dob$ hold. 
We also know that $[M];\ex_s.\bob;[\ex_s.\Llab] = [M];\ex_t.\bob;[\ex_t.\Llab]$ 
and $[\Llab];\ex_s.\bob;[M] \subseteq [\Llab];\ex_t.\bob;[M]$ hold.

\item $\Alab(x)\dmbld/\dmbst \cdot \dmbfull \leadsto \dmbfull \cdot \Alab(x)\dmbld/\dmbst$. 
In this case $\ex_s.\aob = \ex_t.\aob$, $\ex_s.\bob;[M] \subseteq \ex_t.\bob;[M]$, 
$[M];\ex_s.\bob = [M];\ex_t.\bob$, and $\ex_s.\dob = \ex_t.\dob$ hold.

\item $\Wlab/\Llab/\Alab/\dmbfull \cdot \dmbld \leadsto \dmbld \cdot \Wlab/\Llab/\Alab/\dmbfull$. 
In this case $\ex_s.\aob = \ex_t.\aob$, $[M];\ex_s.\bob;[M] \subseteq [M];\ex_t.\bob;[M]$, and $\ex_s.\dob = \ex_t.\dob$ hold.

\item $\dmbld/\Alab/\dmbfull \cdot \dmbst \leadsto \dmbst \cdot \dmbld/\Alab/\dmbfull$. 
In this case $\ex_s.\aob = \ex_t.\aob$, $[M];\ex_s.\bob;[M] = [M];\ex_t.\bob;[M]$, and 
$\ex_s.\dob = \ex_t.\dob$ hold. 

\end{itemize}

Hence $[M];\ex_s.\obi;[M] \subseteq [M];\ex_t.\obi;[M]$ holds. 

We also know that $\ex_s.\lrf = \ex_t.\lrf$ and $\ex_s.\co = \ex_t.\co$ hold.

We also know that $\irr(\ex_t.\ob)$ holds. 

Therefore $\irr(\ex_t.\ob)$ also holds.

\bigskip

We know that $\ex_t.\rmw = \ex_s.\rmw$, $\ex_s.\lrf = \ex_t.\lrf$, $\ex_s.\fr = \ex_t.\fr$, $\ex_s.\co = \ex_t.\co$ hold.
We also know that $\ex_t$ satisfies (atomic). 
Therefore $\ex_s$ also satisfies (atomic).

\bigskip

We already know $\ex_s.\co = \ex_t.\co$ and therefore $\behav(\ex_s) = \behav(\ex_t)$.
\end{proof}

\subsection{Safe eliminations}
\label{app:rarelim}

We prove the following theorem for (RAR), (RAA), and (AAA) safe eliminations in \cref{fig:armatrans}(a).
\[
\inarr{
\srcp \leadsto \tgtp \implies \forall \ex_t \in \denot{\tgtp}~\exists \ex_s \in \denot{\srcp}.~\behav(\ex_t) = \behav(\ex_s) 
}
\]

\begin{proof}

We know $\ex_t$ is ARMv8 consistent. 
We define $\ex_s$ where $a\cdot b \leadsto a$ where 

(RAR) $a=\Rlab(X,v')$ and $b=\Rlab(X,v)$ or 

(RAA) $a=\Alab(X,v')$ and $b=\Rlab(X,v)$ or 

(AAA) $a=\Alab(X,v')$ and $b=\Alab(X,v)$.

\medskip

$\ex_s.\E = \ex_t.\E \cup \set{b}$

$\ex_s.\lpo = (\ex_t.\lpo \cup \set{(a,b)})^+$

$\ex_s.\lrf = \ex_t.\lrf \cup \set{(w,b) \mid \ex_t.\lrf(w,a)}$

$\ex_s.\co = \ex_t.\co$

Moreover, $[\set{a}];\ex_s.\lpo_\imm;[\set{b}];\ex_s.\dob \implies [\set{a}];\ex_t.\dob$.

We show that $\ex_s$ is ARMv8 consistent.

Assume $\ex_s$ is not consistent.

\medskip

(internal)

Asume a $\ex_s.\poloc \cup \ex_s.\ca \cup \ex_s.\lrf$ cycle. 

It implies a $\ex_t.\poloc \cup \ex_t.\ca \cup \ex_t.\lrf$ cycle as $[\set{b}];\ex_s.\fr$ implies $[\set{a}];\ex_s.\fr$, and $[\set{a}];\ex_t.\fr$. 

Therefore a contradiction and $\ex_s$ satisfies (internal).

\bigskip

(external)



We know $\dom(\ex_s.\dob);[\set{b}] \implies \dom(\ex_s.\dob);[\set{a}] = \dom(\ex_t.\dob);[\set{a}]$ hold.


Moreover, $[\set{b}].\ex_s.\dob \implies [\set{a}].\ex_t.\dob$.

Also in case of (AAA), $\codom([\set{b}];\ex_s.\bob) = \codom([\set{a}];\ex_s.\bob) \setminus \set{b} = \codom([\set{a}];\ex_t.\bob)$ hold.

Hence $\ex_s.\ob \subseteq \ex_t.\ob$. 

We know $\irr(\ex_t.\ob)$ holds.

Therefore a contradiction and $\ex_s$ satisfies (external).

\bigskip

(atomicity)

From definition $\ex_s.\rmw = \ex_t.\rmw$, $\ex_s.\coe = \ex_t.\coe$, and $\ex_s.\fre = \ex_t.\fre$ hold.

Therefore $\ex_s$ preserves atomicity as $\ex_t$ preserves atomicity.

\bigskip

Moreover, $\behav(\ex_s) = \behav(\ex_t)$ holds as $\ex_s.\co = \ex_t.\co$ holds.

\end{proof}

\subsection{Access strengthening}

We prove the following theorem for (R-A) from \cref{fig:armatrans}(a).
\[
\inarr{
\srcp \leadsto \tgtp \implies \forall \ex_t \in \denot{\tgtp}~\exists \ex_s \in \denot{\srcp}.~\behav(\ex_t) = \behav(\ex_s) 
}
\]

\begin{proof}

We know $\ex_t$ is ARMv8 consistent. 
We define $\ex_s$ where $a \leadsto b$ where $a=\Rlab(X,v)$ and $b=\Alab(X,v)$.

$\ex_s.\E = \ex_t.\E \cup \set{a} \setminus \set{b}$

$\ex_s.\lpo = \ex_t.\lpo \cup \set{(e,a) \mid \ex_t.\lpo(e,b)} \cup \set{(a,e) \mid \ex_t.\lpo(b,e)}$

$\ex_s.\lrf = \ex_t.\lrf \cup \set{(w,a) \mid \ex_t.\lrf(w,b)}$

$\ex_s.\co = \ex_t.\co$

We show that $\ex_s$ is ARMv8 consistent.

Assume $\ex_s$ is not consistent.

\medskip

(internal)

Asume a $\ex_s.\poloc \cup \ex_s.\ca \cup \ex_s.\lrf$ cycle. 

It implies a $\ex_t.\poloc \cup \ex_t.\ca \cup \ex_t.\lrf$ cycle which is a contradiction and hence $\ex_s$ satisfies (internal).

(external)


We know $\dom(\ex_s.\ob);[\set{a}] = \dom(\ex_s.\ob);[\set{b}]$ and $[\set{a}];\codom(\ex.\lpo) = [\set{b}];\codom(\ex_t.\bob)$. 

Hence $\ex_s.\ob \subseteq \ex_t.\ob$. 

We know $\irr(\ex_t.\ob)$ holds.

Therefore a contradiction and $\ex_s$ satisfies (external).

\bigskip

(atomicity)

From definition $\ex_s.\rmw = \ex_t.\rmw$, $\ex_s.\coe = \ex_t.\coe$, and $\ex_s.\fre = \ex_t.\fre$ hold.

Therefore $\ex_s$ preserves atomicity as $\ex_t$ preserves atomicity.

\bigskip

Moreover, $\behav(\ex_s) = \behav(\ex_t)$ holds as $\ex_s.\co = \ex_t.\co$ holds.

\end{proof}

\clearpage


\section{Fence Elimination}
\label{sec:felim}

%
%
%

\subsection{Fence Elimination in x86}
\label{app:xfenceelim}

%
%
%

%
%

We restate the theorem on x86 fence elimination.

\xfenceelim*

\begin{proof}

We show:
\[
\inarr{
\srcp \leadsto \tgtp \implies \forall \ex_t \in \denot{\tgtp}~\exists \ex_s \in \denot{\srcp}.~\behav(\ex_t) = \behav(\ex_s) 
}
\]

Given a $\ex_t \in \denot{\tgtp}$ we define $\ex_s \in \srcp$ by introducing the corresponding fence event $\event$ such that  
for all events $w \in \ex_s.\WUlab \cup \ex_s.\F$,
\begin{itemize}
\item if $(w,\event) \in \ex_s.\mo^?;\ex_s.\xhb$ holds then $\ex_s.\mo(w,\event)$. 
\item Otherwise $\ex_s.\mo(\event,w)$.
\end{itemize}

We know $\ex_t$ is consistent. 

Now we show $\ex_s$ is consistent.

We prove by contradiction.

(irrHB)
Assume $\ex_s$ has $\ex_s.\xhb$ cycle. 
We know the incoming and outgoing edges to $\event$ are $\ex_s.\lpo$ edges and therefore 
$\ex_t.\xhb$ already has a cycle. 
However, we know $\ex_t.\xhb$ is irreflexive. Hence a contradiction and $\ex_s.\xhb$ is irreflexive.

\bigskip

(irrMOHB)
Assume $\ex_s$ has $\ex_s.\mo;\ex_s.\xhb$ cycle. 
We already know that $\ex_t.\mo;\ex_t.\xhb$ is irreflexive. 
Therefore the cycle contains $\event$. 
Two possiblilities: 

\Case{$\event \in \dom(\ex_s.\xhb)$ and $\event \in \codom(\ex_s.\mo)$}

Suppose $\ex_s.\xhb(\event,w)$ and $\ex_s.\mo(w,\event)$. 
However, from definition we already know $\ex_s.\xhb(\event,w) \implies \ex_s.\mo(\event,w)$ when $w \in \ex_s.\WUlab \cup \ex_s.\F$. 
Hence a contradiction and $\ex_s.\mo;\ex_s.\xhb$ is irreflexive in this case.

\Case{$\event \in \codom(\ex_s.\xhb)$ and $\event \in \dom(\ex_s.\mo)$}

Suppose $\ex_s.\xhb(w,\event)$ and $\ex_s.\mo(\event,w)$. 
However, from definition we already know $\ex_s.\xhb(w,\event) \implies \ex_s.\mo(w,\event)$ when $w \in \ex_s.\WUlab \cup \ex_s.\F$. 
Hence a contradiction and $\ex_s.\mo;\ex_s.\xhb$ is irreflexive in this case.

\bigskip

(irrFRHB)
We know $\ex_t$ does not have a $\ex_t.\fr;\ex_t.\xhb$ cycle.
We also know $\fr \subseteq (\WUlab \times \WUlab)$ and hence event $\event \in \F$ does not introduce 
any new $\ex_s.\fr;\ex_\xhb$ cycle. 
Therefore $\ex_s.\fr;\ex_\xhb$ is irreflexive. 

\bigskip

(irrFRMO)

We know $\ex_t$ does not have a $\ex_t.\fr;\ex_t.\mo$ cycle.
We also know $\fr \subseteq (\WUlab \times \WUlab)$ and hence event $\event \in \F$ does not introduce 
any new $\ex_s.\fr;\ex_\mo$ cycle. 
Therefore $\ex_s.\fr;\ex_\mo$ is irreflexive. 

\bigskip

(irrFMRP)

Assume $\ex_s$ has a $\ex_s.\fr;\ex_s.\mo;\ex_s.\rfe;\ex_s.\lpo$ cycle in $\ex_s$ cycle.

In that case the cycle is of the form:

$[\ex_s.\RUlab];\ex_s.\fr;[\ex_s.\WUlab];\ex_s.\mo;[\ex_s.\WUlab];\ex_s.\rfe;[\ex_s.\RUlab];\ex_s.\lpo;[\ex_s.\RUlab]$.

We know $\event \in \F$ and therefore does not introduce this cycle in $\ex_s$.

In that case $\ex_t$ already has a $\ex_t.\fr;\ex_t.\mo;\ex_t.\rfe;\ex_t.\lpo$ cycle which is a contradiction.

Hence $\ex_s.\fr;\ex_s.\mo;\ex_s.\rfe;\ex_s.\lpo$ cycle in $\ex_s$ is irreflexive.

\bigskip

(irrUF)

Assume $\ex_s$ has a $\ex_s.\fr;\ex_s.\mo;[\ex_s.\Ulab \cup \ex_s.\F];\ex_s.\lpo$ cycle.

Two possiblities 

\Case{$\ex_s.\fr;\ex_s.\mo;[\ex_s.\Ulab];\ex_s.\lpo$}

It implies a $\ex_t.\fr;\ex_t.\mo;[\ex_t.\Ulab];\ex_t.\lpo$ cycle. 

However, we know $\ex_t$ satisfies (irrUF) and hence a contradiction. 

\Case{$\ex_s.\fr;\ex_s.\mo;[\ex_s.\F];\ex_s.\lpo$}

It implies a $[\ex_s.\RUlab];\ex_s.\fr;[\ex_s.\WUlab];\ex_s.\mo;[\ex_s.\F];\ex_s.\lpo;[\ex_s.\RUlab]$ cycle 
created by the introduced event $\event \in \F$.

It implies $[\ex_s.\RUlab];\ex_s.\fr;[\ex_s.\WUlab];\ex_s.\mo;[\set{\event}];\ex_s.\lpo;[\ex_s.\RUlab]$

From definition, we know $[\ex_s.\WUlab];\ex_s.\mo;[\set{\event}]$ when $[\ex_s.\WUlab];\ex_s.\mo^?;\ex_s.\xhb;[\set{\event}]$ holds.

Thus $[\ex_s.\RUlab];\ex_s.\fr;[\ex_s.\WUlab];\ex_s.\mo;[\set{\event}];\ex_s.\lpo;[\ex_s.\RUlab]$

$\implies [\ex_s.\RUlab];\ex_s.\fr;[\ex_s.\WUlab];\ex_s.\mo^?;[\ex_s.\WUlab];\ex_s.\xhb;[\set{\event}];\ex_s.\lpo;[\ex_s.\RUlab]$

\begin{flalign*}
\implies & [\ex_s.\RUlab];\ex_s.\fr;[\ex_s.\WUlab];\ex_s.\mo^?;[\ex_s.\WUlab]; & \\
& (\ex_s.\xhb^?;[\ex_s.\WUlab];\ex_s.\rfe;\ex_s.\lpo \cup \ex_s.\lpo);[\set{\event}];\ex_s.\lpo;[\ex_s.\RUlab]
\end{flalign*}

\begin{flalign*}
\implies &  [\ex_s.\RUlab];\ex_s.\fr;[\ex_s.\WUlab];\ex_s.\mo^?;[\ex_s.\WUlab];\ex_s.\xhb^?;[\ex_s.\WUlab];\ex_s.\rfe;\ex_s.\lpo;[\set{\event}];\ex_s.\lpo;[\ex_s.\RUlab] & \\
& \cup [\ex_s.\RUlab];\ex_s.\fr;[\ex_s.\WUlab];\ex_s.\mo^?;[\ex_s.\WUlab];\ex_s.\lpo;[\set{\event}];\ex_s.\lpo;[\ex_s.\RUlab] &
\end{flalign*}

\begin{flalign*}
\implies &  [\ex_s.\RUlab];\ex_s.\fr;[\ex_s.\WUlab];\ex_s.\mo^?;[\ex_s.\WUlab];\ex_s.\xhb^?;[\ex_s.\WUlab];\ex_s.\rfe;\ex_s.\lpo;[\ex_s.\RUlab] & \\
& \cup [\ex_s.\RUlab];\ex_s.\fr;[\ex_s.\WUlab];\ex_s.\mo^?;[\ex_s.\WUlab];\ex_s.\lpo;[\set{\event}];\ex_s.\lpo;[\ex_s.\RUlab] &
\end{flalign*}

Now we consider two subcases:

\Subcase{$[\ex_s.\RUlab];\ex_s.\fr;[\ex_s.\WUlab];\ex_s.\mo^?;[\ex_s.\WUlab];\ex_s.\xhb^?;[\ex_s.\WUlab];\ex_s.\rfe;\ex_s.\lpo;[\ex_s.\RUlab] $}

$\implies [\ex_t.\RUlab];\ex_t.\fr;[\ex_t.\WUlab];\ex_t.\mo^?;[\ex_t.\WUlab];\ex_t.\xhb^?;[\ex_t.\WUlab];\ex_t.\rfe;\ex_t.\lpo;[\ex_t.\RUlab]$

$\implies  [\ex_t.\RUlab];\ex_t.\fr;\ex_t.\mo^?;\ex_t.\rfe;\ex_t.\lpo;[\ex_t.\RUlab]$

This is a contradiction as $\ex_t$ satisfies (irrFMRP).

\Subcase{$[\ex_s.\RUlab];\ex_s.\fr;[\ex_s.\WUlab];\ex_s.\mo^?;[\ex_s.\WUlab];\ex_s.\lpo;[\set{\event}];\ex_s.\lpo;[\ex_s.\RUlab]$}

Now we consider the $[\ex_s.\WUlab];\ex_s.\lpo;[\set{\event}];\ex_s.\lpo;[\ex_s.\RUlab]$ subsequence.

Possible cases:

\Subsubcase{$[\ex_s.\Wlab];\ex_s.\lpo;[\set{\event}];\ex_s.\lpo;[\ex_s.\Rlab]$ }

It implies $[\ex_t.\Wlab];\ex_t.\lpo;[\ex_t.\F];\ex_t.\lpo;[\ex_t.\Rlab]$ from the definition.

$\implies [\ex_t.\Wlab];\ex_t.\mo;[\ex_t.\F];\ex_t.\lpo;[\ex_t.\Rlab]$.

In that case there exists a $\ex_t.\fr;\ex_t.\mo;[\ex_t.\F];\ex_t.\lpo$ cycle. 

This is a contradiction as $\ex_t$ satisfies (irrFMRP).

\smallskip

\Subsubcase{$[\ex_s.\WUlab];\ex_s.\lpo;[\set{\event}];\ex_s.\lpo;[\ex_s.\Ulab]$ }

It implies $[\ex_t.\WUlab];\ex_t.\lpo;[\ex_t.\Ulab]$.

It implies $[\ex_t.\WUlab];\ex_t.\mo;[\ex_t.\Ulab]$ as $\ex_t$ satisfies (irrMOHB).

In this case $[\ex_s.\RUlab];\ex_s.\fr;[\ex_s.\WUlab];\ex_s.\mo^?;[\ex_s.\WUlab];\ex_s.\lpo;[\set{\event}];\ex_s.\lpo;[\ex_s.\RUlab]$

$\implies [\ex_t.\RUlab];\ex_t.\fr;[\ex_t.\WUlab];\ex_t.\mo^?;[\ex_t.\WUlab];\ex_t.\mo;[\ex_t.\Ulab]$

$\implies [\ex_t.\RUlab];\ex_t.\fr;\ex_t.\mo;[\ex_t.\Ulab]$

Hence a contradiction as $\ex_t$ satisfies (irrFRMO).

\smallskip

\Subsubcase{$[\ex_s.\Ulab];\ex_s.\lpo;[\set{\event}];\ex_s.\lpo;[\ex_s.\Rlab]$}

It implies $[\ex_t.\Ulab];\ex_t.\lpo;[\ex_t.\Rlab]$ and in consequence a 

$[\ex_t.\Rlab];\ex_t.\fr;[\ex_t.\WUlab];\ex_t.\mo^?;[\ex_t.\Ulab];\ex_t.\lpo;[\ex_t.\Rlab]$ cycle.

Now, $[\ex_t.\Rlab];\ex_t.\fr;[\ex_t.\WUlab];\ex_t.\mo^?;[\ex_t.\Ulab];\ex_t.\lpo;[\ex_t.\Rlab]$

$\implies [\ex_t.\Rlab];\ex_t.\fr;[\ex_s.\Ulab];\ex_t.\lpo;[\ex_t.\Rlab] \cup [\ex_t.\Rlab];\ex_t.\fr;\ex_t.\mo;[\ex_t.\Ulab];\ex_t.\lpo;[\ex_t.\Rlab]$

$\implies [\ex_t.\Rlab];\ex_t.\fr;\ex_t.\xhb;[\ex_t.\Rlab] \cup [\ex_t.\Rlab];\ex_t.\fr;\ex_t.\mo;[\ex_t.\Ulab];\ex_t.\lpo;[\ex_t.\Rlab]$

Hence a contradiction as $\ex_t$ satisfies (irrFRHB) and (irrUF).

\bigskip

$\behav(\ex_s) = \behav(\ex_t)$ holds as $\rloc{\ex_s.\mo} = \rloc{\ex_t.\mo}$.
\end{proof}

\subsection{Fence Elimination in ARMv8}
\label{app:armv8felim}

\begin{observation}
\label{obs:armv8fenceelim}
Let $\mbbp$ be an ARMv8 program generated from an 
x86 program following the mappings in \cref{tab:xarm}. 
In this case for all consistent execution $\ex \in \denot{\mbbp}$ 
the followings hold:

\begin{enumerate}
\item  A non-RMW load event is immediately followed by a $\dmbld$ event.

\item A non-RMW store event is immediately preceeded by a $\dmbst$ event, 

\item An RMW is immediately preceeded by a $\dmbfull$ event, 

\item An RMW is immediately followed by a $\dmbfull$ event, 
\end{enumerate}
\end{observation}

We restate \cref{th:dmbfullelim}.

\dmbfullelim*

To prove \cref{th:dmbfullelim}, we show:
\[
\inarr{
\srcp \leadsto \tgtp \implies \forall \ex_t \in \denot{\tgtp}~\exists \ex_s \in \denot{\srcp}.~\behav(\ex_t) = \behav(\ex_s) 
}
\]

\begin{proof}

Given a target execution $\ex_t \in \denot{\tgtp}$ we define a source execution $\ex_s \in \srcp$ by introducing the corresponding fence event $\event \in \dmbfull$. 

We know target execution $\ex_t$ satisfies (internal) and (atomic). 
From definition, source execution $\ex_s$ also supports (internal) and (atomic) as the respective relations remain unchanged. 

We now prove that $\ex_s$ satisfies (external). 


We prove by contradiction. 

Assume $\ex_s$ violates (external).

From definition we know that $\ex_s.\obs = \ex_t.\obs$, 
$\ex_s.\dob = \ex_t.\dob$, $\ex_s.\aob = \ex_t.\aob$. 

In that case there exists events $(a,b) \in \ex_s.\bob$ but $(a,b) \notin \ex_t.\bob$.

Considering possible cases of $a$ and $b$:

\Case{$(a,b) \in [\ex_s.\Rlab] \times [\ex_s.\E]$} Two subcases:

\Subcase{$a \notin \dom(\ex_s.\rmw)$}

It implies $(a,b) \in [\ex_t.\Rlab];\ex_t.\lpo;[\ex_t.\dmbld];\ex_t.\lpo;[\ex_t.\E]$ from Observation (1) in \cref{obs:armv8fenceelim}.

$\implies (a,b) \in [\ex_t.\Rlab];\ex_t.\bob;[\ex_t.\E]$

Hence a contradiction and $\ex_s$ violates (external).

\Subcase{$a \in \dom(\ex_s.\rmw)$}

It implies $(a,b) \in [\ex_t.\Rlab];\ex_t.\lpo;[\ex_t.\dmbfull];\ex_t.\lpo;[\ex_t.\E]$

$\implies (a,b) \in [\ex_t.\Rlab];\ex_t.\bob;[\ex_t.\E]$

Hence a contradiction and $\ex_s$ violates (external).

\Case{$(a,b) \in [\ex_s.\Wlab] \times [\ex_s.\Wlab]$}

$\implies [\ex_t.\Wlab];\ex_t.\lpo;[\ex_t.\dmbst];\ex_t.\lpo;[\ex_t.\Wlab]$ from Observation (2) in \cref{obs:armv8fenceelim}.

$\implies [\ex_t.\Wlab];\ex_t.\bob;[\ex_s.\Wlab]$ 

This is a contradiction and hence $\ex_s$ satisfies (external). 

\medskip

\Case{$(a,b) \in [\ex_s.\Wlab] \times [\ex_s.\Rlab]$}

It implies $(a,b) \in [\ex_t.\Wlab];\ex_t.\lpo;[\ex_t.\dmbfull];\ex_t.\lpo;[\ex_t.\Rlab]$ from the condition in \cref{th:armv8dmbfullelim}.

$\implies [\ex_t.\Wlab];\ex_t.\bob;[\ex_s.\Rlab]$ 

This is a contradiction and hence $\ex_s$ satisfies (external).

\bigskip

As a result, $\ex_s$ also satisfies (external) and is ARMv8 consistent.

\medskip

Moreover, we know that $\ex_s.\co = \ex_t.\co$. Hence $\behav(\ex_s)=\behav(\ex_t)$.

\end{proof}

%
%

%
%

We restate \cref{th:dmbstelim}.

\dmbstelim*

To prove \cref{th:dmbstelim}, we show:
\[
\inarr{
\srcp \leadsto \tgtp \implies \forall \ex_t \in \denot{\tgtp}~\exists \ex_s \in \denot{\srcp}.~\behav(\ex_t) = \behav(\ex_s) 
}
\]

\begin{proof}

Given a target execution $\ex_t \in \denot{\tgtp}$ we define a source execution $\ex_s \in \srcp$ by introducing the corresponding fence event $\event \in \dmbfull$. 

We know target execution $\ex_t$ satisfies (internal) and (atomic). 
From definition, source execution $\ex_s$ also supports (internal) and (atomic) as the respective relations remain unchanged. 

We now prove that $\ex_s$ satisfies (external) by showing $\ex_t.\ob = \ex_s.\ob$.

From definition we know that $\ex_t.\obs = \ex_s.\obs$, $\ex_t.\dob = \ex_s.\dob$, $\ex_t.\aob = \ex_s.\aob$. 

In that case there exists events $(a,b) \in \ex_s.\bob$ but $(a,b) \notin \ex_t.\bob$.

Considering possible cases of $a$ and $b$:

\Case{$(a,b) \in [\ex_s.\Rlab] \times [\ex_s.\E]$}

It implies $(a,b) \in [\ex_t.\Rlab];\ex_t.\lpo;[\ex_t.\dmbld \cup \ex_t.\dmbfull];\ex_t.\lpo;[\ex_t.\E]$ from Observation (1) and (4) in \cref{obs:armv8fenceelim}.

$\implies (a,b) \in [\ex_t.\Rlab];\ex_t.\bob;[\ex_t.\E]$

Hence a contradiction and $\ex_s$ violates (external).

\Case{$(a,b) \in [\ex_s.\Wlab] \times [\ex_s.\Wlab]$}

$\implies [\ex_t.\Wlab];\ex_t.\lpo;[\ex_t.\dmbst \cup \ex_t.\dmbfull];\ex_t.\lpo;[\ex_t.\Wlab]$ from the condition in \cref{th:dmbfullelim}.

$\implies [\ex_t.\Wlab];\ex_t.\bob;[\ex_s.\Wlab]$ 

This is a contradiction and hence $\ex_s$ satisfies (external). 

\Case{$(a,b) \in [\ex_s.\Wlab] \times [\ex_s.\Rlab]$}

It implies $(a,b) \in [\ex_s.\Wlab];\ex_s.\lpo;[\ex_s.\dmbfull];\ex_s.\lpo;[\ex_s.\Rlab]$ as $\ex_s.\bob(a,b)$ holds. 

It implies $(a,b) \in [\ex_t.\Wlab];\ex_t.\lpo;[\ex_t.\dmbfull];\ex_t.\lpo;[\ex_t.\Rlab]$ 

$\implies [\ex_t.\Wlab];\ex_t.\bob;[\ex_s.\Rlab]$ 

This is a contradiction and hence $\ex_s$ satisfies (external). 
\end{proof}

We restate \cref{th:dmbelim}.

\armvsfenceelim*

To prove \cref{th:dmbelim}, we show:
\[
\inarr{
\srcp \leadsto \tgtp \implies \forall \ex_t \in \denot{\tgtp}~\exists \ex_s \in \denot{\srcp}.~\behav(\ex_t) = \behav(\ex_s) 
}
\]

\begin{proof}
From the mapping scheme and the constraint in \cref{th:dmbelim}, in all cases there is a pair of $\dmb$ fences between the 
access pairs and therefore one of the fences is eliminable. 
\end{proof}



\subsection{Fence Weakening in ARMv8}
\label{app:armv8fweaken}

%
%
%
%

%
%

We restate \cref{th:dmbfullweaken}.

\dmbfullweaken*

To prove \cref{th:dmbfullweaken}, we show:
\[
\inarr{
\srcp \leadsto \tgtp \implies \forall \ex_t \in \denot{\tgtp}~\exists \ex_s \in \denot{\srcp}.~\behav(\ex_t) = \behav(\ex_s) 
}
\]

\begin{proof}

Given a target execution $\ex_t \in \denot{\tgtp}$ we define a source execution $\ex_s \in \srcp$. 

From definition we know that $\ex_t.\obs = \ex_s.\obs$, $\ex_t.\dob = \ex_s.\dob$, $\ex_t.\aob = \ex_s.\aob$. 

We know target execution $\ex_t$ satisfies (internal) and (atomic). 
From definition, source execution $\ex_s$ also supports (internal) and (atomic) as the respective relations remain unchanged. 

We now prove that $\ex_s$ satisfies (external). 


We consider following possibilities:

\Case{$(a,b) \in [\Rlab] \times [\E]$} 

In this case $(a,b) \in [\ex_s.\Rlab];\ex_s.\lpo;[\ex_s.\dmbfull];\ex_s.\lpo;[\ex_s.\E]$ 

and $(a,b) \in [\ex_t.\Rlab];\ex_t.\lpo;[\ex_t.\dmbld];\ex_t.\lpo;[\ex_t.\E]$.

It implies both $\ex_s.\bob(a,b)$ and $\ex_t.\bob(a,b)$ hold.

\Case{$(a,b) \in [\Wlab] \times [\Wlab]$} 

In this case $(a,b) \in [\ex_s.\Wlab];\ex_s.\lpo;[\ex_s.\dmbfull];\ex_s.\lpo;[\ex_s.\Wlab]$ 

and $(a,b) \in [\ex_s.\Wlab];\ex_s.\lpo;[\ex_s.\dmbst];\ex_s.\lpo;[\ex_s.\Wlab]$

It implies both $\ex_s.\bob(a,b)$ and $\ex_t.\bob(a,b)$ hold.

We know that $\ex_t.\ob$ is acyclic and hence $\ex_s.\ob$ is also acyclic.

As a result, $\ex_s$ also satisfies (external) and is ARMv8 consistent.

\medskip

Moreover, we know that $\ex_s.\co = \ex_t.\co$. Hence $\behav(\ex_s)=\behav(\ex_t)$.

\end{proof}

\clearpage


\section{Proofs and Algorithms of Robustness Analysis}
\label{app:robustness}

\subsection{SC robust against x86}




\mk*

In this case $R = [\RUlab];\lpo \cup \lpo;[\WUlab] \cup \poloc \cup \lpo;[\Flab];\lpo$.

\begin{proof}


Both x86A and SC satisfies atomicity. 

It remains to show that $(\ex.\lpo \cup \ex.\lrf \cup \ex.\fr \cup \ex.\co)$ is acyclic by contradiction.

Assume  $(\ex.\lpo \cup \ex.\lrf \cup \ex.\fr \cup \ex.\co)$ creates a cycle.

It implies $(\ex.\lpo;\ex.\eco)^+$ creates a cycle.

It implies $(([\RUlab];\lpo \cup \lpo;[\WUlab] \cup \poloc \cup \fence);\ex.\eco)^+$ has a cycle.

Considering incoming and outgoing $\eco$ edges to $\poloc$:

\begin{itemize}

\item $\ex.\rfe;[\Rlab];\ex.\poloc;[\Rlab];\ex.\fre \implies \ex.\co$

\item $[\WUlab];\ex.\poloc;[\Rlab];\ex.\fre \implies \ex.\co$

\item $\ex.\rfe;[\Rlab];\ex.\poloc;[\WUlab] \implies \ex.\co$

\end{itemize}

It implies $((\lpo \setminus \awr \cup \fence \rfe \cup \coe \cup \fre)$ has a cycle.

It implies $(\lpo \setminus \awr \cup \fence \cup \rfe \cup \co \cup \fr)$ has a cycle as $\coi \cup \fri \subseteq \lpo \setminus \awr$.

However, we know $(\lpo \setminus \awr \cup \fence \rfe \cup \co \cup \fr)$ is acyclic and therefore a contradiction.

Hence $\ex$ satisfies $\acy(\ex.\lpo \cup \ex.\lrf \cup \ex.\fr \cup \ex.\co)$.

\end{proof}

%
%
%
%
%
%
%
%
%
%
%
%
%
%
%
%
%
%
%
%
%
%
%
%
%
%
%
%
%
%

\subsection{ SC, x86 robustness against ARMv8}




\mk*

In this case $R = \poloc \cup (\aob \cup \dob \cup \bob)^+$.

\begin{proof}

Both SC and ARMv8 satisfies atomicity.

It remains to show $(\ex.\lpo \cup \ex.\lrf \cup \ex.\fr \cup \ex.\co)$ is acyclic by contradiction.

Assume  $(\ex.\lpo \cup \ex.\lrf \cup \ex.\fr \cup \ex.\co)$ creates a cycle. 

If the cycle has one or no $\epo$ edge then the cycle violates (sc-per-loc). 

Otherwise, the cycle contains two or more $\epo$ edges. 

It implies $(\ex.\epo;\ex.\eco)^+$ creates a cycle.

It implies $((\ex.\poloc \cup (\ex.\aob \cup \ex.\bob \cup \ex.\bob)^+);\ex.\eco)^+$ creates a cycle.

Considering $\ex.\poloc$ with incoming and outgoing $\ex.\eco$, possible cases:

(1) $[\Rlab];\ex.\poloc;[\Rlab];\ex.\fre;[\Wlab] \implies [\Rlab];\ex.\fre$

(2) $[\Wlab];\ex.\poloc;[\Rlab];\ex.\fre;[\Wlab] \implies [\Wlab];\ex.\coe$

(3) $[\Wlab];\ex.\rfe;[\Rlab];\ex.\poloc;[\Wlab] \implies [\Wlab];\ex.\coe$

(4) $(\ex.\coe \cup \ex.\fre);[\Wlab];\ex.\poloc;[\Wlab] \implies \ex.\coe \cup \ex.\fre$

Therefore a $((\ex.\poloc \cup (\ex.\aob \cup \ex.\bob \cup \ex.\bob)^+);\ex.\eco)^+$ cycle 
implies $((\ex.\aob \cup \ex.\bob \cup \ex.\bob)^+;\ex.\eco)^+$ cycle.

It implies an $\ex.\ob$ cycle which violates (external) and therefore a contradiction.
\end{proof}

\subsection{Proof of x86A robustness against ARMv8}
\label{app:axrobust}

%


\mk*

In this case $R=\poloc \cup (\aob \cup \bob \cup \dob)^+ \cup \awr$ 



\begin{proof}

Suppose $((\ex.\lpo \setminus \awr) \cup \fence \cup \ex.\rfe \cup \ex.\co \cup \ex.\fr)$ is a cycle.


It implies $((\ex.\lpo \setminus \awr) \cup \ex.\fence \cup \ex.\rfe \cup \ex.\coe \cup \ex.\fre)$ is a cycle as $\coi \subseteq (\ex.\lpo \setminus \awr)$ and $\fri \subseteq (\ex.\lpo \setminus \awr)$.

It implies $((\ex.\lpo \setminus \awr);\eco \cup \ex.\fence;\ex.\eco;\cup \rloc{\ex.\awr};\ex.\eco \cup \rnloc{\ex.\awr};\ex.\eco)$ cycle.

Now $\rnloc{\ex.\awr} \implies \rnloc{[\Wlab];(\ex.\lpo \setminus \awr);[\Wlab]};\ex.\eco$. 

Therefore it implies $((\ex.\lpo \setminus \awr);\ex.\eco \cup \ex.\fence \cup \rloc{\ex.\awr};\ex.\eco)$ cycle.

Following the definition of $\epo$ 

It implies $((\poloc \cup (\ex.\aob \cup \ex.\dob \cup \ex.\bob)^+);\ex.\eco)^+$ cycle.

Considering the incoming and outgoing edges for $\ex.\poloc$:

$[\Rlab];\ex.\poloc;[\Rlab];\ex.\fre \implies [\Rlab];\ex.\fre$

$[\Wlab];\ex.\rfe;[\Rlab];\ex.\poloc;[\Wlab] \implies [\Wlab];\ex.\coe$

$(\ex.\fre \cup \ex.\coe);[\Wlab];\ex.\poloc;[\Wlab] \implies (\ex.\fre \cup \ex.\coe)$

$[\Wlab];\ex.\poloc;[\Rlab];\ex.\fre \implies [\Wlab];\ex.\coe$

It implies  $(\ex.\aob \cup \ex.\dob \cup \ex.\bob)^+;\ex.\eco)^+$ creates a cycle. 

It implies $\ex.\ob$ creates a cycle which is a contradiction.
 
Therefore $\ex$ is x86A consistent.
\end{proof}

\subsection{SC, x86A, ARMv8, ARMv7mca robust against ARMv7}

%


\mk*

\subsubsection{SC-robust against ARMv7}

In this case $R=\poloc \cup \fence$.

\begin{proof}

Both SC and ARMv7 satisfies atomicity.

It remains to show that $(\ex.\lpo \cup \ex.\lrf \cup \ex.\fr \cup \ex.\co)$ is acyclic by contradiction.

Assume  $(\ex.\lpo \cup \ex.\lrf \cup \ex.\fr \cup \ex.\co)$ creates a cycle.

If the cycle has one or no $\epo$ edge then the cycle violates (sc-per-loc). 

Otherwise, the cycle contains two or more $\epo$ edges. 

It implies $(\ex.\epo;\ex.\eco)^+$ creates a cycle.

It implies $((\ex.\poloc \cup \ex.\fence);\ex.\eco)^+$ creates a cycle.


Considering the incoming and outgoing edges for $\ex.\poloc$:

$[\Rlab];\ex.\poloc;[\Rlab];\ex.\fre \implies [\Rlab];\ex.\fre$

$[\Wlab];\ex.\rfe;[\Rlab];\ex.\poloc;[\Wlab] \implies [\Wlab];\ex.\coe$

$(\ex.\fre \cup \ex.\coe);[\Wlab];\ex.\poloc;[\Wlab] \implies (\ex.\fre \cup \ex.\coe)$

$[\Wlab];\ex.\poloc;[\Rlab];\ex.\fre \implies [\Wlab];\ex.\coe$

It implies  $(\ex.\fence;\ex.\eco)^+$ creates a cycle.

Now we consider $[\codom(fence)];\ex.\eco;[\dom(fence)]$ path.

Possible cases:

\Case{$[\Rlab];\ex.\eco;[\Rlab]$}

It implies $\ex.\fre;\ex.\rfe$.

\smallskip

\Case{$[\Rlab];\ex.\eco;[\Wlab]$}

It implies $\ex.\fre$

\smallskip

\Case{$[\Wlab];\ex.\eco;[\Wlab]$}

It implies $\ex.\coe$

\smallskip

\Case{$[\Wlab];\ex.\eco;[\Rlab]$}

It implies $\ex.\coe;\ex.\rfe$

Thus an $(\ex.\fence;\ex.\eco)^+$ cycle implies 

a $((\ex.\coe \cup \ex.\fre);\ex.\rfe^?;\ex.\fence)^+$ cycle.

It implies a $\prop^+$ cycle which violates (propagation).

Hence a contradiction and therefore SC is preserved.

%
%
%
%
%
%
%
%
%
%
%
%

\end{proof}

\subsubsection{ x86A robust against ARMv7}

%


\mk*

In this case $R = \poloc \cup \fence \cup \awr$.


\begin{proof}

Suppose $((\ex.\lpo \setminus \awr) \cup \ex.\rfe \cup \ex.\co \cup \ex.\fr)$ is a cycle.

It implies $((\ex.\lpo \setminus \awr) \cup \ex.\rfe \cup \ex.\co \cup \ex.\fr)$ is a cycle.

It implies $((\ex.\lpo \setminus \awr) \cup \ex.\fence \cup \ex.\rfe \cup \ex.\coe \cup \ex.\fre)$ is a cycle as $\coi \subseteq \aww$ and $\fri \subseteq (\ex.\lpo \setminus \awr)$.

It implies $((\ex.\lpo \setminus \awr);\ex.\eco \cup \ex.\fence;\ex.\eco \cup \rloc{\ex.\awr};\ex.\eco \cup \rnloc{\ex.\awr};\ex.\eco)$ cycle.

Now $\rnloc{\ex.\awr} \implies \rnloc{[\Wlab];(\ex.\lpo \setminus \awr);[\Wlab]};\ex.\eco$. 

Therefore it implies $((\ex.\lpo \setminus \awr);\ex.\eco \cup \ex.\fence;\ex.\eco \cup \rloc{\ex.\awr};\ex.\eco)$ cycle.

It implies $((\poloc \cup \fence);\ex.\eco)^+$ cycle following the definition of $\epo$. 

Considering the incoming and outgoing edges for $\ex.\poloc$:

$[\Rlab];\ex.\poloc;[\Rlab];\ex.\fre \implies [\Rlab];\ex.\fre$

$[\Wlab];\ex.\rfe;[\Rlab];\ex.\poloc;[\Wlab] \implies [\Wlab];\ex.\coe$

$(\ex.\fre \cup \ex.\coe);[\Wlab];\ex.\poloc;[\Wlab] \implies (\ex.\fre \cup \ex.\coe)$

$[\Wlab];\ex.\poloc;[\Rlab];\ex.\fre \implies [\Wlab];\ex.\coe$

It implies  $(\ex.\fence;\ex.\eco)^+$ creates a cycle. 

It implies $\ex.\prop$ creates a cycle which is a contradiction.
 
Therefore $\ex$ is x86A consistent.
\end{proof}

\subsubsection{ARMv7 robust against ARMv8}

%


\mk*

In this case $R= \poloc \cup [\Wlab];\lpo \cup \fence$.

\begin{proof}
%
%

We show $\ex$ is ARMv8 consistent.

(internal)

Assume a $(\ex.\poloc \cup \ex.\fr \cup \ex.\co \cup \ex.\lrf)$ cycle.

However, $\ex$ satisfies (sc-per-loc) and hence a contradiction. 

Therefore, $\ex$ satisfies (internal).

\medskip

(external)

Assume a $\ex.\ob$ cycle.

It implies $(\ex.\obs;(\ex.\aob \cup \ex.\bob \cup \ex.\dob))^+$ creates cycle.

From the definition, 

$(\ex.\aob \cup \ex.\bob \cup \ex.\dob) \subseteq \poloc \cup \fence \cup [\Wlab];\lpo$ and therefore 

$((\ex.\rfe \cup \ex.\coe \cup \ex.\fre);(\ex.\poloc \cup \fence))^+$ creates cycle.

It implies $\prop$ creates a cycle which violates (propagation).

Therefore a contradiction and $\ex$ satisfies (external).

\medskip

(atomicity)

ARMv7 execution $\ex$ satisfies (atomicity).

\medskip

Therefore $\ex$ has only ARMv8 execution.

\end{proof}

\subsubsection{ARMv7-mca robust against ARMv7}
%
%

%

\mk*

In this case $R=[\Rlab];\poloc \cup \fence;[\Rlab]$.

\begin{proof}



We show that $\ex$ satisfies (mca).

Assume $\ex$ violates (mca) and therefore a $\wo^+$ cycle.

It implies a $(\ex.\rfe;[\Rlab];\ex.\ppo;[\Rlab];\ex.\fre)^+$ cycle.

However, $[\Rlab];\ex.\ppo;[\Rlab] \subseteq \poloc \cup \fence$.

$[\Rlab];\ex.\ppo;[\Rlab] \subseteq \poloc$ violates (sc-per-loc) and therefore a contradiction.

Otherwise it implies a $(\ex.\rfe;[\Rlab];\ex.\fence;[\Rlab];\ex.\fre)^+$ cycle. 

It implies a $\ex.\prop$ cycle which violates (propagation) 

Therefore a contradiction and hence $\ex$ is ARMv7-mca consistent.
\end{proof}

\end{document}